\documentclass[12pt]{article}
\usepackage{amsmath,bm,amsfonts,color,amsthm, amssymb}
\usepackage{graphicx}
\usepackage{natbib}
\usepackage{xcolor}
\usepackage{enumitem}
\usepackage{subcaption,siunitx,booktabs}
\usepackage[citecolor=blue,urlcolor=blue,allcolors=blue]{hyperref}
\usepackage{arydshln}

\usepackage{titlesec}
\titlelabel{\thetitle.\quad}

\usepackage{tikz}
\usepackage{transparent}

\newenvironment{psmallmatrix}
  {\left(\begin{smallmatrix}}
  {\end{smallmatrix}\right)} 

\usetikzlibrary{positioning}
\usetikzlibrary{shadows}
\usepgflibrary{shapes}

\usepackage{mathtools}

\DeclarePairedDelimiter\floor{\lfloor}{\rfloor}

\usepackage{apptools}
\AtAppendix{\counterwithin{lemma}{section}}

\AtAppendix{\counterwithin{proposition}{section}}

\AtAppendix{\counterwithin{Corollary}{section}}

\usepackage{titlesec}

\titleformat*{\section}{\large\bfseries}
\titleformat*{\subsection}{\large\bfseries}
\titleformat*{\subsubsection}{\large\bfseries}
\titleformat*{\paragraph}{\large\bfseries}
\titleformat*{\subparagraph}{\large\bfseries}

\usepackage{titlesec}
\titlelabel{\thetitle\quad}

\theoremstyle{definition}

\newtheorem{mytheorem}{Theorem}
\numberwithin{mytheorem}{section}
\usepackage{setspace}
\newcommand*{\QEDA}{\hfill\ensuremath{\square}}

\newtheorem{myremark}{Remark}
\numberwithin{myremark}{section}

\numberwithin{mycorollary}{section}

\newtheorem{myassumption}{Assumption}
\numberwithin{myassumption}{section}

\numberwithin{definition}{section}

\begin{document}

\def\spacingset#1{\renewcommand{\baselinestretch}%
{#1}\small\normalsize} \spacingset{1}

\title{\bf Properties of Test Statistics for Nonparametric Cointegrating Regression Functions Based on Subsamples}
  \author{Sepideh Mosaferi\thanks{
    Contact the corresponding author at \href{mailto:smosaferi@umass.edu}{smosaferi@umass.edu}}\hspace{.2cm}\\
    University of Massachusetts Amherst\\
    and \\
    Mark S. Kaiser \\
    Iowa State University \\
    and \\
    Daniel J. Nordman \\
    Iowa State University }

  \date{}
    
  \maketitle


\bigskip
\begin{abstract}
Nonparametric cointegrating regression models have been extensively used in financial markets, stock prices, heavy traffic, climate data sets, and energy markets. Models with parametric regression functions can be more appealing in practice compared to non-parametric forms, but do result in  potential functional misspecification. Thus, there exists a vast literature on developing a model specification test for parametric forms of regression functions. In this paper, we develop two test statistics which are applicable for the endogenous regressors driven by long memory and semi-long memory input shocks in the regression model. 
The limit distributions of the test statistics under these two scenarios are complicated and cannot be effectively used in practice.  To overcome this difficulty, we use the subsampling method and compute the test statistics on smaller blocks of the data to construct  their empirical distributions. Throughout, Monte Carlo simulation studies are used to illustrate the properties of test statistics. We also provide an empirical example of relating gross domestic product to total output of carbon dioxide in two European countries. 
\end{abstract}

\noindent%

\spacingset{1.5} 

\section{Introduction} \label{Intro}

In this article we consider the following nonlinear regression model
\begin{equation} \label{reg}
y_k=f(x_k)+u_k, \qquad k=1,...,N,
\end{equation}
where $f(.)$ is an unknown real function, and $x_k$ and $u_k$ are regressors and regression errors, respectively. In econometrics, when $x_k$ is a nonstationary time series 
(\ref{reg}) is called a nonlinear cointegrating regression.  In the literature, $x_k$ has often been assumed to be a short memory process uncorrelated with $u_k$, entailing so-called \textit{exogeneity}. Extending this to a case where $x_k$ is driven by long memory (LM) or semi-long memory (SLM) innovations depending on $u_k$, so-called \textit{endogeneity}, which has received less attention in the literature though may be anticipated in many applications.

A number of tests has been proposed to check the adequacy of the form of the regression in (\ref{reg}).
 Under the assumptions of $E(u_k|x_k)=0$ and independent observations $\{(x_k,y_k)\}$,  \cite{hardle1993comparing} (H-M) derived a test statistic which involves the $L_2$-distance between the nonparametric and parametric fits. They approximated the asymptotic behavior of the test statistic by a Gaussian distribution with mean that converges to infinity. For the same situation, \cite{horowitz2001adaptive} proposed a rate-optimal test statistic, which is uniformly consistent, and its asymptotic distribution has a mean of zero and variance of one.
 \cite{wang2012specification}  constructed a so-called \textit{self-normalized U (SNU)} test statistic based on martingale differences and although this statistic does not explicitly incorporate a nonparametric estimate of $f(x)$, it does apply a kernel weight function to residuals from the fitted hypothesized model, namely $y_k - g(x_k, \hat{\bm{\theta}})$.  These authors proposed a self-normalized version of the test statistic to remove the effect of  nuisance parameters and demonstrated that its limit distribution follows a standard normal variate. The SNU test offers an attractive procedure in applications, but it is hard to extend its theory to a case involving  endogeneity in the regressor $x_k$.
 
Subsequently, \cite{wang2016nonparametric} extended the setting of \cite{wang2012specification} to allow the equation error $u_k$ to be serially dependent and the regressors to be endogenous and driven by LM innovations.  These authors proposed a \textit{modified H-M (MHM)} test statistic in the form of \cite{hardle1993comparing} (H-M) and showed the limit law of the statistic involves the local time of fractional Brownian motion, and thus depends on a fractional differencing parameter $d$. Consequently, this statistic does not easily lend itself to use in applications.   
In the so-called semi-long memory (SLM) case, \cite{Mosaferi} have used the idea of tempering the LM innovations for the regressors, and considered a  test statistic with the form of \cite{wang2016nonparametric}. With this modification, the limit distribution for the test involves the local time of standard Brownian motion and is free of the unknown fractional differencing parameter. However, the limit distribution of the test statistic still does not have a simple form, which hampers its use in practice.  \cite{wang2020measure} proposed a so-called \textit{Portmanteau (P)} test statistic that is appropriate under LM and endogeneity when the equation errors $u_k$ are assumed to follow an autoregressive process.  When the order of the process is known the statistic has a simple chi-squared limiting distribution. 

To the best of our knowledge, there has not been a successful test to effectively examine the form of the regression function in (\ref{reg}) when the regressors are endogenous with LM or SLM structure, and the error terms $u_k$ are general.
A central purpose of this article is to modify the subsampling method of \cite{politis1994large} so that it can be used to determine reference distributions for the SNU and MHM test statistics to make their use more practical  for the class of endogenous regressors with LM or SLM input shocks.
Both of the aforementioned test statistics do not assume any particular structures for the error process $u_k$, and their limiting distributions have complex forms and are impractical for use in applications.

The crux of the subsampling method is to recompute the test statistics on smaller blocks or ``\textit{subsamples}" of the observed data to construct empirical distributions of the test estimates across data subsamples. A complication, however, is that standard subsampling uses 
a common form of sequential data blocks, which are motivated by stationary time series (cf.~\citealp{politis1999subsampling}),  and such subsamples require modification to handle the   non-stationary processes considered here.  
Under appropriate conditions, these empirical distributions can then approximate the sampling distributions of the test statistics to make them applicable for actual problems.

The remainder of the article is organized as follows.  
 Our models and main assumptions are presented in Section \ref{Models}, and  Section \ref{Tests} explains the SNU, MHM, and P test statistics.    
In Section \ref{Method}, we describe and establish the subsampling methodology for the SNU and MHM test statistics. Through the use of   Monte Carlo simulation studies we examine the behaviors of the test statistics in terms of size and power in Section \ref{Simulation}. 
In Section \ref{Carbon}, we apply the procedures in testing hypothesized forms of cointegrated regressions to relate carbon dioxide emissions (CO$_2$) to gross domestic product (GDP) in two developed countries. Concluding remarks are contained in Section \ref{Conclusion}.
 Technical  proofs and additional simulation results are given in the supplementary material. 
Throughout the paper, we use  $\xrightarrow{D}$ and  $\xrightarrow{P}$ to denote  convergence in distribution and probability, repsectively. Also,  i.i.d.~means independent and identically distributed, $\floor{.}$ is the floor function, and $f(x) \sim g(x)$ denotes asymptotic equivalence of two generic non-zero functions, i.e.,  the ratio $f(x)/g(x)\rightarrow 1$ as $x\to \infty$.  
We denote the 
fractional differencing parameter for LM or SLM processes as  $d$, and the tempering parameter for SLM processes as $\lambda$.

\section{Covariate Process Models and Assumptions} \label{Models}

In this section we present two models to be used for the covariate processes $x_k$ in (\ref{reg}), and several technical assumptions that will be needed to verify the efficacy of subsampling to approximate sampling distributions of statistics that incorporate these processes.

In model (\ref{reg}), we let the regressors $x_k=\sum_{j=1}^{k}X(j)$  be a partial sum 
of input shocks $X(j)$, with possible LM input shocks  denoted as
 $X(j)\equiv X_d(j)$, or possible SLM input shocks denoted as $X(j)\equiv X_{d, \lambda}(j)$.    
To define the shocks, for each integer $k \geq 0$, let $\phi(d,k) \sim k^{d-1} \rho(k)$ to denote coefficients $\phi(d,0) \neq 0$, and $\rho(k)$ is a function slowly varying at $\infty$.  Here, $d \in (0,1/2)$ is called the fractional differencing parameter.  Then define, \begin{itemize}
\item \textbf{LM:} $X_d(j)=\sum_{k=0}^{\infty} \phi(d,k) \xi(j-k)$, and
\item \textbf{SLM:} $X_{d,\lambda}(j)=\sum_{k=0}^{\infty} e^{-\lambda k} \phi(d,k) \xi(j-k)$, 
\end{itemize}
where $\{\xi(j)\}$ is an i.i.d. noise with $\mathbb{E}\xi(0)=0$ and $\mathbb{E}\xi^2(0)=1$. In the above, $\lambda >0$ represents a tempering parameter in the SLM case.

To incorporate endogeneity, we take $\xi(k)$ from the above expressions and let $\eta_k=(\xi(k),\epsilon(k))'$ be a sequence of random vectors with $\mathbb{E}(\eta_0)=0$ and $\mathbb{E}(\eta_0 \eta_0')=\Sigma$ such that
\begin{equation*}
\Sigma \equiv \begin{pmatrix}1 & \mathbb{E}(\xi(0) \epsilon(0)) \\
\mathbb{E}(\epsilon(0) \xi(0)) & \mathbb{E}(\epsilon(0)^2) \end{pmatrix},
\end{equation*} 
where $\mathbb{E}(\xi(0) \epsilon(0)) \neq 0$.
We assume the characteristic function $\varphi(t)$ of $\xi(0)$ satisfies the integrability condition $\int_\mathbb{R} (1+|t|) |\varphi(t)| dt < \infty$, which ensures smoothness in the corresponding density. Now, we set some assumptions as follows. 

\begin{myassumption} \label{assum.temp}
The tempering parameter $\lambda \equiv \lambda_N>0$ in SLM depends on $N$ and satisfies $\lambda \rightarrow 0$ and $N \lambda \rightarrow \infty$ as $N \rightarrow \infty$.  This is the strongly tempered case (see \citealp{sabzikar2020}). 
\end{myassumption} 
\begin{myassumption} \label{assum.corr}
For equation errors in the model, let $u_k=\sum_{j=0}^{\infty} \psi_j \eta_{k-j}$ for $\psi_j=(\psi_{j1},\psi_{j2})$, and assume
\begin{itemize}
\item[(a)] $\sum_{j=0}^{\infty}\psi_j \neq 0$ where $\sum_{j=0}^{\infty}j^{1/4}(|\psi_{j1}|+|\psi_{j2}|)< \infty$ holds.
\item [(b)] $\mathbb{E}||\eta_0||^{\alpha} < \infty$ for some $\alpha>2$ holds.
\end{itemize}
\end{myassumption}
Assumption \ref{assum.corr} implies that $\mathbb{E}(u_0^2)=\sum_{j=0}^{\infty} \psi_j \Sigma \psi_j'$ and $cov(u_k,x_k) \neq 0$. 
\begin{myassumption} \label{assum.D[0,1].WP}
If $x_k = \sum_{j=1}^{k} X_{d}(j)$ from a LM process, 
$\{x_k\}_{k \geq 1}$ is a stochastic process such that the following weak convergence applies on $D[0,1]$: as $N\to \infty$, 
\begin{equation}\label{strongxN}
\frac{x_{\floor{Nt}}}{d_N} \xrightarrow{D} B_{d+1/2}(t), \; \mbox{$t\in [0,1]$ \quad\text{under LM}},
\end{equation}
 where scaling $d_N:=[\mathbb{E}(x^2_N)]^{1/2}$  asymptotically has the form of $d_N \sim \rho(N) N^{d+1/2} \sqrt{c_d}$ as $N \rightarrow \infty$ with $c_d= \frac{1}{d(1+2d)} \int_{0}^{\infty} \{x(x+1)\}^{-(1-d)}dx$. Above $B_{d+1/2}(t)$ denotes  fractional Brownian motion with parameter $d+1/2$; see \cite{wang2016nonparametric}.  
\end{myassumption}

\begin{myassumption} \label{assum.D[0,1].MK}
If $x_k = \sum_{j=1}^{k} X_{d,\lambda}(j)$ from a SLM process,
$\{x_{k}\}_{k \geq 1}$ is a stochastic process such that the following weak convergence applies on $D[0,1]$: as $N\to \infty$,
\begin{equation} \label{weak.lambda}
\frac{x_{\floor{Nt}}}{d_{N}} \xrightarrow{D} B(t),  \; \mbox{$t\in [0,1]$ \quad \text{under SLM}},
\end{equation}
  where scaling $d_{N}:=[\mathbb{E}(x^2_{N})]^{1/2}$  asymptotically has the form of $d_{N} \sim \sqrt{N}/\lambda^d$ as $N \rightarrow \infty$.  Above $B_{}(t)$ denotes standard Brownian motion; see \cite{Mosaferi}.
\end{myassumption}

Assumptions \ref{assum.temp}-\ref{assum.corr} are basic, while Assumptions \ref{assum.D[0,1].WP}-\ref{assum.D[0,1].MK} are mild for prescribing limit behaviors in standardized partial sums of regressors.     
Additionally, note that in Assumption \ref{assum.D[0,1].MK} under SLM, we have standard Brownian motion $B(t)$ in (\ref{weak.lambda}), which does not depend on the fractional differencing parameter $d$ and is applicable for any $d>0$. This is unlike  Assumption \ref{assum.D[0,1].WP}, in which fractional Brownian motion $B_{d+1/2}(t)$ appears in (\ref{strongxN}).  In addition,  by tempering the coefficients $\phi(d,k)$ under SLM setting, we can extend the range of the fractional differencing parameter $d$ from $(0, 1/2)$ to $(0, \infty)$. 

\begin{myremark}
A tempered linear process that we consider is the autoregressive tempered fractionally integrated moving average process,  denoted as ARTFIMA. ARTFIMA models extend  earlier work on tempered fractionally integrated (TFI) models given by \cite{giraitis2000stationary} and can capture aspects of low frequency activity of time series better than ARFIMA models.
\end{myremark}
\begin{myremark}
In practice under the SLM setting, when the tempering parameter $\lambda$ is fixed and does not depend on the sample size $N$, one can use Whittle estimation or maximum likelihood estimation to estimate both the unknown parameters $d$ and $\lambda$. These estimators are strongly consistent under quite general conditions (see \citealp{sabzikar2019parameter}). On the other hand, when the parameter $\lambda$ depends on the sample size $N$, estimating $\lambda$ is a complex problem and can be pursued by developing confidence intervals (see Remark 3.7 in \cite{sabzikar2020} for a related argument as well as Section 5 in \cite{sabzikar2018invariance}). 
\end{myremark}

\section{Background on Test Statistics} \label{Tests}

We consider in detail several versions of three test statistics to be applied to the hypothesis that  $f(x) = g(x,\theta)$ in (\ref{reg}) for some known parametric function $g$. 

\subsection{Modified H{\"a}rdle and Mammen (MHM) test statistic}\label{Test2016} 

The MHM test is intended for situations that include LM shocks to the covariate process and endogeneity, and has a kernel-smoothed form given  by
\begin{equation} \label{test2016}
T_N := \int_{\mathbb{R}} \Big\{\sum_{k=1}^{N} K\Big[\frac{x_k-x}{h} \Big] \hat{u}_k\Big\}^2 \pi(x) dx,
\end{equation}
where $\hat{u}_k = y_k-g(x_k,\hat{\theta}_N)$ denote residuals based on a parametric estimator $\hat{\theta}_N$  of $\theta$  obtained through some optimization procedure such as minimizing $Q_N(\theta)=\sum_{k=1}^{N}(y_k-g(x_k,\theta))^2$. 
Note that the estimated regression is from a parametric procedure, but the test statistic (\ref{test2016}) depends on choosing a kernel smoothing function $K$ not involved in that estimation.   In (\ref{test2016})
  $\pi(x)$ denotes a positive weight function, which is integrable such that $K(x)\pi(x)$ has a compact support. We will  use  a Gaussian kernel for $K$ in what follows.
   
Under LM input shocks,  \cite{wang2016nonparametric} established a limit distribution for a normalized
version of $T_N$ given by $\tau_N^{-1}T_N$, which uses scaling $\tau_N:=Nh/d_N$ with $d_N$ as defined in Assumption \ref{assum.D[0,1].WP}.  The result involves certain bandwidth conditions   
 that we also suppose. Assume that, for the bandwidth $h$, it holds that $\tau_N \rightarrow \infty$ and $\tau_N h^{2\gamma} \rightarrow 0$ as $N\to\infty$, where $\gamma \in (0,1]$.  Also assume that for a small enough  $\delta_0$, $\tau_N N^{-\delta_0} \rightarrow \infty$.  Then, under the hypothesis that $f(x)=g(x, \theta)$, the normalized test statistic converges as 
\begin{equation} \label{selfnormTest.WP}
\tau_N^{-1} T_N \xrightarrow{D} d^2_{(0)} L_{B_{d+1/2}}(1,0), \quad \text{as} \quad N \rightarrow \infty,
\end{equation} 
where $L_{B_{d+1/2}}(1,0)$ denotes a local time random variable with respect to fractional Brownian motion $B_{d+1/2}$ and   
$d^2_{(0)}=\mathbb{E}(u^2_0) \int_{\mathbb{R}} K^2(s)ds \int_{\mathbb{R}}\pi(x)dx$ is a process constant; see Theorem 3.1 of \cite{wang2016nonparametric}.  The limit distribution in   (\ref{selfnormTest.WP}) depends on the fractional differencing parameter $d$ under LM and is not simple to use directly.

If we consider SLM shocks rather than LM ones, \cite{Mosaferi} have shown that the same test statistic form $T_N$  in (\ref{test2016}) with a modified  normalization factor $\tau_N$ has a somewhat simpler limit distribution.   Under SLM, the scaling becomes
$\tau_N:=Nh/d_{N}=\sqrt{N} \lambda^d h$ with $d_{N}$ as in Assumption \ref{assum.D[0,1].MK}.  Under the same bandwidth conditions from the LM case (e.g., $\tau_N h^{2\gamma} \rightarrow 0$, $\tau_N N^{-\delta_0} \rightarrow \infty$) and the hypothesis that $f(x) = g(x, \theta)$  the normalized test statistic $\tau_N^{-1} T_{N}$ converges as, 
\begin{equation} \label{selfnormTest.MK}
\tau^{-1}_N T_{N} \xrightarrow{D} d^2_{(0)} L_{B}(1,0), \quad \text{as} \quad N \lambda \rightarrow \infty,
\end{equation}
where   $d^2_{(0)}$ is as in (\ref{selfnormTest.WP}) and  $L_{B}(1,0)$ denotes a local time random variable with respect to standard Brownian motion $B$; see
  Theorem 5.1 of \cite{Mosaferi}. In contrast to (\ref{selfnormTest.WP}), the limit distribution with SLM processes does not depend on the  fractional differencing parameter $d$ and thus is simpler.  This limit does, however, still 
involve $d^2_{(0)}$ as well as a local time process and so is still not  easy to use in practice. 
Thus, we use the subsampling technique to approximate the  distribution of MHM test statistic under both LM and SLM versions.

\subsection{Self-Normalized U (SNU) test statistic} \label{Test2012}

\cite{wang2012specification} proposed a test statistic for assessing the   hypothesis  that $f(x) = g(x,\theta)$ in (\ref{reg}) for some known specified $g$ with unknown parameter $\theta$.
Although the limit theory for the test statistic was originally developed for nonstationary covariate processes as near unit root autoregressions that are driven by short memory errors with exogeneity, the test statistic itself may have applicability to more complex situations. 

Similar to the MHM test statistic from (\ref{test2016}),  the SNU test statistic also uses residuals  
 $\hat{u}_k\equiv y_k-g(x_k,\hat{\theta}_N)$ along with a kernel smoothing function $K$, which are combined to define the   statistic,  
\begin{equation} \label{S_N}
S_N \equiv \sum_{k,j=1, k \neq j}^{N} \hat{u}_k \hat{u}_j K\Big[\frac{x_k-x_j}{h} \Big].
\end{equation}
 As for the MHM test statistic, we will use a Gaussian kernel $K$ in all that follows.  
Dependence of $S_N$ on nuisance parameters can be removed by a self-normalization, resulting in,  
\begin{equation} \label{self-normalized}
Z_N \equiv \frac{S_N}{\sqrt{2}V_N}, \qquad  \mbox{for} \quad V_N^2 \equiv  \sum_{k,j=1, k \neq j}^{N} \hat{u}^2_k \hat{u}^2_j K^2\Big[\frac{x_k-x_j}{h}\Big].
\end{equation}

Under the conditions of a short memory covariate process with unit root or near unit root behavior and exogeneity, \cite{wang2012specification} showed that $Z_N \xrightarrow{D} N(0,1)$  as $N \to\infty$. 
For LM and SLM processes with endogeneity, limit distributions can follow under appropriate assumptions (e.g., Assumptions \ref{assum.corr}-\ref{assum.D[0,1].WP} under LM and Assumptions \ref{assum.temp}-\ref{assum.corr}, and \ref{assum.D[0,1].MK} under SLM) which we denote as,
\begin{equation} \label{complex_asymp}
Z_N \xrightarrow{D} Z_{0,LM}, \quad \text{or} \quad Z_N \xrightarrow{D} Z_{0,SLM}, \quad \text{as} \quad N \rightarrow \infty.
\end{equation}
The distributions of these limit variables can be complex because of the involvement of the kernel weights $K((x_k-x_j)/h)$ in the test statistic, and would be difficult to determine in the case of an endogenous regressor.
We assume there exists a continuous limit distribution for $Z_N$ as in (\ref{complex_asymp}), which may differ across LM and SLM processes and may not be $N(0,1)$.
Thus, we use the subsampling technique to approximate the finite reference distribution of $Z_N$ and find its related critical values. 

\subsection{Portmanteau (P) test statistic} \label{Test2020}

\cite{wang2020measure} proposed a test suitable for the framework of endogeneity with LM and SLM regressors that is based on an idea originally suggested by \cite{box1970distribution} and \cite{ljung1978measure}. This statistic was developed under an assumption that $u_k$ in (\ref{reg}) follows an AR(p) structure,
\begin{equation*}
u_k= \psi_1 u_{k-1}+\psi_2 u_{k-2}+...+\psi_p u_{k-p} + \epsilon(k).   
\end{equation*}
Let $\hat{u}_k=y_k-g(x_k,\hat{\theta}_N)$ and take $\hat{\psi}_j$ to be the least squares estimate of $\psi_j$, for $j = 1, . . . , p$, based on the assumed model taking $\hat{u}_k$ as the observed version of $u_k$, namely $\hat{u}_k= \psi_1 \hat{u}_{k-1}+\psi_2 \hat{u}_{k-2}+...+\psi_p \hat{u}_{k-p} + \epsilon(k)$. 

Set $\hat{\epsilon}(k)=\hat{u}_k-\hat{\psi}_1 \hat{u}_{k-1}-\hat{\psi}_2 \hat{u}_{k-2}-...-\hat{\psi}_p \hat{u}_{k-p}$. The P test statistic is then, 
\begin{equation} \label{eq.test2020}
\tilde{U}_N(\mathcal{L}):= N(N+2) \sum_{k=1}^{\mathcal{L}} \frac{\hat{a}_k^2}{N-k}, 
\end{equation}
for some integer $\mathcal{L} \geq 1$, where
\begin{equation*}
 \hat{a}_k=\dfrac{\sum_{t=k+1}^N \hat{\epsilon}(k) \hat{\epsilon}(t-k)}{\sum_{k=1}^{N} \hat{\epsilon}^2(k)}.
\end{equation*}
The limiting distribution of $\tilde{U}_N(\mathcal{L})$ can be approximated by $\chi^2(\mathcal{L}-p)$ for large $\mathcal{L}$.
Although the settings under which (\ref{eq.test2020}) is appropriate are restricted to autoregressive structure in the regression error terms $u_k$ of (\ref{reg}), it has a limit distribution that is easy to use in practice. In the applications, the order of the assumed autoregressive
process must be determined before the test statistic is constructed.

\section{Subsampling Methodology} \label{Method}

Subsampling is an approach for approximating unknown sampling distributions of statistics by dividing data into blocks that preserve the dependence structure of the entire process within each block.  \cite{hall1998sampling} and \cite{lahiri1993moving} have shown that subsampling works asymptotically under certain verifiable conditions, whereas block bootstrap fails for some LM processes.
Establishing the validity of subsampling for time series under strong dependency is not easy because of slowly decaying correlations, but there have been successful efforts in a number of different situations.
\cite{jach2012subsampling} used the subsampling technique to make inference about the mean of heavy-tailed LM time series.
\cite{giordano2017fast} used a fast subsampling method for estimating the distribution of signal to noise ratio statistics in nonparametric time series regression models that encompass both short and long range dependence.

Under a LM linear process which is not necessarily Gaussian, \cite{nordman2005validity} verified the consistency of subsampling for the sample mean with a deterministic normalization.  
\cite{bai2017validity} used the canonical correlation between two blocks instead of the usual $\alpha$-mixing coefficient to establish the validity of subsampling for LM Gaussian subordinated models.
\cite{betken2018subsampling} applied subsampling to a self-normalized change-point test statistic to make inference about structural breaks in long range dependent time series when the limit distribution depends on unknown parameters.  

We note that  all of the previously mentioned work on subsampling with LM assume {\it stationary} series, whereas our inference   involves {\it non-stationary} processes.  This distinction  creates complications for subsampling because standard data blocks  will not provide ``mini-copies" of the original series, which is a fundamental property in many subsampling developments, even under forms of non-stationarity (cf.~\citealp{politis1999subsampling}, p.~101).  Hence, we need to define subsamples in a modified manner to accommodate non-stationarity, which we do presently.

\subsection{Subsample construction} \label{Test2012.sub}

The null limit distributions in (\ref{selfnormTest.WP})-(\ref{selfnormTest.MK}) are theoretically determined by replacing the parametric residuals $\hat{u}_k$ in the test statistic $T_N$ from (\ref{test2016}) with original errors $u_k$.  Essentially then, subsamples should  ideally provide small-scale copies of the original data-level series $\mathcal{S}_N\equiv \{(x_1,u_1),\ldots, (x_N,u_N)\}$ for replicating   the distribution of test statistics under the null hypothesis $f(x)=g(x,\theta)$.  However,  note that  length  $b<N$ data blocks based on time windows
$\{(x_i,u_i),\ldots, (x_{i+b-1},u_{i+b-1})\}$   with starting indices $i=1,\ldots,N-b+1$,  as the standard form of blocks for subsampling (cf.~\citealp{politis1994large}),   would not provide small-scale  copies of 
$\mathcal{S}_N$.  The reason owes to the regressors $x_k $ being  non-stationary partial sums  
(e.g., $x_k=\sum_{j=1}^{k}X_d(j)$  or  $x_k=\sum_{j=1}^{k}X_{d,\lambda}(j)$).  

This issue can be repaired by uniformly ``re-setting" the regressors in the $i$-th data block as $ \{(x_i-x_{i-1},u_i),\ldots, (x_{i+b-1}-x_{i-1},u_{i+b-1})\}$ by  shifting these by the regressor $x_{i-1}$   preceding the block, $i=1,\ldots,N-b+1$; we define $x_0=0$ here.  Such   blocks do represent  distributional copies of $\mathcal{S}_N$ of length $b$.   As the pure errors $u_k$ are not observed in practice, though, we replace each $u_k$ with a residual, say 
$\tilde{u}_k \equiv y_k - \hat{f}(x_k)$, based on a full data estimator $\hat{f}(x)$ of the trend function $f(x)$.
We then define subsamples through data blocks as
\begin{equation}
\label{eqn:block}
\mathcal{S}_{i,b}\equiv \{(x_{i}-x_{i-1},\tilde{u}_{i}), ..., (x_{i+b-1}-x_{i-1},\tilde{u}_{i+b-1})\}, \quad   i=1,...,N-b+1.
\end{equation}

For clarity, in defining residuals $\tilde{u}_{k}\equiv y_k - \hat{f}(x_k)$ for use in subsamples, the  same parametric residuals 
$\tilde{u}_{k} = \hat{u}_k = y_k-g(x_k,\hat{\theta}_N)$ may be applied as appearing in the test statistic $T_N$ in (\ref{test2016}) (i.e., setting $\hat{f}(x_k) = g(x_k,\hat{\theta}_N)$).  Another possibility is the use of nonparametric residuals defined by $\tilde{u}_{k} = y_k-\hat{f}(x_k)$, where $\hat{f}(x)= \sum_{t=1}^N y_t K( (x- x_t)/h)/\sum_{t=1}^N  K( (x- x_t)/h)$ denotes a standard (full-data-based) kernel-smoothing estimator of $f(x)$; see \cite{wang2016nonparametric} for properties  under LM.    
If the underlying trend $f(x_t)$ diverges dramatically as a function of regressors $x_t$ with increasing time $t$, then the residuals $\tilde{u}_{k}$ in subsampling, particularly in subsamples $\mathcal{S}_{i,b}$ with starting point $i$ close to the maximal index $N-b+1$ in
(\ref{eqn:block}), can naturally become less precise approximations of the pure errors $u_k$.  Hence, for greater flexibility and generality in the following we consider an implementation of subsampling whereby distributional estimators can potentially be based on the first $M$ subsamples of length $b$, rather than simply
all $N-b+1$ subsamples,  for some selected number $M \leq N-b+1$ satisfying $M\to \infty$ with $b/M\to 0$ as $N\to \infty$. 

Sections~\ref{Test2016.sub} and \ref{Test2012.sub} provide subsampling results for the MHM and SNU test statistics, respectively.
To establish subsampling under both LM and SLM structures, we also use a mild mixing assumption, described next.  Under short range dependence, a standard $\alpha$-mixing condition is often used to validate subsampling (e.g., \citealp{politis1994large}). On the other hand, under long range dependence, $\alpha$-mixing may not hold. For LM Gaussian subordinated processes, \cite{bai2017validity} instead verified subsampling under a maximal correlation condition between blocks. 
Therefore, in order to unify conditions for verifying subsampling under either LM or SLM cases, 
we employ a between-block mixing coefficient denoted by $\alpha_{\ell,b}$, where $\ell$ is the distance between two blocks and $b$ is the block size (cf.~\citealp{politis1994large}, Sec.~3.1), which is defined as  
\begin{equation} \label{rho} 
\alpha_{\ell,b}=\sup_{j \in \mathbb{Z}} \{|P(A \cap B)-P(A)P(B)|, A \in \mathcal{F}_{j+1}^{j+b}, B \in \mathcal{F}_{j+\ell+1}^{j+\ell+b}\}, \quad \ell,b \geq 1,
\end{equation}
where $\mathcal{F}_m^n$ denotes the sigma field generated by
$\{z_m,...,z_n\}$ for underlying shocks/errors $z_n \equiv (X(n),u_n)$ and integers $n,m \in \mathbb{Z}$, $n \geq m$.

If a process is $\alpha$-mixing, 
as often associated with weak dependence, 
then the strong mixing coefficient $\alpha({\ell})\equiv \sup_{b \geq 1} \alpha_{\ell, b}$ goes to zero as $\ell \rightarrow \infty$ by definition, as does $\alpha_{\ell,b}$ for any block sequence $b$. However, for certain LM processes, \cite{bai2017validity}  
established subsampling by imposing an implicit condition of $\sum_{\ell=1}^{N} \rho_{\ell,b}=o(N)$, along a block size satisfying $b \rightarrow \infty$ with $b/N \to 0$ as $N \to \infty$, where $\rho_{\ell,b}$ corresponds to a type of $\rho$-mixing (or correlation-based) coefficient. Such $\rho$-mixing-based coefficients are generally stronger than those in (\ref{rho}) by $\alpha_{\ell,b} \leq \rho_{\ell,b} \leq 1$ for all $\ell , b \geq 1$; see \cite{bradley2005basic} for details on mixing. In this sense, we may accommodate both weak and strong forms of dependence in the following mild condition, based on $\alpha_{\ell,b}$, 
for establishing subsampling: 

  \begin{description}
\item
\noindent \textit{(Subsampling Condition)} \label{condition}
For a block size satisfying $b^{-1}+b/N \rightarrow 0$ as $N \rightarrow \infty$ and for any arbitrary $\epsilon \in (0,1)$, it holds that  $
\max_{[N \epsilon] \leq \ell \leq N}\alpha_{\ell,b}=o(1)$.\end{description}

\noindent 
This simply states that length $b$ stretches of underlying errors which are ``hugely distant," namely separated by at least $[N \epsilon]$ lags as a large distance  relative to a block size $b$, should act approximately independent.  This condition is natural for many error processes and entails that data blocks in subsampling can provide a type of replication (i.e., well-separated blocks are approximately independent).         
The condition is equivalent to $\sum_{\ell=1}^{N} \alpha_{\ell,b}=o(N)$ as $N \rightarrow \infty$, which is perhaps less transparent but also common in subsampling.

\subsection{Subsampling approximation of modified H-M  statistic}\label{Test2016.sub}

 To compute subsampling distribution estimators for the MHM test statistic, 
we use the first $M \leq N-b+1$ subsamples $\mathcal{S}_{i,b}$, $i=1,\ldots,M$ of length $b$ from (\ref{eqn:block}).  For each subsample, 
  we compute a version of the normalized test statistic $\tau_N^{-1} T_N$ as  
\begin{equation*}
\tau_b^{-1} T_{i,b} \equiv  \frac{d_b}{b h_b}\int_{\mathbb{R}} \Big\{\sum_{j=1}^{b} K\Big[\frac{(x_{i+j-1}-x_{i-1})-x}{h_b} \Big] \tilde{u}_{i+j-1} \Big\}^2 \pi(x) dx, 
\end{equation*} 
where the scaling $\tau_b \equiv b h_b/d_b$ represents the length $b$ analog of  $\tau_N \equiv N h/d_N$.  The empirical distribution of the subsample test statistics is given by

\begin{align*}
\hat{F}_{M,b}(x) & \equiv \frac{1}{M} \sum_{i=1}^{M} \mathbb{I} \{\tau^{-1}_b T_{i,b} \leq x \},  \quad x \in \mathbb{R}, 
\end{align*} 
where $\mathbb{I}$ denotes an indicator function, and this provides a subsampling approximation of the  distribution of the test statistic $\tau_N^{-1}T_N$ under the null hypothesis $f(x) = g(x,\theta)$.  We denote this null target distribution as $F_{N,H}(x)\equiv \text{Prob}_{H} \{\tau_N^{-1}T_N \leq x \}$, $x\in\mathbb{R}$, which converges as prescribed in      
 (\ref{selfnormTest.WP}) or  (\ref{selfnormTest.MK}) under LM or SLM, respectively. 
Theorem~\ref{theo.sub.2016} next justifies the use of subsampling estimation for the MHM test statistics.
\begin{mytheorem} \label{theo.sub.2016}
Suppose Assumptions \ref{assum.corr}-\ref{assum.D[0,1].WP} with  (\ref{selfnormTest.WP}) for LM series  or Assumptions \ref{assum.temp}-\ref{assum.corr} and \ref{assum.D[0,1].MK} with  (\ref{selfnormTest.MK}) for SLM series.   
Further assume the \textit{subsampling condition}  
along with $M^{-1}+b/M \to 0$ and $\max_{1 \leq t \leq M}  \tau_b  |f(x_t) - \hat{f}(x_t)|^2 = o_p(1)   $  as $N \to \infty$, where $M$ is the number of subsamples and $\hat{f}$ is a full data estimator of the trend $f$ in subsampling.  Then,     
\begin{equation*}
\sup_{x>0} |\hat{F}_{M,b}(x)-F_{N,H}(x)| \xrightarrow{P} 0 \quad \mbox{ as $N \to \infty$};
\end{equation*}
further, under a nonparametric estimator $\hat{f}$, the above convergence holds even if the hypothesis $f(x)=g(x,\theta)$ fails to be true.
\end{mytheorem}
From Theorem~\ref{theo.sub.2016},   subsampling    captures the   null  distribution of test statistics under either LM or SLM structures, and so provides a tractable reference for assessing evidence.  While residuals in subsampling may involve either parametric or non-parametric estimators of $f$, non-parametric-based residuals yield valid subsampling estimators even when hypothesis $f(x)=g(x,\theta)$ fails, 
which can facilitate  control of size; see the numerical studies of Section~\ref{Simulation}.  The  condition  $\max_{1 \leq t \leq M}  \tau_b  |f(x_t) - \hat{f}(x_t)|^2 = o_p(1)   $    in Theorem~\ref{theo.sub.2016} is generally mild (recall $\tau_b \equiv (b h_b /d_b $)) and this allows  added flexibility in  coordinating the block length $b$ and number $M$ of subsamples. 

For example, under parametric estimation with $\hat{f}(x)=g(x,\hat{\theta}_N)$, we might expect 
$\max_{1 \leq t \leq M} \tau_b  |f(x_t) - \hat{f}(x_t)|^2 =  O_p(  [b d_N ]/[N d_b]  h_b d_M^\delta)$ to hold for some $\delta>0$, based on typical parametric  rates 
$\|\hat{\theta}_N-\theta \|^2 = O_p( d_N/N)$ (cf.~\citealp{wang2016nonparametric}) combined with partial sum growth $\max_{1 \leq t \leq M} |x_t| = O_p(d_m)$ (Assumptions~\ref{assum.D[0,1].WP}-\ref{assum.D[0,1].MK}). Note  $[b d_N ]/[N d_b]\to 0$ and $h_b\to 0$, while $\delta>0$ here is a constant related to growth of $g(x,\theta)$, where $\delta=0$
for bounded functions.  Under non-parametric estimation,
the rate  $O_p( d_N/N )$ is replaced 
by $O_p( d_N/[N h] )$ (cf.~\citealp{wang2016nonparametric}, Sec.~2).  In any theoretical case, $\max_{1 \leq t \leq M} \tau_b  |f(x_t) - \hat{f}(x_t)|^2 =o_p(1)$ can hold  readily under  
appropriate   $M$ and $b$.  In practice, a block length   $b$ proportional to $N^{1/2}$ can provide a starting point, as this has been suggested in other subsampling developments under strong dependence 
(cf.~\citealp{bai2017validity,hall1998sampling,Zhangetal}).

\subsection{Subsampling approximation of self-normalized U statistic} \label{Test2012.sub}

To describe the subsampling estimator for the SNU test  statistic we apply the same   subsamples $\mathcal{S}_{i,b}$, $i=1,\ldots,N-b+1$, from (\ref{eqn:block}) based on (modified) length $b$ data blocks.  For purposes of generality, we again consider subsampling estimators defined by the first $M$ subsamples for some number $M$   ($M\leq N-b+1$) satisfying $M\to \infty$ with $b/M \to 0$ as $N\to \infty$.    
On each subsample $\mathcal{S}_{i,b}$,  a  statistic $Z_{i,b}$ is computed as the length $b$ analog  of the test statistic $Z_N$ in  (\ref{self-normalized}).  These subsample statistics lead to a subsampling estimator,     
\begin{equation*} 
\hat{F}_{M,b}(x)\equiv\frac{1}{M} \sum_{i=1}^M  \mathbb{I}\{Z_{i,b} \leq x \}, \quad  x \in \mathbb{R},
\end{equation*}
for approximating the distribution of the  test statistic $Z_N$ under the hypothesis $f(x)=g(x,\theta)$; we denote this target distribution as $F_{N,H}(x)\equiv \text{Prob}_{H} \{Z_N \leq x \}$, $x\in\mathbb{R}$.

Theorem \ref{theo.sub.2012} establishes   subsampling for approximating the SNU statistic in testing.

\begin{mytheorem} \label{theo.sub.2012}
Suppose Assumptions \ref{assum.corr}-\ref{assum.D[0,1].WP} under LM series or Assumptions \ref{assum.temp}-\ref{assum.corr} and \ref{assum.D[0,1].MK} under SLM series with (\ref{complex_asymp}). 
 Further assume the \textit{subsampling condition}  
along with $M^{-1}+b/M \to 0$ and $\max_{1 \leq t \leq M} |f(x_t) - \hat{f}(x_t)|^2  b^2  h_b /d_b  = o_p(1)   $  as $N \to \infty$, where $M$ is the number of subsamples and $\hat{f}$ is a full data estimator of the trend $f$ in subsampling.  Then,     
\begin{equation*}
\sup_{x>0} |\hat{F}_{M,b}(x)-F_{N,H}(x)| \xrightarrow{P} 0 \quad \mbox{ as $N \to \infty$};
\end{equation*}
further, under a nonparametric estimator $\hat{f}$, the above convergence holds even if the hypothesis $f(x)=g(x,\theta)$ fails to be true.     
\end{mytheorem}
 Under Theorem~\ref{theo.sub.2012},   subsampling  is valid for estimating the null sampling distribution of the SNU test statistic  under both LM or SLM process forms.   Theorem~\ref{theo.sub.2012} may be viewed as the subsampling counterpart of Theorem~\ref{theo.sub.2016}, and both results apply under similar assumptions.  The rate condition 
in Theorem~\ref{theo.sub.2012} regarding estimation $\hat{f}$ of the trend function is the analog 
of the same condition in Theorem~\ref{theo.sub.2016}, where the scaling $\tau_b$  in the MHM test statistic is replaced by  similar underlying scaling for the SNU test statistic.

\section{Simulation Studies} \label{Simulation}

In this section, we investigate  the behaviors  of the three test statistics through the use of  Monte Carlo studies. To keep the computational burden feasible, we assume the sample size is $N=500$ and use $M=2000$ Monte Carlo replications in each situation examined. Further simulation results for a range of sample sizes $N=\{50,100,200,500\}$ are given in the Section B of the supplementary material. We generate data sets under the model with the form of (\ref{reg}) as $y_k = f (x_k) + \sigma u_k$ and assume $\sigma=0.2$. 

The regressor process $x_k$ is defined for both LM and SLM settings following  Section \ref{Models}.
Let $u_k$ follow an AR(1) structure such that $u_k=\psi u_{k-1}+\epsilon(k)$ with $\psi=0.25$. 
In Section \ref{sec.extension}, we extend the structure of $u_k$ to other processes such as MA(1).
Also, let 
$\mathbb{E}(\epsilon(0)^2)=1$ and $\mathbb{E}(\xi(0) \epsilon(0)) \equiv r=\{0.5,1\}$ in the $\Sigma$ matrix.
Therefore, we have $(\xi(k), \epsilon(k))$ are i.i.d. $N(0,\begin{psmallmatrix}1 & r \\ r & 1 \end{psmallmatrix})$. Additional results for $r=\{-1,-0.5\}$ are included in the supplementary material.
A similar set-up has been previously used by \cite{wang2016nonparametric}, \cite{wang2020measure}, and \cite{Mosaferi}.

In construction of the test statistics (\ref{test2016}) and (\ref{self-normalized}), we use a Gaussian kernel, specifically,
\begin{equation}\label{Gkern}
K(x) = \frac{1}{\sqrt{2 \pi}} \exp \left( - \frac{1}{2} x^2 \right).
\end{equation}
Also in (\ref{test2016}) we take $\pi(x) = \mathbb{I}(-100 \leq x \leq 100)$ where $\mathbb{I}(\cdot)$ is the indicator function, and when $K$ is applied to $(x_k-x)/h$, we take the bandwidth $h$ as $h=N^{-1/3}$.
For generation of covariates, we examine values of the fractional differencing parameter as $d = \{0.1,0.2,0.3,0.4\}$ and of the tempering parameter as $\lambda=\{N^{-1/3}, N^{-1/4}, N^{-1/5}, N^{-1/6} \}$.  

\subsection{Properties of self-normalized U and modified H-M statistics} \label{test_properties}
We first examine several properties of the SNU and MHM test  statistics that do not require the use of subsampling.  The asymptotic distribution of the SNU test statistics is normal under short memory processes and exogeneity.  We conducted a Monte Carlo simulation study as described previously using LM, SLM, and endogeneity, and determined p-values for the SNU test statistic using a standard normal reference distribution. For this exercise we used the hypothesized model, $f(x_k) = \theta_0 + \theta_1 x_k$, and for data generation used $(\theta_0,\theta_1)=(0.0,1.0)$.

 Empirical distributions of the p-values are shown in Section B of the supplementary material for LM and SLM cases for several values of $d$ and $r$ that control endogeneity and memory properties.  The observed distributions  depart from what would be expected for samples from a uniform distribution on the unit interval, particularly as $d$ and $r$ increase together, indicating that a standard normal calibration is not appropriate for p-values with the SNU test statistic under LM.  Distributions for the SLM cases appear a bit more uniform, but even these p-values are clearly far from uniformity for larger values of both $d$ and $r$.
 These numerical results suggest that p-values for the SNU test statistic cannot effectively be approximated by a standard normal limit for LM or SLM cases, unlike the short-memory case, which motivates alternative subsampling approximations.

\begin{figure}[!h]
\centering
\begin{tabular}{ c }
 \includegraphics[width=0.8\textwidth]{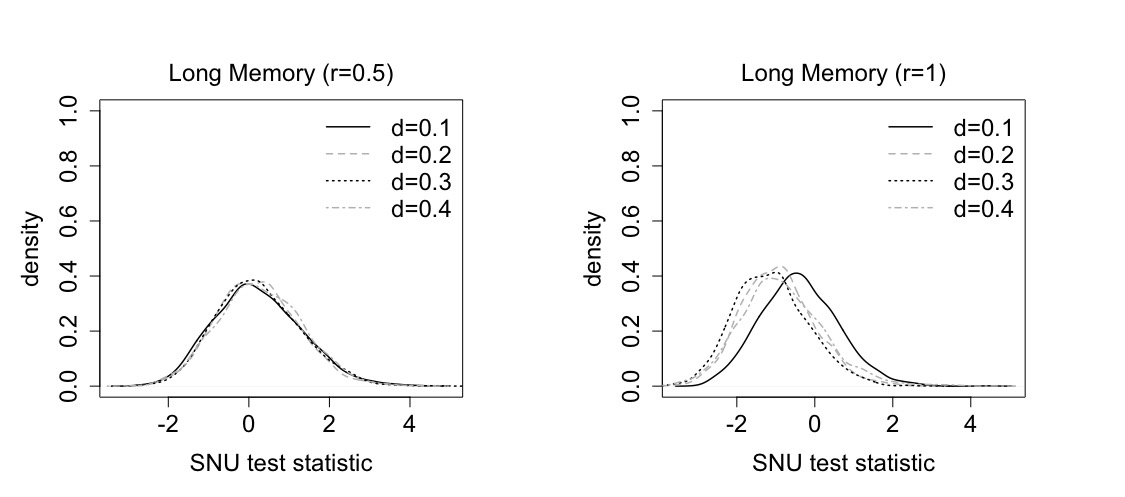} \\
 \includegraphics[width=0.8\textwidth]{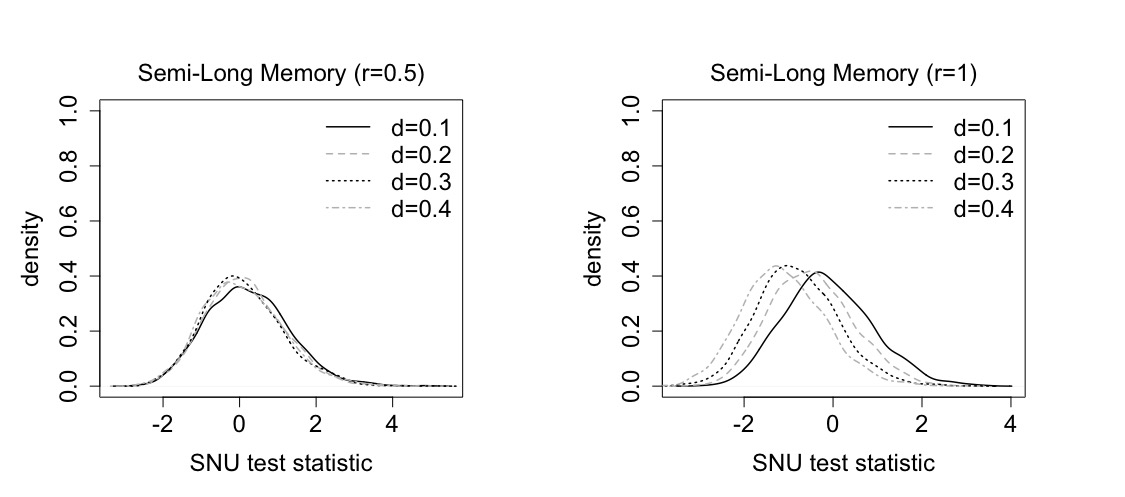}
 \end{tabular}
\caption{Comparison of Monte Carlo densities of SNU test statistic. The results are for LM (top) and SLM (bottom) cases for different values of $d$. We assume the bandwidth is $h=N^{-1/3}$ and the tempering parameter for the SLM is $\lambda=N^{-1/6}$.}
\label{Fig.Test2012}
\end{figure}

\begin{figure}[!h]
\centering
\begin{tabular}{ c }
 \includegraphics[width=0.8\textwidth]{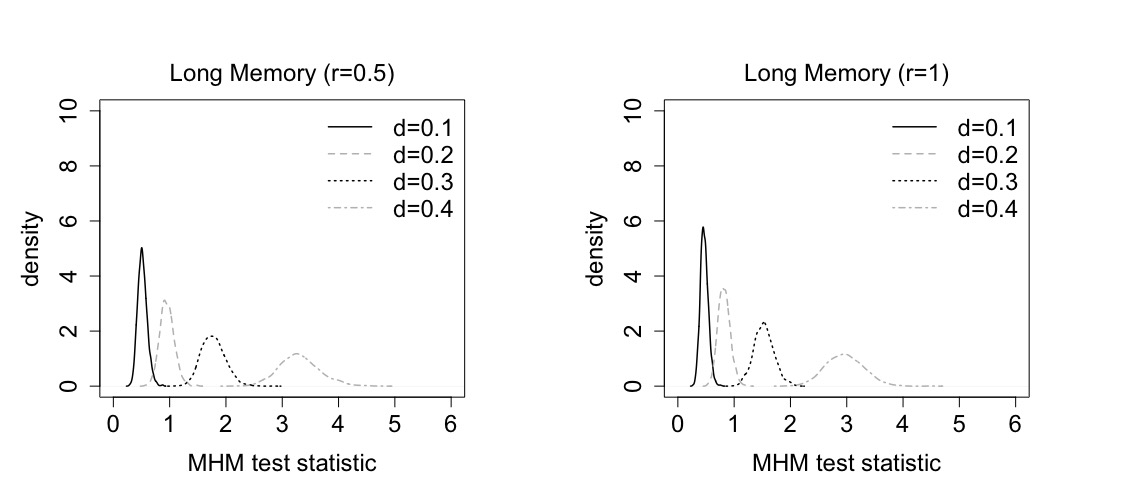} \\
 \includegraphics[width=0.8\textwidth]{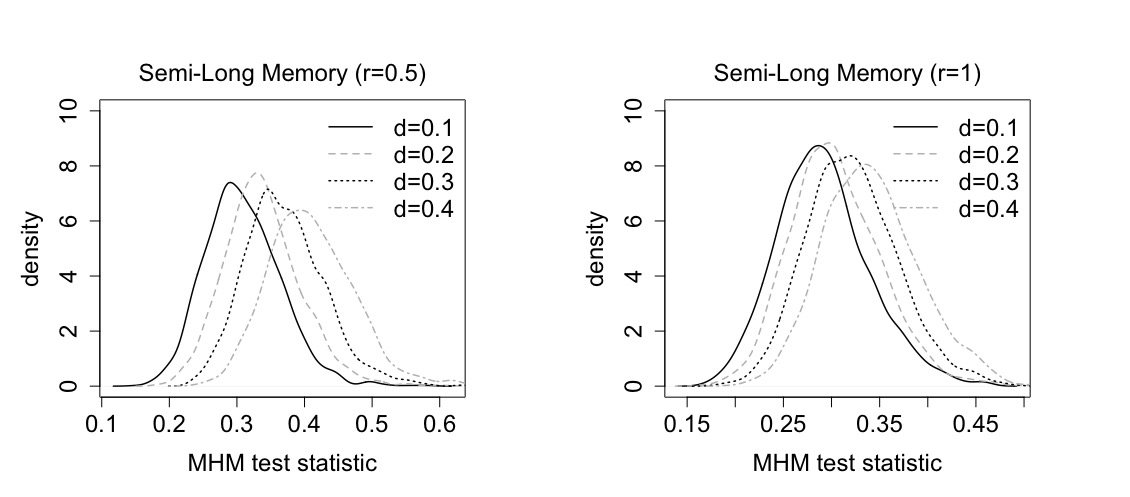}
 \end{tabular}
\caption{Comparison of Monte Carlo densities of the MHM test statistic. The results are for LM (top) and SLM (bottom) cases for different values of $d$. We assume the bandwidth is $h=N^{-1/3}$ and the tempering parameter for the SLM is $\lambda=N^{-1/6}$.}
\label{Fig.Test2016}
\end{figure}

Using the same simulation design just described, the SNU and MHM test statistics were computed for a variety of values of $r$ and $d$.  Of particular interest was the influence of the differencing parameter $d$ and the level of endogeneity $r$ on the realized distributions of these statistics. Smoothed empirical densities of the SNU statistic are shown in Figure \ref{Fig.Test2012} and we see that when there is a substantial level of endogeneity ($r=1.0$) the  differencing parameter $d$ begins to influence the distributions, pulling them quite far from a center near zero. Smoothed densities for the MHM statistic are shown in Figure \ref{Fig.Test2016} in which 
the differencing parameter $d$ 
influences all of the distributions, though that effect is considerably more pronounced for the LM version of the statistic than it is for the SLM version. The value of $d$ determines the location of these distributions, with larger values of $d$ corresponding to larger locations.  In total, these results provide further motivation for the use of subsampling to determine reference distributions.

\subsection{Empirical size and power of tests}\label{sizepowersims}

In this section we examine the behavior of the test statistics of interest, using subsampling methodology to approximate reference distributions for the SNU and MHM test statistics. There are two versions of the SNU test statistic represented in what follows, differing  based on whether the residuals used in the construction of subsamples were computed from a parametric or a nonparametric estimate of the regression function (see Section \ref{Test2012.sub}).  The approximate chi-squared limit is used as a reference distribution for the P test statistic.

To use subsampling we need to choose block lengths.  
 \cite{hall1998sampling} suggested a block length $b$ proportional to $N^{1/2}$ in simulations under strong dependence, while \cite{Zhangetal}   showed  a block order $N^{1/2}$ to be optimal for subsampling variance estimation with many LM processes defined by Gaussian subordination.  This block choice also agrees
 with Remark 4.2 of \cite{bai2017validity} as well as  
 subsampling theory of \cite{betken2018subsampling}
   under LM   for $d\in(0,1/2)$.    Based on these guideposts, we examine block lengths $b = [cN^{1/2}]$ for $c = 0.5$, 1, 2, and 4. 
With $N=500$, these turn out to be $11$, $22$, $44$, and $89$ observations long, resulting in $489$, $478$, $456$ and $411$ blocks, respectively.  
Test statistics were computed for each of these blocks for each simulated data set, and used to construct the subsampling distributions $\hat{F}_{M,b}$ as discussed in Section \ref{Method}. We present results using the maximal number $N-b+1$ of subsamples, although findings with smaller numbers of blocks were qualitatively similar and appear in supplementary material.   These blocks  were then taken to approximate the finite sampling distributions of the test statistics for the purpose of determining p-values.

 In all cases, the hypothesis tested was $f(x_k) = \theta_0 + \theta_1 x_k$.  To examine the size associated with these tests, data were generated as previously with $\theta_0=0.0$ and $\theta_1=1.0$, as well as $r=0.5$ and $r=1.0$.
The SNU and MHM test statistics differ in the range of values possible, which influences determination of p-values as being from two-sided or one-sided test procedures.  To determine the p-value for a given simulated data set, we used the following procedure for a nominal test level of $\alpha$.  Here, $Z_N$ denotes the actual value of the SNU test statistic computed from the entire data set, and similarly for the MHM test $\tau^{-1}_N T_{N}$ under LM or under SLM.

\begin{itemize}
\item[(i)]   For the SNU test statistic, we calculated a p-value by the proportion of subsample statistics $Z_{i,b}$ exceeding the observed  statistic $Z_N$ in absolute value:
\begin{align*}
P_{m} \equiv   \frac{1}{N-b+1} \sum_{i=1}^{N-b+1} \mathbb{I}\{|Z_N|< |Z_{i,b}|\}.
\end{align*}
\item[(ii)] For the MHM test statistic (which is non-negative), we computed a p-value $P_m \equiv 1 - \hat{F}_{N-b+1,b}( \tau_N^{-1}T_{N})$ as the proportion of  
subsample statistics $\tau_b^{-1}T_{i,b}$ exceeding the observed  statistic $\tau_N^{-1}T_{N}$.    
\end{itemize}
Therefore, for a given nominal size level of $\alpha$, the observed size across simulation runs is given as  $\sum_{m=1}^{2000}  \mathbb{I}\{P_m \leq \alpha\}/2000$.

In conducting the simulations reported in this section we identified a difficulty with the MHM statistic that appeared to be in the form of a finite-sample upward bias.  An implication was that using the MHM statistic resulted in a nearly $100\%$ rejection of the hypothesis when the hypothesis was, in fact, the true data generating mechanism and the test was conducted at a nominal $0.05$ level.  In order to mitigate the problem we suggest a de-biased version of the MHM test statistic based on subsampling, where the related steps are given in the Section C of the supplementary material. All further numerical results related to the MHM statistic are based on that de-biased version.
In part due to this correction, we present results separately for a combination of the SNU and P statistics, and the MHM statistic alone.

Empirical sizes of the SNU and P test statistics at the nominal level $5\%$ are given in Table \ref{Tab.size_nonintegrable}.  Tests conducted with the SNU statistic and subsampling range from  a bit conservative to a bit liberal, though there does not appear to be a clearly identifiable pattern across values of $d$ and $r$ that is consistent for all memory processes.  It does seem  that, in general, size for the SNU statistic decreases as block size increases within any level of $d$, $r$, and memory process; as may be expected under LM/SLM, nominal size is often better maintained with longer block sizes for subsampling in Table \ref{Tab.size_nonintegrable}, where non-parametric residuals may also help toward improving size control over parametric residuals in the subsampling approximation.  Empirical size estimates for the P statistic hover around the nominal level of $0.05$. 
\begin{table}[!ht]
\footnotesize
\caption{Empirical size of the SNU and P test statistics at $5\%$ level. The values in the curly brackets correspond to the 4 block sizes in the paper, and we assume $h=N^{-1/3}$.} \label{Tab.size_nonintegrable}
\centering
\setlength{\tabcolsep}{1pt} 
\renewcommand{\arraystretch}{1.15} 
\begin{tabular}{@{} ccccc @{}}
\toprule
test & $d$ $\symbol{92}$ $\hat{f}(.)$ & LM & SLM $(\lambda=N^{-1/3})$ & SLM $(\lambda=N^{-1/6})$ \\\midrule
 && \multicolumn{3}{c}{$r=0.5$} \\
 SNU & $0.1$ $\symbol{92}$ parametric & $\{0.151, 0.122, 0.099, 0.096\}$ & $ \, \{0.132, 0.105, 0.086, 0.079\}$ & 
 $\{0.148, 0.117, 0.093, 0.087\}$ \\
 
& $0.1$ $\symbol{92}$ non-parametric & $\{0.146, 0.116, 0.058, 0.016\}$ & $\{0.128, 0.103, 0.056, 0.015\}$ & $\{0.140, 0.119, 0.060, 0.019\}$  \\[0.5cm]

P & $0.1$ $\symbol{92}$ parametric & $(0.051,0.052,0.059)$ & $(0.050,0.052,0.057)$ & $(0.049,0.051,0.057)$ \\
 
 & $\mathcal{L}=(6,12,18)$ \\ \midrule
 
SNU & $0.4$ $\symbol{92}$ parametric & $\{0.144, 0.108, 0.076, 0.069\}$ & $\{0.108, 0.083, 0.070, 0.064 \}$ & $\{0.121, 0.095, 0.079, 0.073 \}$ \\
 
& $0.4$ $\symbol{92}$ non-parametric & $\{0.146, 0.103, 0.041, 0.012\}$ & $\{0.113, 0.078, 0.028, 0.005\}$ & $\{0.125, 0.085, 0.038, 0.011\}$  \\[0.5cm]

P & $0.4$ $\symbol{92}$ parametric & $(0.051,0.054,0.061)$ & $(0.051,0.053,0.059)$ & $(0.051,0.052,0.058)$ \\
 
& $\mathcal{L}=(6,12,18)$ \\ \midrule \midrule
 
 & & \multicolumn{3}{c}{$r=1$} \\
SNU & $0.1$ $\symbol{92}$ parametric & $\{0.142, 0.103, 0.086, 0.091\}$ & $\{0.113, 0.085, 0.068, 0.070\}$ & $\{0.110, 0.082, 0.070, 0.066\}$ \\
 
& $0.1$ $\symbol{92}$ non-parametric & $\{0.097, 0.074, 0.038, 0.004\}$ & $\{0.072, 0.061, 0.030, 0.005\}$ & $\{0.064, 0.054, 0.034, 0.007\}$  \\[0.5cm]

P & $0.1$ $\symbol{92}$ parametric & $(0.047,0.051,0.055)$ & $(0.045,0.049,0.053)$ & $(0.044,0.048,0.054)$ \\
 & $\mathcal{L}=(6,12,18)$ \\ \midrule
 
SNU & $0.4$ $\symbol{92}$ parametric & $\{0.236, 0.168, 0.128, 0.109\}$ & $\{0.419, 0.289, 0.217, 0.164\}$ & $\{0.287, 0.198, 0.149, 0.116\}$ \\
 
& $0.4$ $\symbol{92}$ non-parametric & $\{0.265, 0.180, 0.074, 0.016\}$ & $\{0.445, 0.309, 0.139, 0.018\}$ & $\{0.287, 0.200, 0.087, 0.007\}$  \\[0.5cm]

P & $0.4$ $\symbol{92}$ parametric & $(0.050,0.055,0.063)$ & $(0.049,0.052,0.059)$ & $(0.048,0.051,0.056)$ \\
 & $\mathcal{L}=(6,12,18)$ \\ 
\bottomrule
\end{tabular}
\end{table}

To examine power under a local departure from the hypothesis, data were simulated using $f(x)=\theta_0+\theta_1x+\rho_N |x|^{\nu}$ with $\rho_N=1/(N^{1/4+\nu/3}h^{1/4})$, with 
 $\theta_0=0.0$, $\theta_1=1.0$, and $\nu=3$. This same generating mechanism has been used in other studies (e.g., \citealp{wang2012specification}) to study power.
The procedure described previously was used to compute p-values. 
Results for the SNU and P test statistics are given in Table \ref{Tab.power_nonintegrable}. The values of power for the SNU test statistic are excellent and are usually around $100\%$ using either parametric or non-parametric residuals. The values of power for the P test statistic are smaller, particularly for $d=0.1$, but do increase somewhat for $d=0.4$ and are perhaps a bit greater when $r=1.0$ than when $r=0.5$.  This differs from the results on size for the P statistic, in which there was no discernible effect of either $d$ or $r$.  

\begin{table}[!h]
\footnotesize
\caption{Empirical power of the SNU and P test statistics at $5\%$ level. The values in the curly brackets correspond to the 4 block sizes in the paper, and we assume $h=N^{-1/3}$.} \label{Tab.power_nonintegrable}
\centering
\setlength{\tabcolsep}{1pt} 
\renewcommand{\arraystretch}{1.15} 
\begin{tabular}{@{} ccccc @{}}
\toprule
test & $d$ $\symbol{92}$ $\hat{f}(.)$ & LM & SLM $(\lambda=N^{-1/3})$ & SLM $(\lambda=N^{-1/6})$ \\\midrule
 && \multicolumn{3}{c}{$r=0.5$} \\
 SNU & $0.1$ $\symbol{92}$ parametric & $\{1.000, 1.000, 1.000, 0.996\}$ & $\{1.000, 1.000, 1.000, 0.997\}$ & $\{1.000, 1.000, 1.000, 0.998\}$ \\
 
& $0.1$ $\symbol{92}$ non-parametric & $\{1.000, 1.000, 1.000, 1.000\}$ & $\{1.000, 1.000, 1.000, 1.000\}$ & $\{1.000, 1.000, 1.000, 1.000\}$  \\[0.5cm]

P & $0.1$ $\symbol{92}$ parametric & $(0.597,0.614,0.620)$ & $(0.579,0.607,0.622)$ & $(0.570,0.601, 0.616)$ \\
& $\mathcal{L}=(6,12,18)$ \\ \midrule
 
SNU & $0.4$ $\symbol{92}$ parametric & $\{1.000, 1.000, 0.989, 0.923\}$ & $\{1.000, 1.000, 1.000, 0.978\}$ & $\{1.000, 1.000, 1.000, 0.992\}$ \\
 
& $0.4$ $\symbol{92}$ non-parametric & $\{1.000, 1.000, 1.000, 1.000\}$ & $\{1.000, 1.000, 1.000, 1.000\}$ & $\{1.000, 1.000, 1.000, 1.000\}$  \\[0.5cm]

P & $0.4$ $\symbol{92}$ parametric & $(0.888,0.857,0.829)$ & $(0.971,0.956,0.950)$ & $(0.879,0.865,0.855)$ \\
 & $\mathcal{L}=(6,12,18)$ \\ \midrule \midrule
 
 & & \multicolumn{3}{c}{$r=1$} \\
SNU & $0.1$ $\symbol{92}$ parametric & $\{1.000, 1.000, 1.000, 0.996\}$ & $\{1.000, 1.000, 1.000, 1.000\}$ & $\{1.000, 1.000, 1.000, 0.998\}$ \\
 
& $0.1$ $\symbol{92}$ non-parametric & $\{1.000, 1.000, 1.000, 1.000\}$ & $\{1.000, 1.000, 1.000, 1.000\}$ & $\{1.000, 1.000, 1.000, 1.000\}$  \\[0.5cm]

P & $0.1$ $\symbol{92}$ parametric & $(0.643,0.671,0.685)$ & $(0.653,0.686,0.703)$ & $(0.634,0.685,0.708)$ \\
 & $\mathcal{L}=(6,12,18)$ \\ \midrule
 
SNU & $0.4$ $\symbol{92}$ parametric & $\{1.000, 1.000, 0.991, 0.909\}$ & $\{1.000, 1.000, 1.000, 0.978\}$ & $\{1.000, 1.000, 1.000, 0.994\}$ \\
 
& $0.4$ $\symbol{92}$ non-parametric & $\{1.000, 1.000, 1.000, 1.000\}$ & $\{1.000, 1.000, 1.000, 1.000\}$ & $\{1.000, 1.000, 1.000, 1.000\}$  \\[0.5cm]

P & $0.4$ $\symbol{92}$ parametric & $(0.906,0.871,0.846)$ & $(0.976,0.962,0.954)$ & $(0.890,0.881,0.871)$ \\
  & $\mathcal{L}=(6,12,18)$ \\ 
\bottomrule
\end{tabular}
\end{table}

The results on size and power for the de-biased MHM test statistic are given in Table \ref{Tab.size&power_2016}. We observe that the values of size are small, and as the block size increases, the values become smaller.  For a given level of endogeneity, size appears larger for the larger value of $d=0.4$ than for the smaller $d=0.1$.  Power of the de-biased MHM statistic is quite low compared to either of the other two statistics, never reaching $0.40$ and usually being below $0.30$; such low power is an artifact of de-biasing, which necessarily offsets the magnitude of statistics to control size but then reduces power.  Despite the rather lackluster values of Table \ref{Tab.size&power_2016}, the size performance of the MHM statistic has been greatly improved by the de-biasing modification made possible by subsampling. In the next section we attempt to expand the situations under which we can assess the performance of these statistics.

\begin{table}[!h]
\footnotesize
\caption{Empirical size and power for the de-biased MHM test statistic at $5\%$ level. The residuals for the subsamples follow the parametric forms. The values in the curly brackets correspond to the 4 block sizes in the paper, and we assume $h=N^{-1/3}$.} \label{Tab.size&power_2016}
\centering
\setlength{\tabcolsep}{1pt} 
\renewcommand{\arraystretch}{1.15} 
\begin{tabular}{@{} ccccc @{}}
\toprule
size/power & $d$ & LM & SLM $(\lambda=N^{-1/3})$ & SLM $(\lambda=N^{-1/6})$ \\\midrule
 && \multicolumn{3}{c}{$r=0.5$} \\
 size & $0.1$ & $\{0.052, 0.028, 0.021, 0.020\}$ & $\{0.050, 0.023, 0.017, 0.021 \}$ & $\{0.050, 0.026, 0.018, 0.019\}$ \\

power & $0.1$ & $\{0.310, 0.309, 0.306, 0.300\}$ & $\{0.301, 0.301, 0.298, 0.286\}$ & $\{0.338, 0.335, 0.328, 0.315 \}$ \\[0.5cm]

size & $0.4$ & $\{0.060, 0.041, 0.031, 0.025\}$ & $\{ 0.062, 0.037, 0.026, 0.030 \}$ & $\{0.051, 0.026, 0.019, 0.022\}$ \\

power & $0.4$ & $\{0.211, 0.211, 0.209, 0.207 \}$ & $\{0.204, 0.204, 0.203, 0.199\}$ & $\{0.254, 0.254, 0.252, 0.245 \}$ \\\midrule \midrule

 & & \multicolumn{3}{c}{$r=1$} \\
size & $0.1$ & $\{0.050, 0.031, 0.017, 0.017\}$ & $\{0.051, 0.033, 0.024, 0.027\}$ & $\{0.049, 0.025, 0.020, 0.021\}$ \\

power & $0.1$ & $\{0.230, 0.230, 0.293, 0.288\}$ & $\{0.302, 0.301, 0.296, 0.285 \}$ & $\{0.328, 0.327, 0.323, 0.313\}$ \\[0.5cm]

size & $0.4$ & $\{0.133, 0.106, 0.084, 0.070\}$ & $\{0.148, 0.100, 0.086, 0.088\}$ & $\{0.104, 0.070, 0.061, 0.064\}$ \\

power & $0.4$ & $\{0.201, 0.201, 0.201, 0.200\}$ & $\{0.207, 0.207, 0.205, 0.199\}$ & $\{0.244, 0.244, 0.242, 0.235 \}$ \\
 
\bottomrule
\end{tabular}
\end{table}

\subsection{Extended simulations} \label{sec.extension}
We wish to examine behavior of the statistics under a larger flexible set of both integrable and nonintegrable regression functions beyond the local alternative in Section \ref{sizepowersims}. To this end, we simulated data from a set of $7$ nonintegrable functions and a set of $3$ integrable functions.  We used the same sets of functions previously used by \cite{wang2020measure} and will not list them here.  Results are provided in the supplementary material. For the nonintegrable functions, the hypothesis was $y_k = \theta_0+\theta_1x_k$.  For the integrable functions, the hypothesis was $y_k = \exp(-\theta_1 |x_k|) + \sigma u_k$.  In all cases, $u_k$ was taken to follow an AR(1) structure as was used previously in Section \ref{sizepowersims}.  

In general,  the results of these simulations (deferred to the supplementary material) reinforce the patterns seen in Tables \ref{Tab.size_nonintegrable}, \ref{Tab.power_nonintegrable} and \ref{Tab.size&power_2016}.  The P statistic maintains size cross particular situations, while size for the SNU varied from case to case, often being a bit above the nominal level for the parametric residual version and often being below the nominal level for the nonparametric residual version, with nominal size being better maintained over larger block sizes with LM/SLM.  Both the  SNU statistic and the P statistic exhibit good power with one being a bit higher in some cases and the other being higher in other cases.  
The values for the de-biased MHM test statistic show that it pretty uniformly has the greatest size and lowest power among all the statistics considered.

Next, in order to investigate the performance of test statistics for other generating processes for the regression errors $u_k$, we specified data generating mechanisms in which  such errors follow a MA(1) process,  $u_k=\mu+\epsilon(k)+\theta \epsilon(k-1)$ where $(\mu,\theta)'=(0,0.8)$ as opposed to the previously used AR processes for errors. The rest of the simulation set-up is the same as given in Section \ref{sizepowersims}. In particular, we are interested in the relative sizes of the SNU and P statistics, in part to examine the robustness of the P statistic and the flexibility of the SNU statistic. 
 The results for size under the basic linear model used previously are given in Table \ref{Tab.size_MA} with the update of now applying MA(1) errors.
While values for empirical size of the SNU statistic are somewhat elevated over the nominal $0.05$ value, they appear to stay under control far better than those for the P statistic, which explode to a value of $1.0$ no matter what the values of $d$, $r$, or the type of memory process.  It appears that the P statistic is highly sensitive to departures of the equation error process for the $u_k$ in (\ref{reg}) from an AR(p) structure, which may deteriorate performance.

\begin{table}[!h]
\footnotesize
\caption{Empirical size of the SNU and P test statistics at $5\%$ level with MA(1) structure for $u_k$'s. The values in the curly brackets correspond to the 4 block sizes in the paper, and we assume $h=N^{-1/3}$.} \label{Tab.size_MA}
\centering
\setlength{\tabcolsep}{1pt} 
\renewcommand{\arraystretch}{1.15} 
\begin{tabular}{@{} ccccc @{}}
\toprule
test & $d$ $\symbol{92}$ $\hat{f}(.)$ & LM & SLM $(\lambda=N^{-1/3})$ & SLM $(\lambda=N^{-1/6})$ \\\midrule
 && \multicolumn{3}{c}{$r=0.5$} \\
 SNU & $0.1$ $\symbol{92}$ parametric & $\{0.210, 0.157, 0.112, 0.094\}$ & $ \, \{0.188, 0.142, 0.114, 0.109\}$ & 
 $\{0.198, 0.158, 0.127, 0.107\}$ \\
 
& $0.1$ $\symbol{92}$ non-parametric & $\{0.208, 0.170, 0.095, 0.033\}$ & $\{0.188, 0.155, 0.079, 0.031 \}$ & $\{0.198, 0.163, 0.089, 0.033\}$  \\[0.5cm]

P & $0.1$ $\symbol{92}$ parametric & $(1.000, 1.000, 1.000)$ & $(1.000, 1.000, 1.000)$ & $(1.000, 1.000, 1.000)$ \\
 
 & $\mathcal{L}=(6,12,18)$  \\ \midrule
 
SNU & $0.4$ $\symbol{92}$ parametric & $\{0.262, 0.179, 0.108, 0.090\}$ & $\{0.159, 0.109, 0.087, 0.085 \}$ & $\{0.168, 0.121, 0.095, 0.098 \}$ \\
 
& $0.4$ $\symbol{92}$ non-parametric & $\{0.282, 0.214, 0.105,      0.041\}$ & $\{0.168, 0.116, 0.050, 0.016\}$ & $\{0.165, 0.128, 0.065,      0.022 \}$  \\[0.5cm]

P & $0.4$ $\symbol{92}$ parametric & $(1.000, 1.000, 1.000)$ & $(1.000, 1.000, 1.000)$ & $(1.000, 1.000, 1.000)$ \\
 
 & $\mathcal{L}=(6,12,18)$  \\ 
\midrule \midrule
 
 & & \multicolumn{3}{c}{$r=1$} \\
SNU & $0.1$ $\symbol{92}$ parametric & $\{0.146, 0.103, 0.073, 0.075\}$ & $\{0.139, 0.106, 0.090, 0.092\}$ & $\{0.138, 0.098, 0.082, 0.079\}$ \\
 
& $0.1$ $\symbol{92}$ non-parametric & $\{0.098, 0.077, 0.037, 0.007\}$ & $\{0.092, 0.086, 0.040, 0.007\}$ & $\{0.086, 0.076, 0.043, 0.015\}$  \\[0.5cm]

P & $0.1$ $\symbol{92}$ parametric & $(1.000, 1.000, 1.000)$ & $(1.000, 1.000, 1.000)$ & $(1.000, 1.000, 1.000)$ \\
 
 & $\mathcal{L}=(6,12,18)$  \\ \midrule
 
SNU & $0.4$ $\symbol{92}$ parametric & $\{0.191, 0.109, 0.084,     0.077\}$ & $\{0.312, 0.225, 0.185, 0.168\}$ & $\{0.207, 0.154, 0.136,      0.133\}$ \\
 
& $0.4$ $\symbol{92}$ non-parametric & $\{0.197, 0.127, 0.049,       0.013\}$ & $\{0.311, 0.205, 0.070, 0.004\}$ & $\{0.184, 0.131, 0.039,     0.005\}$  \\[0.5cm]

P & $0.4$ $\symbol{92}$ parametric & $(1.000, 1.000, 1.000)$ & $(1.000, 1.000, 1.000)$ & $(1.000, 1.000, 1.000)$  \\
 
 & $\mathcal{L}=(6,12,18)$  \\ 
\bottomrule
\end{tabular}
\end{table}

\section{Application to Carbon Kuznets Curve} \label{Carbon}

The Carbon Kuznets Curve (CKC) hypothesis was proposed by \cite{kuznets1955economic}, and attempts to explain an inverted-U shaped relationship between CO$_2$ emissions and gross domestic product (GDP) of a country as due to changes in air pollution that occur as a country develops technologically. \cite{piaggio2012co2} studied 31 countries over the period 1950--2006 and have shown the relationship between CO$_2$ emissions and GDP varies considerably among countries.  They suggest that the relationship can follow  straight line, quadratic or cubic functional forms, and
that the use of cointegration techniques can help to prevent identification of a specious relationship between CO$_2$ emissions and economic activity (see \cite{enders2008applied} as an example). 

\cite{piaggio2012co2} suggested that the relation between $\log(\text{CO}_2)$ emissions and $\log(\text{GDP})$ may be a straight line for Spain and a quadratic curve for France.
Thus, we fit and test a model with a straight line regression function $f(x_k) = \theta_0 + \theta_1 x_k$ for Spain and a model with a quadratic function $f(x_k)=\theta_0+\theta_1 x_k + \theta_2 x_k^2$ for France using the procedure developed in this article.
The data set that we have used is from 1950 to 2008 and contains 59 observations.  The CO$_2$ emission data come from the Carbon Dioxide Information Analysis Center (\citealp{boden2009global}), and the GDP data come from \cite{Maddison}. 

It is generally accepted that both CO$_2$ emissions  and GDP exhibit  nonstationary behaviors over time, and it has been suggested that an assumption of exogeneity may not hold because of measurement error and other sources of errors; see \cite{wang2018model} for a related discussion. In preparation for the application of the cointegrated regression model (\ref{reg}), we used the \texttt{artfima} package from the statistical software \texttt{R} developed by \cite{sabzikar2019parameter}
to verify that $\log(\text{GDP})$ processes for both countries exhibit SLM behavior (Whittle estimators of the ARTFIMA parameters were $d=1.079$ and $\lambda=0.138$ for Spain, while $d=1.093$ and $\lambda=0.138$ for France).

\begin{figure}[!ht]
\centering
\begin{tabular}{ c }
\includegraphics[width=1.05\textwidth]{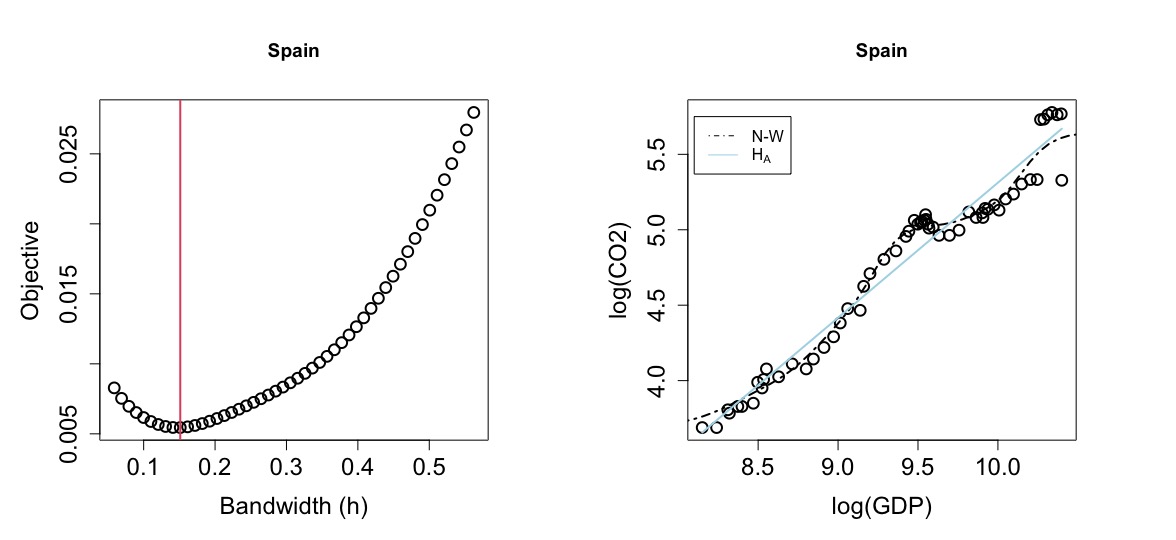} \\
\includegraphics[width=1.05\textwidth]{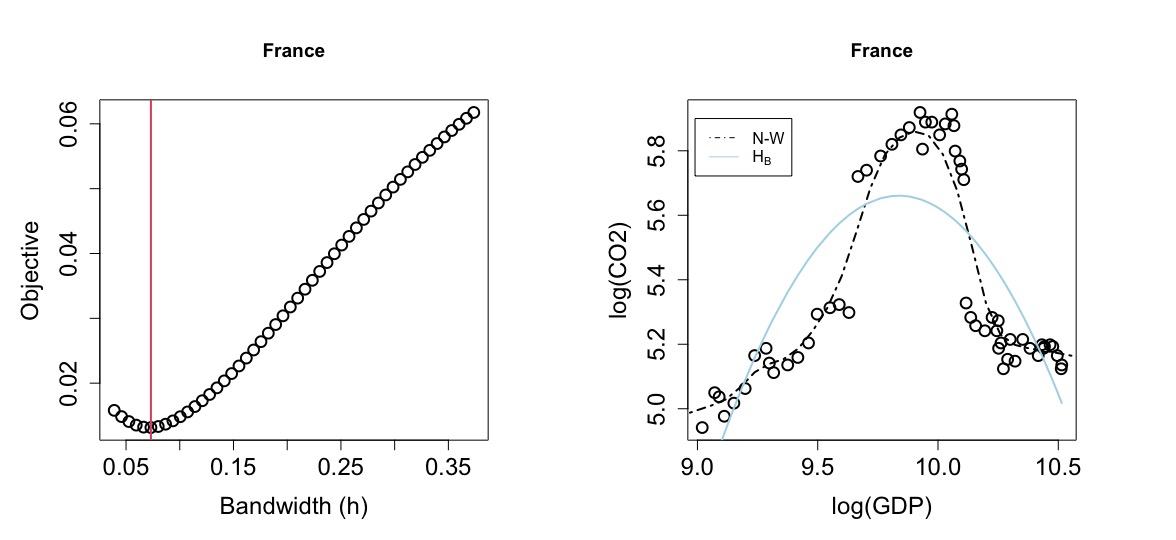}
\end{tabular}
\caption{Plots of LCV bandwidth selection criteria (Left) and log(CO$_2$) emissions versus log(GDP) for Spain and France (Right).  Overlain on the scatterplots are the kernel smooths as dashed curves and parametric fits. }
\label{Fig.countries}
\end{figure}

We also apply a nonparametric regression smoother to these data which will provide a visual comparison with the parametric fits. 
We use a Nadaraya-Watson (N-W) kernel regression estimator of $f(x)$,
\begin{equation*} 
\hat{f}(x)=\frac{\sum_{k=1}^{N}y_k K_h(x_k-x)}{\sum_{k=1}^{N}K_h(x_k-x)},
\end{equation*}
where $K_h(s)=h^{-1}K(s/h)$ for bandwidth $h$.  The kernel function $K$ is chosen to be the Gaussian kernel (\ref{Gkern}). 
We select a suitable bandwidth based on the use of a leave-one-out cross-validation procedure to minimize a least squares deviation criterion that is described more fully in Section D of the supplementary material.  Values of the least squares cross-validation criterion for a range of bandwidths are plotted on the left hand side of  Figure \ref{Fig.countries}. 
The optimal bandwidths are depicted as vertical lines and were chosen to be $0.151$ for Spain and $0.073$ for France.

\begin{figure}[ht]
\centering
\begin{tabular}{ c }
\includegraphics[width=1\textwidth]{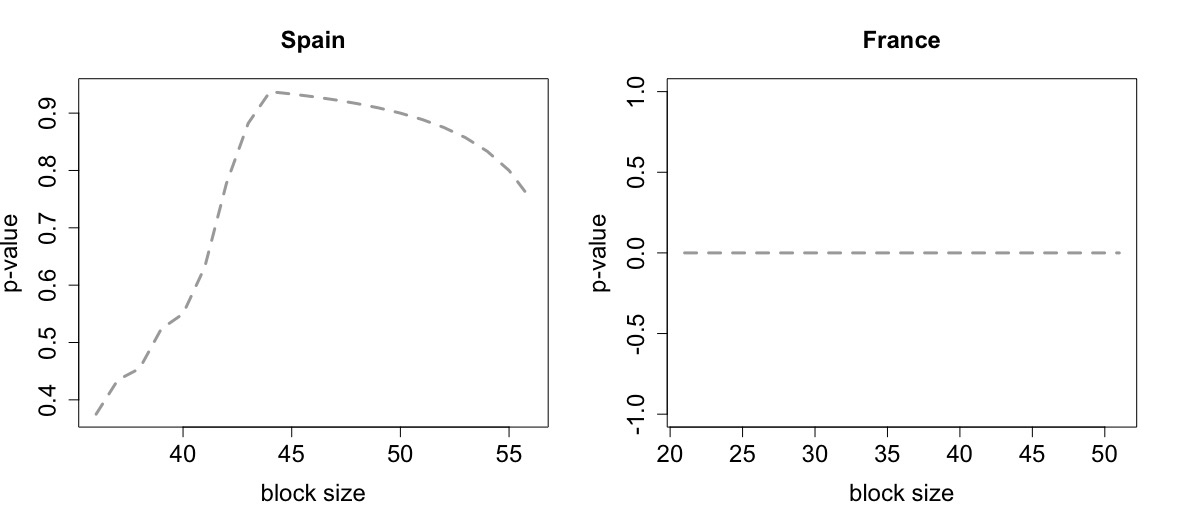} 
\end{tabular}
\caption{Illustration of p-values versus a continuous range of block sizes for Spain and France.}
\label{Fig.pvalue_block}
\end{figure}

We focus on the SNU and P test statistics, excluding the MHM statistic here due to its bias issues. The results of our tests are contained in Table \ref{Tab:applicationpvals}. 
Subsampling block sizes for the SNU test statistic were chosen based on the demonstration of p-values versus a continuous range of block sizes as given in Figure \ref{Fig.pvalue_block}, which provides a useful tool in practice for selecting a block size.  By then applying the block selection rule of minimal volatility (cf.~\citealp{politis1999subsampling}), an appropriate block size can be chosen by a point or region where p-values stabilize visually, for instance $b=44$ for Spain.  
For Spain, a hypothesis of a straight line regression function is not rejected using either the SNU statistic or the P statistic, at least with the larger values of $\mathcal{L}$.  For France, however, the hypothesized quadratic model is solidly rejected using the SNU statistic, but is not even questionable according to any version of the P test procedure. Section B of the supplementary material provides some further numerical study of block selections via minimal volatility, suggesting that these are also in reasonable concert with  appropriate block sizes in Table~\ref{Tab.size_nonintegrable}.

\begin{table}[!ht]
\footnotesize
\caption{Significance levels (p-values) for tests of the functional forms of carbon curves in Spain and France. } 
\label{Tab:applicationpvals}
\centering
\setlength{\tabcolsep}{10pt} 
\renewcommand{\arraystretch}{1.25} 
\begin{tabular}{cccc}
\toprule
&& Spain & France \\
Test Statistic & Related Test Criterion &  Straight Line & Quadratic \\
\midrule
Self-Normalized U (SNU) &  $\hat{b}=44$ & $0.938$ & -- \\ 
 & $\hat{b}=30$ & -- &  $0.000$ \\\midrule
 Portmanteau (P) & $\mathcal{L}=6$ & $0.051$ & $0.840$ \\
$p=1$ & $\mathcal{L}=12$ & $0.249$ & $0.942$ \\
 & $\mathcal{L}=18$ & $0.332$ & $1.000$ \\
\bottomrule
\end{tabular}
\end{table}

The right hand side of Figure \ref{Fig.countries} shows scatterplots of the data, the  nonparametric kernel estimates, and the fitted parametric models.  For Spain in the upper right panel, the fitted straight line regression appears visually reasonable.  But for France in the lower right panel, the fitted quadratic model is clearly an inadequate description of the relation between CO$_2$ and GDP.  This begs the question as to why the P statistic fares so poorly in this application, while in the simulation studies of Section \ref{sizepowersims} it consistently maintained its nominal size, even if it was not always the most powerful against the data generating models examined.  

The difficulty is not with the P test statistic \textit{per se}, but seemingly rather with use of the data to both determine the order of the AR(p) process and then compute the statistic. 
Residuals from a poorly fitting  parametric regression function can exhibit the structure of an AR process even if the true data generating mechanism contains no such structure.  To demonstrate this, we simulated data from a regression model with fixed covariates, independent errors, and a parametric nonlinear response function not entirely different in shape from the nonparametric carbon curve for France (see Section E of the supplementary material).  Applying the P test procedure to these data with a quadratic hypothesis produces results analogous to those seen in Table \ref{Tab:applicationpvals} in that the test is unable to detect the departure from the hypothesized model.  This same phenomenon appears to be occurring with the data from France.  In the simulations, the AR(1) structure of the equation error terms was assumed known and was not subject to this deleterious double-use of the data.

\section{Concluding Remarks} \label{Conclusion}

There has been several attempts in the literature to study the adequacy of the form of the regression functions within the context of cointegration. In particular, it is of great interest to assume the class of endogenous regressors with LM or SLM input shocks as it is expected in real applications. In this article, we have developed and justified a subsampling procedure to determine finite sample reference distributions for the SNU and MHM test statistics. This approximation allows us to be able to effectively study these two statistics under a complex dependent structure in the data sets. In addition, we make comparisons with the P test statistic developed explicitly for regression errors that follow an autoregressive process. 

Simulation studies demonstrate that the MHM test statistic has a bias issue, and we have suggested a potential bias-correction procedure to greatly improve its behavior relative to size. Specifically, the SNU test statistic has a good performance and is flexible towards the generating process for the regression errors. Use of the P test statistic does not require subsampling as its limiting distribution can be approximated by a chi-squared distribution. The major drawbacks to this statistic are a lack of robustness to non-autoregressive error processes and the need to determine the order of the assumed AR structure used in construction.
In the application, we have been able to successfully examine the form of regression functions with the SNU test statistic combined with subsampling.

\section*{Supplementary Materials}

Supplementary material available online includes proofs of the theorems and additional numerical results. A stand-alone package for implementing the methods described
in this paper can be downloaded from \url{https://github.com/SepidehMosaferi/TestStatistics_Subsampling}.

\section*{Acknowledgements}
We would like to thank an associate editor and anonymous reviewers who made excellent comments and
suggestions that helped us to improve the paper.  Nordman's research was partially supported by NSF DMS-2015390.

\section*{Disclosure Statement}
No potential conflict of interest was reported by the author(s).

\bibliographystyle{ims}
\bibliography{Bibliography}

\newpage
\clearpage

\def\spacingset#1{\renewcommand{\baselinestretch}%
{#1}\small\normalsize} \spacingset{1}

\begin{center}

\LARGE{Supplementary Material for \hspace{.2cm}\\ ``Properties of Test Statistics for Nonparametric Cointegrating Regression Functions Based on Subsamples"}
\vspace{1cm}

  \large{\author{Sepideh Mosaferi\hspace{.2cm}\\
    University of Massachusetts Amherst\\
    and \\
    Mark S. Kaiser \\
    Iowa State University \\
    and \\
    Daniel J. Nordman \\
    Iowa State University }}
  \end{center}

\spacingset{1.5} 

\setcounter{secnumdepth}{0}

This supplementary material is structured as follows. We provide detailed proofs of theorems in Section A. Further simulation results are in Section B. Section C covers the related steps for the de-biased MHM test statistic. Section D provides a practical guidance on the selection of bandwidth. Section E covers a simulated example, where the regression errors do not follow an AR structure in order to explain the form of the regression function through the SNU and P test statistics. 

\section{A \quad Technical Details and Proofs} \label{sec:App1}
\renewcommand{\theequation}{A.\arabic{equation}}
\setcounter{equation}{0}
\renewcommand{\thelemma}{A.\arabic{lemma}}
\renewcommand{\thedefinition}{A.\arabic{definition}}

In this section, we provide detailed proofs of theorems.

\vspace{0.25cm}

\noindent \textbf{Proof of Theorem \ref{theo.sub.2016}.}
Recall   the  estimator 
$\hat{F}_{M,b}(x)  \equiv \frac{1}{M} \sum_{i=1}^{M} \mathbb{I} \{\tau^{-1}_b T_{i,b} \leq x \}$,  $x \in \mathbb{R}$, involves  subsamples $\mathcal{S}_{i,b}\equiv \{(x_{i}-x_{i-1},\tilde{u}_i),\ldots,(x_{i+b-1}-x_{i-1},\tilde{u}_{i+b-1})\}$, $i=1,\ldots,M \leq N-b+1$ of length $b$ and  corresponding subsample test statistics
\begin{equation}
\label{eqn:subtest}
\tau_b^{-1} T_{i,b} \equiv  \frac{d_b}{b h_b}\int_{\mathbb{R}} \Big\{\sum_{j=1}^{b} K\Big[\frac{(x_{i+j-1}-x_{i-1})-x}{h_b} \Big] \tilde{u}_{i+j-1} \Big\}^2 \pi(x) dx  
\end{equation}
with the scaling $\tau_b \equiv b h_b/d_b$.  
Above $\tilde{u}_i \equiv  y_i - \hat{f}(x_i)$, $i \geq 1$, denote residuals from a (full data) estimator $\hat{f}$ of the trend $f$, which estimate pure errors as $u_i=y_i-f(x_i)$.  
The target distribution  $F_{N,H_0}(x)\equiv \text{Prob}_{H_0} \{\tau_N^{-1}T_N \leq x \}$, $x\in\mathbb{R}$, of the original-level test statistic $\tau_N^{-1}T_N$ under the null hypothesis $H_0: f(x) = g(x,\theta_0)$ is determined by substituting true errors $u_k$ in (\ref{test2016}) and   converges as prescribed in 
 (\ref{selfnormTest.WP}) or  (\ref{selfnormTest.MK}) under LM or SLM, respectively; see \cite{wang2016nonparametric} and \cite{Mosaferi}. That is, if we let $Z_0$ (say) denotes the continuous limiting variable in (\ref{selfnormTest.WP}) or  (\ref{selfnormTest.MK}), with distribution function denoted as $F_0(x) \equiv \text{Prob}(Z_0 \leq x)$ for $x\in\mathbb{R}$, it holds that
 $\sup_{x\in\mathbb{R}}|F_{N,H_0}(x) - F_0(x)| \rightarrow 0$ as $N\to \infty$.  Hence, it suffices to establish Theorem~\ref{theo.sub.2016} with $F_0(x)$ in place of  $F_{N,H_0}(x)$.
 
We first consider showing that 
\begin{equation}
\label{eqn:subtest2}
\sup_{x\in\mathbb{R}}|\bar{F}_{M,b}(x) -F_{0}(x)  | \xrightarrow{P} 0
\end{equation}
holds   as $N \to \infty$ for a modified subsampling estimator  $\bar{F}_{M,b}(x)  \equiv \frac{1}{M} \sum_{i=1}^{M} \mathbb{I} \{\tau^{-1}_b \bar{T}_{i,b} \leq x \}$,  $x \in \mathbb{R}$, where subsample statistics $\tau_b^{-1}\bar{T}_{i,b}$ are defined by replacing 
residuals $\tilde{u}_j$ with true errors $u_i$ in 
(\ref{eqn:subtest}).   By strict stationarity 
of the modified subsample statistics 
(i.e. $\tau_b^{-1}\bar{T}_{1,b} \stackrel{D}{=} \tau_b^{-1}\bar{T}_{i,b}$ for $i \geq 1$), it holds for  a given real $x$ that      
\begin{equation*}
 \mathbb{E}(\bar{F}_{M,b}(x)) =   
\text{Prob}(\tau_b^{-1}\bar{T}_{1,b} \leq x) \rightarrow  F_0(x) 
\end{equation*}
as $b\rightarrow \infty$.  Likewise, in considering the variance of $\bar{F}_{M,b}(x)$
for a fixed $x$, we have
\begin{align*}
\text{Var}(\bar{F}_{M,b}(x)) & =\frac{1}{M^2} \sum_{\ell=-M}^{M} (M-|\ell|) \text{Cov}\Big[\mathbb{I}\{ \tau_b^{-1}\bar{T}_{1,b} \leq x \}, \mathbb{I}\{ \tau_b^{-1}\bar{T}_{1+\ell,b} \leq x \}\Big] \\
& \leq 2 \epsilon +  \max_{[M \epsilon] \leq \ell \leq M} \alpha_{\ell,b}
\end{align*}
for any given $\epsilon \in (0,1)$, which follows 
by splitting the sum over $|\ell| \leq \lfloor M \epsilon \rfloor$ and $|\ell|\geq \lceil M \epsilon \rceil$ applying the covariance bound
$\big|\text{Cov}\big[\mathbb{I}\{ \tau_b^{-1}\bar{T}_{1,b} \leq x \}, \mathbb{I}\{ \tau_b^{-1}\bar{T}_{1+\ell,b} \leq x \}\big]\big|\leq \min\{1,\alpha_{\ell,b}\}$.
Under    the \textit{subsampling condition}
with $b/M \rightarrow 0$ and $M\to \infty$ (as $N\to \infty$),  it follows that  $\limsup_{N\to \infty}\text{Var}(\bar{F}_{M,b}(x)) \leq 2 \epsilon$ and, as $\epsilon>0$ is arbitrary, we then have $\lim_{N\to \infty}\text{Var}(\bar{F}_{M,b}(x))=0$.   Consequently, we have shown 
$|\bar{F}_{M,b}(x) -F_{0}(x)  | \xrightarrow{P} 0$ holds as $N\to \infty$ for any $x\in\mathbb{R}$.  

From this,  we may establish (\ref{eqn:subtest2}) using the correspondence between convergence in probability and convergence almost surely along subsequences.  Namely,  let $\{r_i\}_{i=1}^{\infty}$ denote  a dense but countable collection  of points in $\mathbb{R}$ (i.e., continuity points of $Z_0$) and let $N_k$ denote an arbitrary sequence of $N$.  As $|\bar{F}_{M,b}(r_1) -F_{0}(r_1)  | \xrightarrow{P} 0$ holds, there exists a subsequence $N_{k,1}$ of $N_k$ such that
$|\bar{F}_{M_{k,1},b_{k,1}}(r_1) -F_{0}(r_1)  | \rightarrow 0$ holds almost surely as $N_{k,1}\rightarrow \infty$. 

Continuing in this fashion, by $|\bar{F}_{M,b}(r_i) -F_{0}(r_i)  | \xrightarrow{P} 0$,  there exists a subsequence of $N_{k,i}$ of $N_{k,i-1}$
such that $|\bar{F}_{M_{k,i},b_{k,i}}(r_{i}) -F_{0}(r_i)  | \rightarrow 0$ almost surely as $N_{k,i}\rightarrow \infty$ for each $i >1$.  As $\{r_i\}$ is countable, there then exists a set of probability 1 and a subsequence $N_{k,k}$ (the diagonal of a matrix with entries $N_{k,i}$, $k,i\geq 1$) where  $|\bar{F}_{M_{k,k},b_{k,k}}(r_{i}) -F_{0}(r_i)  | \rightarrow 0$  holds as $N_{k,k}\to \infty$ for each $i \geq 1$, which entails that   $\sup_{x\in\mathbb{R}}|\bar{F}_{M_{k,k},b_{k,k}}(x) -F_{0}(x)  | \rightarrow 0$ almost surely (i.e., continuous $Z_0$).  As the originating subsequence $N_k$ was arbitrary, the latter implies    (\ref{eqn:subtest2}).

To ease notation in completing the proof, we assume the number of subsamples to be $M-b+1$ rather than $M$; this change is inconsequential as both $\sup_{x\in\mathbb{R}}|\hat{F}_{M,b}(x) - \hat{F}_{M-b+1,b}(x)| \leq 2b/M \rightarrow 0$  and $\sup_{x\in\mathbb{R}}|\bar{F}_{M,b}(x) - \bar{F}_{M-b+1,b}(x)|\leq 2b/M \rightarrow 0$ by $b/M\rightarrow$ as $N\to \infty$.  To finish the proof, let $Y_b^{*}$ denote a random draw from the subsampling distribution $\hat{F}_{M-b+1,b}$, whereby $Y_b^*$ can be defined based on (\ref{eqn:subtest}) as 
\[
Y_b^* \equiv \tau_b^{-1} T_{I^*,b} = \frac{d_b}{b h_b}\int_{\mathbb{R}} \Big\{\sum_{j=1}^{b} K\Big[\frac{(x_{I^*+j-1}-x_{I^*-1})-x}{h_b} \Big] \tilde{u}_{I^*+j-1} \Big\}^2 \pi(x) dx 
\]
using an integer $I^*$ drawn uniformly from $\{1,\ldots,M-b+1\}$.  Correspondingly, define 
\[
Z_b^* \equiv     \frac{d_b}{b h_b}\int_{\mathbb{R}} \Big\{\sum_{j=1}^{b} K\Big[\frac{(x_{I^*+j-1}-x_{I^*-1})-x}{h_b} \Big] u_{I^*+j-1} \Big\}^2 \pi(x) dx 
\]
as a random draw from the modified subsampling distribution $\bar{F}_{M-b+1,b}$, so that  
\begin{equation}
\label{eqn:subtest4}
 |Y_b^* - Z_b^*| \leq 2 |Z_b^* R_b^*|^{1/2} + R_b^*
\end{equation}
holds by the Cauchy-Schwarz inequality applied for a remainder
\[
R_b^* \equiv     \frac{d_b}{b h_b}\int_{\mathbb{R}} \Big\{\sum_{j=1}^{b} K\Big[\frac{(x_{I^*+j-1}-x_{I^*-1})-x}{h_b} \Big] [f(x_{I^*+j-1}) -\hat{f}(x_{I^*+j-1})] \Big\}^2 \pi(x) dx 
\]
involving estimated trends $\hat{f}$.  

To establish Theorem~\ref{theo.sub.2016}, it now suffices to show 
\begin{equation}
\label{eqn:subtest3}
 \mathbb{E}_*|R_b^*|   \xrightarrow{P} 0
\end{equation}
as $N\to \infty$, where $\mathbb{E}_*$ denotes expectation with respect to the resampling  (i.e., $I^*$) conditional on the data.  Then, from  (\ref{eqn:subtest2}) and (\ref{eqn:subtest3}), we have that, for any subsequence $N_k$ of $N$, there exists a further subsequence $N_\ell$ of $ N_k$ such that  $Z_{b_\ell}^* \xrightarrow{D} Z_0$ and  
  $R_{b_\ell}^* \xrightarrow{D} 0$ hold as $N_\ell \to \infty$ almost surely.  By Slutsky's theorem with (\ref{eqn:subtest4}), it then follows that
  $Y_{b_\ell}^* \xrightarrow{D} Z_0$  as $N_\ell \to \infty$ almost surely, so that $\sup_{x} |\hat{F}_{M_\ell-b_\ell+1,b_\ell}(x) - F_0(x) |   \rightarrow 0$  as $N_\ell \to \infty$ almost surely.  As the subsequence $N_k$ was arbitrary, this gives the convergence in probability in Theorem~\ref{theo.sub.2016}.

To show (\ref{eqn:subtest3}), we first expand
\begin{eqnarray}
\nonumber &&\mathbb{E}_*|R_b^*| \\
\nonumber &=& \frac{1}{M-b+1}\sum_{i=1}^{M-b+1}\frac{d_b}{b h_b}\int_{\mathbb{R}} \Big\{\sum_{j=1}^{b} K\Big[\frac{(x_{i+j-1}-x_{i-1})-x}{h_b} \Big] [f(x_{i+j-1}) -\hat{f}(x_{i+j-1})]  \Big\}^2 \pi(x) dx\\ \label{eqn:subtest5}&\leq&  
\Delta_N  \max_{1 \leq j \leq M }[f(x_{j}) -\hat{f}(x_j)]^2 
\end{eqnarray}
for
\[
\Delta_N \equiv \frac{1}{M-b+1}\sum_{i=1}^{M-b+1}\frac{d_b}{b h_b}\int_{\mathbb{R}} \Big\{\sum_{j=1}^{b} K\Big[\frac{(x_{i+j-1}-x_{i-1})-x}{h_b} \Big]  \Big\}^2 \pi(x) dx.
\]
Using stationarity of the adjusted subsamples,
along with $\sup_x \mathbb{E} \left( \sum_{j=1}^b K\Big[\frac{(x_{j}-x)}{h_b} \Big] \right)^2 \leq C [b h_b/d_b]^2$ (e.g.~Lemma~8.2 of \cite{wang2016nonparametric}) and $\int \pi(x) dx <\infty$,
we have that
\[
  \mathbb{E}\Delta_N = \frac{d_b}{b h_b} \int_{\mathbb{R}} \mathbb{E}\left(  \sum_{j=1}^{b} K\Big[\frac{(x_{j}-x)}{h_b} \Big] \right)^2 \pi(x) dx   = O(b h_b/d_b).   
\]
It follows that $ [d_b/(b h_b)] \Delta_N = O_P(1)$, while  $[(b h_b)/d_b] \max_{1 \leq j \leq M }[f(x_{j}) -\hat{f}(x_j)]^2 =o_P(1)$ by assumption, which then establishes  (\ref{eqn:subtest3}) from (\ref{eqn:subtest5}).

\QEDA

\noindent\textbf{Proof of Theorem \ref{theo.sub.2012}.}
The proof is similar to the proof of Theorem \ref{theo.sub.2016}, and we briefly overview the details.
For each subsample  $i=1,\ldots,M$, recall  subsample  analog  
$Z_{i,b} \equiv S_{i,b}/\sqrt{2 V_{i,b}}$ of the test statistic 
$Z_N \equiv  S_N/\sqrt{2V_N}$ from (\ref{self-normalized}), where
\begin{eqnarray} \label{self-normalized1}
  \tau_{b}^{-1/2}S_{i,b} &\equiv &\sum_{k,j=1, k \neq j}^{b} \tilde{u}_{k+i-1} \tilde{u}_{j+i-1} K\Big[\frac{x_{k+i-1}-x_{j+i-1}}{h} \Big], \\\nonumber \tau_b^{-1}V_{i,b}^2& \equiv&  \sum_{k,j=1, k \neq j}^{b} \tilde{u}^2_{k+i-1} \tilde{u}^2_{j+i-1} K^2\Big[\frac{x_{k+i-1}-x_{j+i-1}}{h}\Big],
\end{eqnarray}
where  $\tilde{u}_i \equiv  y_i - \hat{f}(x_i)$, $i \geq 1$, are residuals from a (full data) estimator $\hat{f}$ of the trend $f$ that again estimate pure errors as $u_i=y_i-f(x_i)$ and we define a scaling term $\tau_b \equiv b^2 h_b/d_b$.  

By substituting   error terms $u_i$ for residuals $\tilde{u}_i$, we may define counterpart versions of subsample statistics  $Z_{i,b} \equiv S_{i,b}/\sqrt{2 V_{i,b}}$ as, say, $\bar{Z}_{i,b}\equiv \bar{S}_{i,b}/\sqrt{2 \bar{V}_{i,b}^2}$ with $\bar{S}_{i,b}$ and $\bar{V}^2_{i,b}$.   The resulting subsample copies $(\tau_{b}^{-1/2}\bar{S}_{i,b}, \tau_b^{-1}\bar{V}_{i,b}^2)$ have
 the same distribution for $i=1,\ldots,M$ by stationarity.  Furthermore,
a subsample copy $(\tau_{b}^{-1/2}\bar{S}_{1,b}, 2\tau_b^{-1}\bar{V}_{1,b}^2) \xrightarrow{D} (Z_1,Z_2)$ converges in distribution as $b\to \infty$, involving a bivariate pair of random variables, and the associated statistic $\bar{Z}_{1,b} \equiv \bar{S}_{1,b}/\sqrt{2 \bar{V}_{1,b}^2} \xrightarrow{D} Z_0 \equiv Z_1/\sqrt{Z_2}$ converges to the continuous limit distribution described in (\ref{complex_asymp}).  

If we let $M-b+1$ denote the number of subsamples and  if $(Y_{1,b}^*,Y_{2,b}^*)\equiv (\tau_{b}^{-1/2}\bar{S}_{I^*,b}, 2\tau_b^{-1}\bar{V}_{I^*,b}^2)$ denotes a randomly selected subsampling pair, defined by a uniform $I^*$ draw from $\{1,\ldots,M-b+1\}$, then the subsampling distribution induced by    $(Y_{1,b}^*,Y_{2,b}^*)$ converges to the distribution of  $(Z_1,Z_2)$ (in probability) under the mixing/subsampling   condition.
If we similarly define, with the same variable $I^*$, a subsampling pair, say $(W_{1,b}^*,W_{2,b}^*)\equiv (\tau_{b}^{-1/2}S_{I^*,b}, 2\tau_b^{-1}V_{I^*,b}^2)$ based on a random draw from the paired original subsampling statistics $$\{(\tau_{b}^{-1/2}S_{1,b}, 2\tau_b^{-1}V_{1,b}^2),\ldots, (\tau_{b}^{-1/2}S_{M-b+1,b}, 2\tau_b^{-1}V_{M-b+1,b}^2)\}$$ (i.e., computed from residuals $\tilde{u}_i$), then it is enough to show that
 $\tau_{b}^{-1/2}\mathbb{E}_* |W_{1,b}^* -  Y_{1,b}^*|
+ \tau_{b}^{-1}\mathbb{E}_* |W_{2,b}^* -  Y_{2,b}^*|  \xrightarrow{P} 0$ as $N\to \infty$, where $\mathbb{E}_*$ denotes expectation with respect to the resampling  (i.e., $I^*$) conditional on the data.  

By Slutsky's theorem and the probabilistic convergence of the distribution of $Y_{1,b}^*/\sqrt{Y_{2,b}^*}$ to the distribution of the target limit $Z_0 \equiv Z_1/\sqrt{Z_2}$, we can then conclude that
the distribution of $W_{1,b}^*/\sqrt{W_{2,b}^*}$ converges to the distribution of   $Z_0$ (in probability).   This yields  convergence in probability of the subsampling distribution estimator
$\hat{F}_{M-b+1,b}$, as $\hat{F}_{M-b+1,b}$ is the distribution of $W_{1,b}^*/\sqrt{W_{2,b}^*}$.   

To show $\tau_{b}^{-1/2}\mathbb{E}_* |W_{1,b}^* -  Y_{1,b}^*|\xrightarrow{P} 0$, we write 
\[ 
 \tau_{b}^{-1/2}\mathbb{E}_* |W_{1,b}^* -  Y_{1,b}^*| =
 \frac{\tau_{b}^{-1/2}}{M-b+1} \sum_{i=1}^{M-b+1} |S_{i,b}-\bar{S}_{i,b}|
 \leq \big[ \delta+\delta_N^2\big] \Delta_N\]
for $\delta_N \equiv  \max_{1 \leq j \leq M}| f(x_j)-\hat{f}(x_j)|$ and
\[
\Delta_N\equiv \frac{2\tau_{b}^{-1/2}}{M-b+1} \sum_{i=1}^{M-b+1}\sum_{k,j=1, k \neq j}^{b} \big(1+|u_{k+i-1}| + |u_{j+i-1}| \big)K\Big[\frac{x_{k+i-1}-x_{j+i-1}}{h} \Big].
\]

Using  $ \mathbb{E} \left( \sum_{j=i+1}^b K\Big[\frac{(x_{j}-x_i)}{h_b} \Big] \right)^2 \leq C [b h_b/d_b]^2$ (e.g.~Lemma~8.2 of \cite{wang2016nonparametric}), $\mathbb{E}|u_i|^2=\mathbb{E}|u_1|^2<\infty$ and Holder's  inequality, we may bound $
\mathbb{E}\Delta_N \leq   \tau_{b}^{-1/2} b^2 h_b/d_b =\tau_{b}^{1/2} $
so that $\Delta_N=O_p(\tau_{b}^{1/2})$.  Hence, we have  $\tau_{b}^{-1/2}\mathbb{E}_* |W_{1,b}^* -  Y_{1,b}^*| = O_P(\Delta_n)O_P(\delta_N + \delta_N^2) = o_P(1)$ using that
$\delta_N^2 \tau_b = o_P(1)$ by assumption.  This establishes $\tau_{b}^{-1}\mathbb{E}_* |W_{1,b}^* -  Y_{1,b}^*|\xrightarrow{P} 0$.  An analogous argument also shows $\tau_{b}^{-1/2}\mathbb{E}_* |W_{2,b}^* -  Y_{2,b}^*|\xrightarrow{P} 0$.

\QEDA

\section{B \quad Further Simulation Results} \label{sec:App2}
\renewcommand{\theequation}{B.\arabic{equation}}
\setcounter{equation}{0}
\renewcommand{\thelemma}{B.\arabic{lemma}}
\renewcommand{\thedefinition}{B.\arabic{definition}}
\setcounter{table}{0}
\renewcommand{\thetable}{B.\arabic{table}}
\setcounter{figure}{0}
\renewcommand{\thefigure}{B.\arabic{figure}}
 
In this section, we provide further simulation results divided into six subsections.

\subsection{B.1 \quad Validity of Normal Assumption for the SNU Test Statistic}

\begin{figure}[ht]
\centering
\begin{tabular}{ c }
\includegraphics[width=0.68\textwidth]{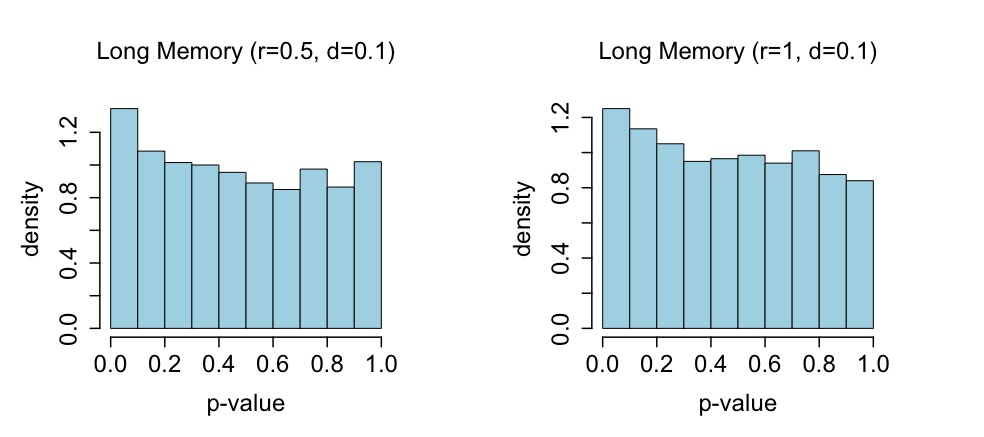} \\
\includegraphics[width=0.68\textwidth]{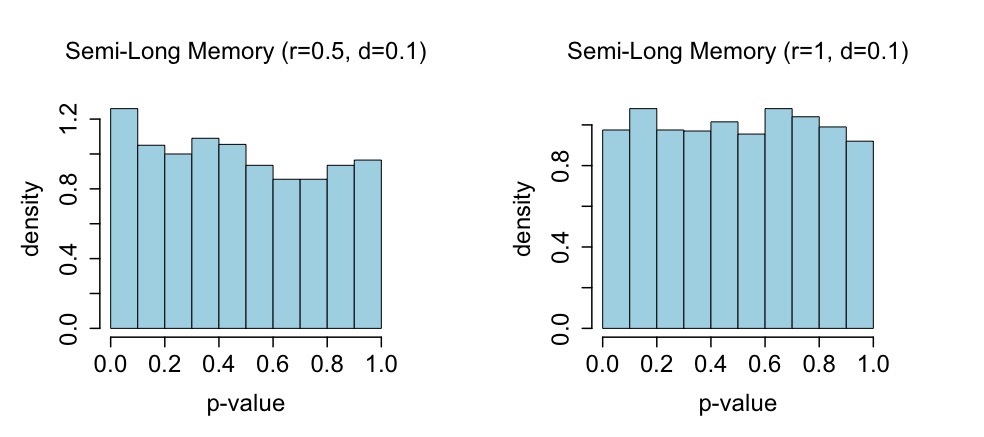} \\
\includegraphics[width=0.68\textwidth]{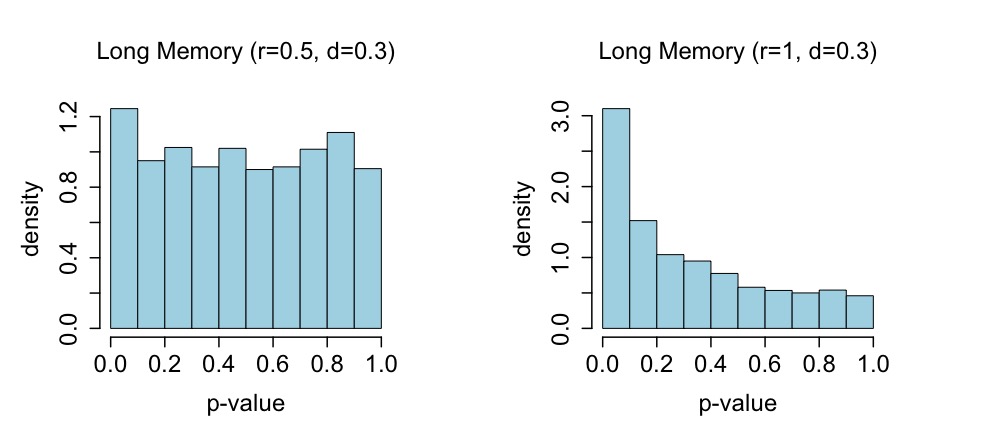} \\
 \includegraphics[width=0.68\textwidth]{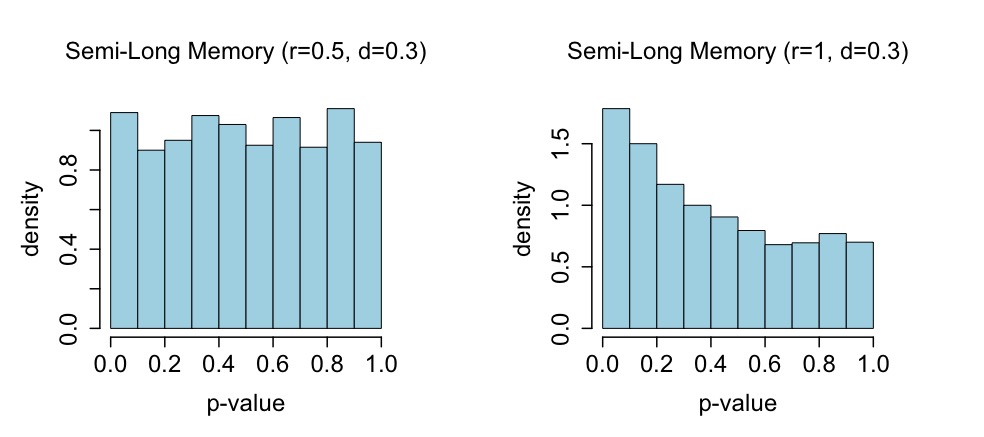}
 \end{tabular}
\caption{Comparison of Monte Carlo histograms of p-values for SNU test statistic based on normal assumption. The results are for LM and SLM assuming $d=0.1$ and $0.3$ under the basic linear model. The bandwidth is $h=N^{-1/3}$ and the tempering parameter for the SLM is $\lambda=N^{-1/6}$.}
\label{Fig.Histograms1}
\end{figure}

\begin{figure}[ht]
\centering
\begin{tabular}{ c }
\includegraphics[width=0.68\textwidth]{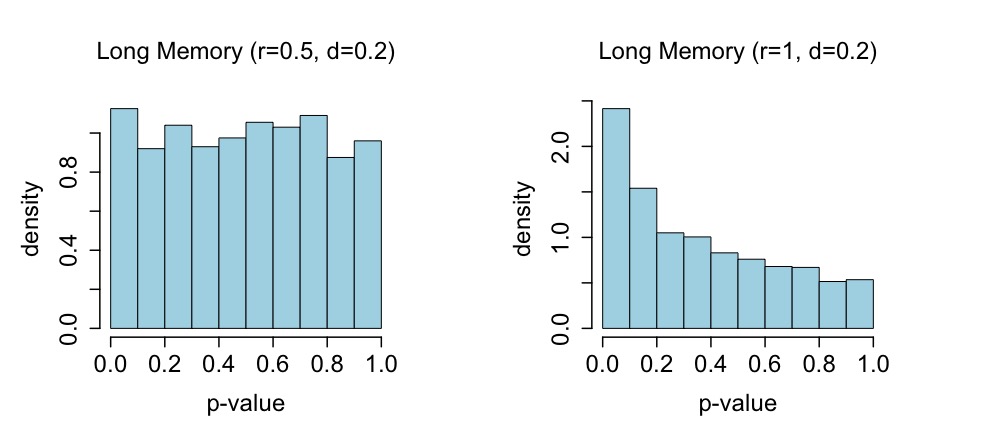} \\
\includegraphics[width=0.68\textwidth]{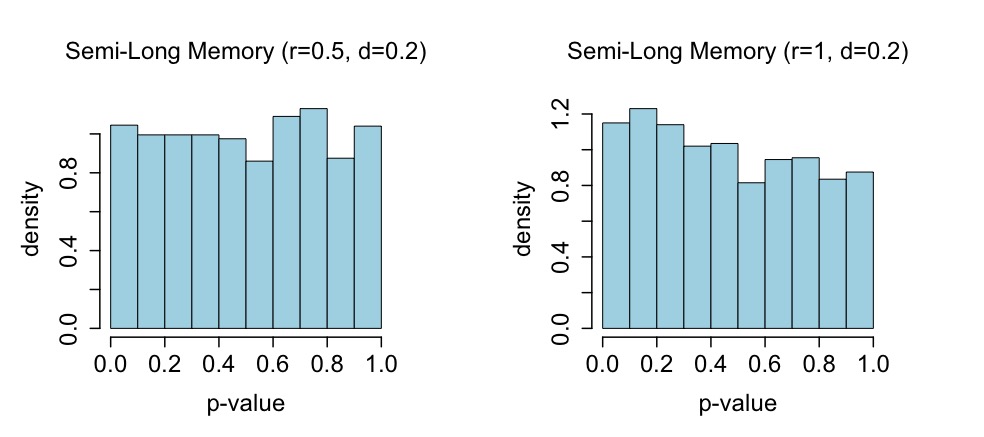} \\
\includegraphics[width=0.68\textwidth]{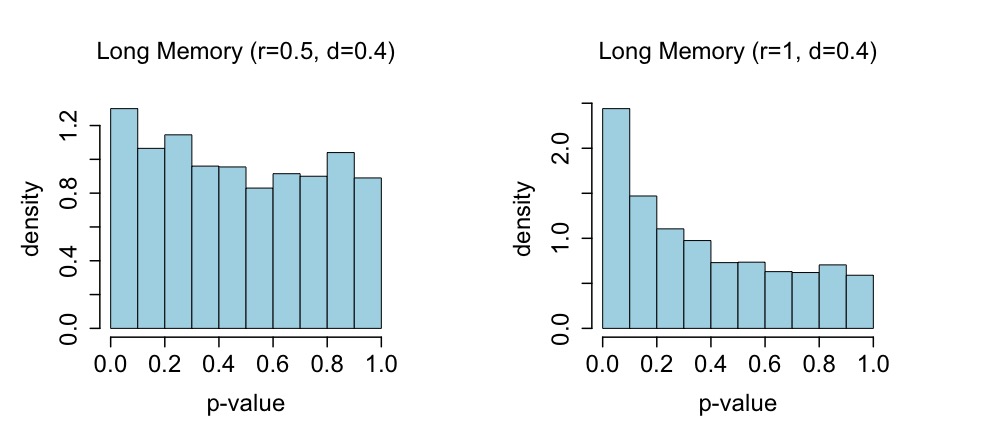} \\
 \includegraphics[width=0.68\textwidth]{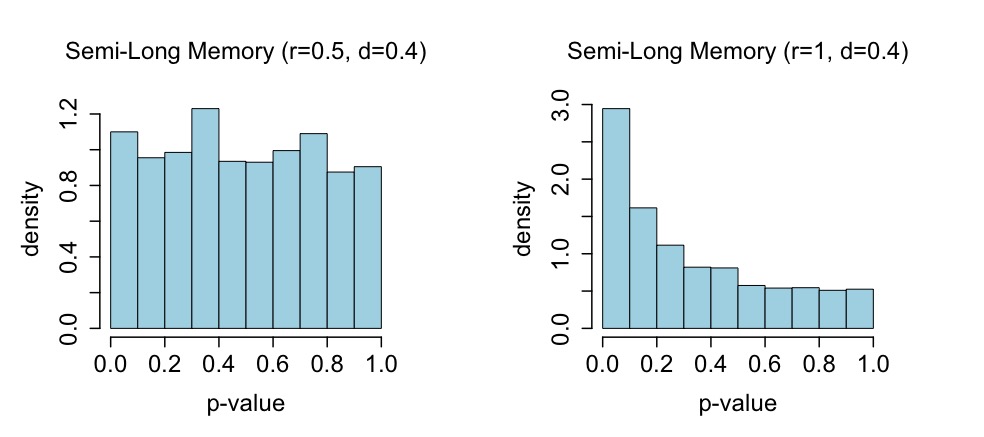}
 \end{tabular}
\caption{Comparison of Monte Carlo histograms of p-values for SNU test statistic based on normal assumption. The results are for LM and SLM assuming $d=0.2$ and $0.4$ under the basic linear model. The bandwidth is $h=N^{-1/3}$ and the tempering parameter for the SLM is $\lambda=N^{-1/6}$.}
\label{Fig.Histograms2}
\end{figure}

We provide Monte Carlo histograms of p-values for the SNU test statistic for $d=\{0.1,0.2,0.3,0.4\}$ under the null hypothesis of $y_k=\theta_0+\theta_1 x_k + \sigma u_k$. The histograms are given in Figures \ref{Fig.Histograms1} and \ref{Fig.Histograms2}. The results confirm that the distribution of SNU test statistic does not follow a standard normal. 

\subsection{B.2 \quad Nonintegrable and Integrable Regression Functions}

We study the size and power of test statistics under a variety of nonintegrable and integrable regression functions as follows. 

\begin{itemize}
\item \textbf{Nonintegrable regression function:}
\begin{align} 
y_k & = \theta_0+\theta_1 x_k+ \sigma u_k, 
\label{eq:nonintegrablefs1} \\
y_k & = \theta_0+ \theta_1 x_k + 0.5 |x_k|^2 \mathbb{I}(|x_k| \leq 10) + \sigma u_k, 
\label{eq:nonintegrablefs2} \\
y_k & = \theta_0 + \theta_1 x_k + 20 \exp(-|x_k|^2) + \sigma u_k, 
\label{eq:nonintegrablefs3} \\
y_k & = \theta_0 + \theta_1 x_k + 0.1 |x_k| + \sigma u_k, 
\label{eq:nonintegrablefs4}\\
y_k & = \theta_0 + \theta_1 x_k + 0.1 |x_k|^2 + \sigma u_k.
\label{eq:nonintegrablefs5}
\end{align}
\item \textbf{Integrable regression function:}
\begin{align} 
y_k & = \exp(-\theta_1 |x_k|) + \sigma u_k, 
\label{eq:integrablefs1} \\
y_k & = \exp(-\theta_1 |x_k|) + 0.5 |x_k|^2 \mathbb{I}(|x_k| \leq 10) + \sigma u_k, \label{eq:integrablefs2}\\
y_k & = \exp(- \theta_1 |x_k|) + 20 \exp(-|x_k|^2) + \sigma u_k, \label{eq:integrablefs3}\\
y_k & = \exp(- \theta_1 |x_k|) + 0.1 |x_k| + \sigma u_k, 
\label{eq:integrablefs4}\\
y_k & = \exp(- \theta_1 |x_k|) + 0.1 |x_k|^2 + \sigma u_k.
\label{eq:integrablefs5}
\end{align}
\end{itemize}

For the nonintegrable regression functions, the generating model in (\ref{eq:nonintegrablefs1}) is explained in Section \ref{sizepowersims} of the main manuscript. This model is used for calculating the size, and the results are in the manuscript. We use the generating models in (\ref{eq:nonintegrablefs2})--(\ref{eq:nonintegrablefs5}) to calculate the power of test statistics, and the results are given in Tables \ref{Tab.power_16}--\ref{Tab.power_19}. 

For the integrable regression functions, the generating model in (\ref{eq:integrablefs1}) is used for calculating the size of the tests, and the results are listed in Table \ref{Tab.size_20}.  
The empirical powers for all the test statistics under the generating models (\ref{eq:integrablefs2})--(\ref{eq:integrablefs5}) are listed in Tables \ref{Tab.power_21}--\ref{Tab.power_24}. Note that models in  (\ref{eq:integrablefs4}) and (\ref{eq:integrablefs5}) are not integrable unlike models given in (\ref{eq:integrablefs2}) and (\ref{eq:integrablefs3}). Since the null model in (\ref{eq:integrablefs1}) is integrable, we have used the title of \textit{integrable} to refer to the models.

\subsection{B.3 \quad Size for Integrable Regression Function with MA(1) Process}

In Table \ref{Tab.size_20_MA}, we provide the empirical sizes for the SNU and P test statistics for the integrable regression function (\ref{eq:integrablefs1}) when $u_k$'s follow an MA(1) process. The results show that the values of size are relatively small for the SNU test statistic, which is not the case for the P test statistic.

\subsection{B.4 \quad Size for Thinned Block Sizes}

In Table \ref{Tab.size_thin}, we provide the values of empirical size for the SNU and de-biased MHM test statistics under the generating models (\ref{eq:nonintegrablefs1}) and (\ref{eq:integrablefs1}),  where the number of blocks is thinned to $M=N^{0.9}$ and the level of endogeneity is equal to $0.5$ ($r=0.5$); results with this number of subsamples are quantitatively similar to results presented earlier with the full subsample number of these generating models.

\newpage
\clearpage

\begin{table}[ht]
\footnotesize
\caption{Empirical power of the test statistics at $5\%$ level for the model given in expression (\ref{eq:nonintegrablefs2}). The values in the curly brackets correspond to the 4 block sizes in the paper, and we assume $h=N^{-1/3}$.} \label{Tab.power_16}
\centering
\setlength{\tabcolsep}{1pt} 
\renewcommand{\arraystretch}{1.15} 
\begin{tabular}{@{} ccccc @{}}
\toprule
test & $d$ $\symbol{92}$ $\hat{f}(.)$ & LM & SLM $(\lambda=N^{-1/3})$ & SLM $(\lambda=N^{-1/6})$ \\\midrule
 && \multicolumn{3}{c}{$r=0.5$} \\
 SNU & $0.1$ $\symbol{92}$ parametric & $\{0.995, 0.957, 0.886, 0.769\}$ & $ \, \{0.998, 0.973, 0.915, 0.828\}$ & $\{0.998, 0.978, 0.920, 0.842\}$ \\
 
& $0.1$ $\symbol{92}$ non-parametric & $\{1.000, 0.997, 0.991, 0.975\}$ & $\{1.000, 0.999, 0.997, 0.988\}$ & $\{1.000, 0.999, 0.997, 0.992\}$  \\[0.5cm]

de-biased MHM & $0.1$ $\symbol{92}$ parametric & $\{0.255, 0.226, 0.126, 0.098\}$ & $\{0.200, 0.169, 0.101, 0.077\}$ & $\{0.181, 0.153, 0.088, 0.071 \}$ \\[0.5cm]

P & $0.1$ $\symbol{92}$ parametric & $(0.875,0.914,0.905)$ & $(0.881,0.906,0.908)$ & $(0.890,0.915,0.921)$ \\
 
 & $\mathcal{L}=(6,12,18)$ \\ \midrule
 
SNU & $0.4$ $\symbol{92}$ parametric & $\{0.930, 0.812, 0.649, 0.489\}$ & $\{ 0.993, 0.955, 0.875, 0.756\}$ & $\{0.997, 0.973, 0.912, 0.821 \}$ \\
 
& $0.4$ $\symbol{92}$ non-parametric & $\{0.954, 0.927, 0.862, 0.767\}$ & $\{0.996, 0.991, 0.979, 0.952\}$ & $\{0.999, 0.998, 0.992, 0.979\}$  \\[0.5cm]

de-biased MHM & $0.4$ $\symbol{92}$ parametric & $\{0.654, 0.619,       0.520, 0.457\}$ & $\{0.354, 0.316, 0.185, 0.147\}$ & $\{0.261, 0.217, 0.114, 0.089 \}$ \\[0.5cm]

P & $0.4$ $\symbol{92}$ parametric & $(0.663,0.726,0.751)$ & $(0.677,0.719,0.719)$ & $(0.751,0.796,0.793)$ \\
 
& $\mathcal{L}=(6,12,18)$ \\ \midrule \midrule
 
 & & \multicolumn{3}{c}{$r=1$} \\
SNU & $0.1$ $\symbol{92}$ parametric & $\{0.993, 0.955, 0.883, 0.778\}$ & $\{0.998, 0.977, 0.921, 0.837\}$ & $\{1.000, 0.976, 0.922, 0.854\}$ \\
 
& $0.1$ $\symbol{92}$ non-parametric & $\{0.996, 0.993, 0.987, 0.967\}$ & $\{0.999, 0.997, 0.994, 0.985\}$ & $\{1.000, 0.998, 0.999, 0.991\}$  \\[0.5cm]

 de-biased MHM & $0.1$ $\symbol{92}$ parametric & $\{0.256, 0.222,       0.117, 0.093\}$ & $\{0.191, 0.162, 0.092, 0.075\}$ & $\{0.188,       0.164, 0.087, 0.071\}$ \\[0.5cm]

P & $0.1$ $\symbol{92}$ parametric & $(0.885,0.911,0.900)$ & $(0.879,0.909,0.904)$ & $(0.894,0.928,0.925)$ \\
 & $\mathcal{L}=(6,12,18)$ \\ \midrule
 
SNU & $0.4$ $\symbol{92}$ parametric & $\{0.925, 0.789, 0.631, 0.480\}$ & $\{0.989, 0.947, 0.859, 0.751\}$ & $\{0.992, 0.967, 0.899, 0.802\}$ \\
 
& $0.4$ $\symbol{92}$ non-parametric & $\{0.955, 0.924, 0.860, 0.754\}$ & $\{0.994, 0.987, 0.975, 0.940\}$ & $\{0.995, 0.991, 0.986, 0.972\}$  \\[0.5cm]

de-biased MHM & $0.4$ $\symbol{92}$ parametric & $\{0.637, 0.607,       0.505, 0.432\}$ & $\{0.356, 0.306, 0.193, 0.161 \}$ & $\{0.257,       0.225, 0.126, 0.102\}$ \\[0.5cm]

P & $0.4$ $\symbol{92}$ parametric & $(0.642,0.723,0.753)$ & $(0.666, 0.704,0.722)$ & $(0.751,0.779,0.787)$ \\
 & $\mathcal{L}=(6,12,18)$ \\ 
\bottomrule
\end{tabular}
\end{table}

\begin{table}[ht]
\footnotesize
\caption{Empirical power of the test statistics at $5\%$ level for the model  given in expression (\ref{eq:nonintegrablefs3}). The values in the curly brackets correspond to the 4 block sizes in the paper, and we assume $h=N^{-1/3}$.} \label{Tab.power_17}
\centering
\setlength{\tabcolsep}{1pt} 
\renewcommand{\arraystretch}{1.15} 
\begin{tabular}{@{} ccccc @{}}
\toprule
test & $d$ $\symbol{92}$ $\hat{f}(.)$ & LM & SLM $(\lambda=N^{-1/3})$ & SLM $(\lambda=N^{-1/6})$ \\\midrule
 && \multicolumn{3}{c}{$r=0.5$} \\
 SNU & $0.1$ $\symbol{92}$ parametric & $\{0.954, 0.877, 0.759, 0.675\}$ & $ \, \{0.972, 0.922, 0.815, 0.738\}$ & $\{0.976, 0.933, 0.829, 0.743\}$ \\
 
& $0.1$ $\symbol{92}$ non-parametric & $\{0.959, 0.950, 0.921, 0.856\}$ & $\{0.974, 0.969, 0.950, 0.913\}$ & $\{0.980, 0.976, 0.956, 0.921\}$  \\[0.5cm]

de-biased MHM & $0.1$ $\symbol{92}$ parametric & $\{0.612, 0.558,       0.443, 0.434\}$ & $\{0.556, 0.484, 0.356, 0.350\}$ & $\{0.540, 0.471, 0.340, 0.330\}$ \\[0.5cm]

P & $0.1$ $\symbol{92}$ parametric & $(0.842, 0.875, 0.874)$ & $(0.824, 0.874, 0.871)$ & $(0.832, 0.875, 0.875)$ \\
 
 & $\mathcal{L}=(6,12,18)$ \\ \midrule
 
SNU & $0.4$ $\symbol{92}$ parametric & $\{0.766, 0.603, 0.440, 0.396\}$ & $\{0.915, 0.797, 0.648, 0.573 \}$ & $\{0.953, 0.879, 0.746, 0.666\}$ \\
 
& $0.4$ $\symbol{92}$ non-parametric & $\{0.795, 0.725, 0.602, 0.440\}$ & $\{0.931, 0.902, 0.822, 0.726\}$ & $\{0.960, 0.943, 0.905, 0.842\}$  \\[0.5cm]

de-biased MHM & $0.4$ $\symbol{92}$ parametric & $\{0.815, 0.797,       0.718, 0.693\}$ & $\{0.637, 0.568, 0.446, 0.445\}$ & $\{0.582, 0.513, 0.378, 0.383\}$ \\[0.5cm]

P & $0.4$ $\symbol{92}$ parametric & $(0.729, 0.733, 0.717)$ & $(0.744, 0.773, 0.779)$ & $(0.762, 0.812, 0.810)$ \\
 
  & $\mathcal{L}=(6,12,18)$ \\ \midrule \midrule
 
 & & \multicolumn{3}{c}{$r=1$} \\
SNU & $0.1$ $\symbol{92}$ parametric & $\{0.948, 0.878, 0.761, 0.672\}$ & $\{0.970, 0.920, 0.822, 0.741\}$ & $\{0.976, 0.928, 0.847, 0.748\}$ \\
 
& $0.1$ $\symbol{92}$ non-parametric & $\{0.938, 0.925, 0.905, 0.835\}$ & $\{0.962, 0.955, 0.940, 0.897\}$ & $\{0.965, 0.962, 0.952, 0.914\}$  \\[0.5cm]

de-biased MHM & $0.1$ $\symbol{92}$ parametric & $\{0.620, 0.564,      0.442, 0.422\}$ & $\{0.560, 0.498, 0.382, 0.361\}$ & $\{0.549, 0.485, 0.365, 0.347\}$ \\[0.5cm]

P & $0.1$ $\symbol{92}$ parametric & $(0.831, 0.875, 0.875)$ & $(0.823, 0.862, 0.859)$ & $(0.823, 0.873, 0.870)$ \\
 & $\mathcal{L}=(6,12,18)$ \\ \midrule
 
SNU & $0.4$ $\symbol{92}$ parametric & $\{0.728, 0.545, 0.369, 0.328\}$ & $\{0.904, 0.790, 0.630, 0.518\}$ & $\{0.941, 0.863, 0.758, 0.670\}$ \\
 
& $0.4$ $\symbol{92}$ non-parametric & $\{0.770, 0.691, 0.562, 0.404\}$ & $\{0.918, 0.885, 0.803, 0.705\}$ & $\{0.946, 0.927, 0.895, 0.828\}$  \\[0.5cm]

de-biased MHM & $0.4$ $\symbol{92}$ parametric & $\{0.812, 0.787,       0.708, 0.686\}$ & $\{0.650, 0.588, 0.467, 0.450 \}$ & $\{0.602,       0.538, 0.401, 0.389\}$ \\[0.5cm]

P & $0.4$ $\symbol{92}$ parametric & $(0.734, 0.749, 0.726)$ & $(0.746, 0.786, 0.786)$ & $(0.773, 0.824, 0.830)$ \\
 & $\mathcal{L}=(6,12,18)$ \\ 
\bottomrule
\end{tabular}
\end{table}

\begin{table}[ht]
\footnotesize
\caption{Empirical power of the test statistics at $5\%$ level for the model  given in expression (\ref{eq:nonintegrablefs4}). The values in the curly brackets correspond to the 4 block sizes in the paper, and we assume $h=N^{-1/3}$.} \label{Tab.power_18}
\centering
\setlength{\tabcolsep}{1pt} 
\renewcommand{\arraystretch}{1.15} 
\begin{tabular}{@{} ccccc @{}}
\toprule
test & $d$ $\symbol{92}$ $\hat{f}(.)$ & LM & SLM $(\lambda=N^{-1/3})$ & SLM $(\lambda=N^{-1/6})$ \\\midrule
 && \multicolumn{3}{c}{$r=0.5$} \\
 SNU & $0.1$ $\symbol{92}$ parametric & $\{0.845, 0.815, 0.744, 0.672\}$ & $ \, \{ 0.751, 0.740, 0.721, 0.681 \}$ &  $\{0.756, 0.739, 0.717, 0.676\}$ \\
 
& $0.1$ $\symbol{92}$ non-parametric & $\{0.825, 0.801, 0.791, 0.790\}$ & $\{0.753, 0.742, 0.725, 0.691\}$ & $\{0.755, 0.742, 0.720, 0.694\}$  \\[0.5cm]

de-biased MHM & $0.1$ $\symbol{92}$ parametric & $\{0.419, 0.405,       0.389, 0.349\}$ & $\{0.427, 0.411, 0.386, 0.348\}$ & $\{0.436, 0.416,       0.394, 0.352 \}$ \\[0.5cm]

P & $0.1$ $\symbol{92}$ parametric & $(0.560, 0.579, 0.591)$ & $(0.653, 0.655, 0.648)$ & $(0.642, 0.641, 0.633)$ \\
 
 & $\mathcal{L}=(6,12,18)$ \\ \midrule
 
SNU & $0.4$ $\symbol{92}$ parametric & $\{0.759, 0.704, 0.626, 0.548\}$ & $\{0.770, 0.754, 0.730, 0.670 \}$ & $\{0.762, 0.748, 0.726, 0.676 \}$ \\
 
& $0.4$ $\symbol{92}$ non-parametric & $\{0.739, 0.706, 0.679, 0.633\}$ & $\{0.770, 0.755, 0.730, 0.702\}$ & $\{0.764, 0.747, 0.726, 0.703\}$  \\[0.5cm]

de-biased MHM & $0.4$ $\symbol{92}$ parametric & $\{0.333, 0.325,       0.307, 0.280 \}$ & $\{0.386, 0.369, 0.344, 0.310\}$ & $\{0.414,       0.393, 0.370, 0.327 \}$ \\[0.5cm]

P & $0.4$ $\symbol{92}$ parametric & $(0.730, 0.722, 0.711)$ & $(0.714, 0.694, 0.676)$ & $(0.690, 0.679,  0.670)$ \\
 
  & $\mathcal{L}=(6,12,18)$ \\ \midrule \midrule
 
 & & \multicolumn{3}{c}{$r=1$} \\
SNU & $0.1$ $\symbol{92}$ parametric & $\{0.841, 0.812, 0.746, 0.676\}$ & $\{0.740, 0.730, 0.706, 0.659\}$ & $\{0.741, 0.733, 0.714, 0.661\}$ \\
 
& $0.1$ $\symbol{92}$ non-parametric & $\{0.820, 0.788, 0.780, 0.782\}$ & $\{0.724, 0.719, 0.704, 0.680\}$ & $\{0.720, 0.714, 0.707, 0.682\}$  \\[0.5cm]

de-biased MHM & $0.1$ $\symbol{92}$ parametric & $\{0.451, 0.436,       0.412, 0.378\}$ & $\{0.438, 0.423, 0.405, 0.364\}$ & $\{0.439, 0.424,      0.404, 0.361\}$ \\[0.5cm]

P & $0.1$ $\symbol{92}$ parametric & $(0.572, 0.604, 0.608)$ & $(0.663, 0.659, 0.653)$ & $(0.657, 0.651, 0.645)$ \\
 & $\mathcal{L}=(6,12,18)$ \\ \midrule
 
SNU & $0.4$ $\symbol{92}$ parametric & $\{0.756, 0.706, 0.635, 0.549\}$ & $\{0.784, 0.752, 0.724, 0.655\}$ & $\{0.776, 0.745, 0.721, 0.670\}$ \\
 
& $0.4$ $\symbol{92}$ non-parametric & $\{0.740, 0.696, 0.682, 0.637\}$ & $\{0.787, 0.756, 0.717, 0.685\}$ & $\{0.774, 0.748, 0.714, 0.683\}$  \\[0.5cm]

de-biased MHM & $0.4$ $\symbol{92}$ parametric & $\{0.383, 0.373,       0.357, 0.327 \}$ & $\{0.422, 0.402, 0.383, 0.348\}$ & $\{0.435,       0.416, 0.393, 0.359\}$ \\[0.5cm]

P & $0.4$ $\symbol{92}$ parametric & $(0.736, 0.737, 0.727)$ & $(0.705, 0.690, 0.680)$ & $(0.694, 0.676, 0.665)$ \\
 & $\mathcal{L}=(6,12,18)$ \\ 
\bottomrule
\end{tabular}
\end{table}

\begin{table}[ht]
\footnotesize
\caption{Empirical power of the test statistics at $5\%$ level for the model  given in expression (\ref{eq:nonintegrablefs5}). The values in the curly brackets correspond to the 4 block sizes in the paper, and we assume $h=N^{-1/3}$.} \label{Tab.power_19}
\centering
\setlength{\tabcolsep}{1pt} 
\renewcommand{\arraystretch}{1.15} 
\begin{tabular}{@{} ccccc @{}}
\toprule
test & $d$ $\symbol{92}$ $\hat{f}(.)$ & LM & SLM $(\lambda=N^{-1/3})$ & SLM $(\lambda=N^{-1/6})$ \\\midrule
 && \multicolumn{3}{c}{$r=0.5$} \\
 SNU & $0.1$ $\symbol{92}$ parametric & $\{1.000, 1.000, 1.000, 0.997\}$ & $ \, \{1.000, 1.000, 1.000, 1.000\}$ & $\{1.000, 1.000, 1.000, 0.999\}$ \\
 
& $0.1$ $\symbol{92}$ non-parametric & $\{1.000, 1.000, 1.000, 1.000\}$ & $\{1.000, 1.000, 1.000, 1.000\}$ & $\{1.000, 1.000, 1.000, 1.000\}$  \\[0.5cm]

de-biased MHM & $0.1$ $\symbol{92}$ parametric & $\{0.280, 0.280,       0.277, 0.270\}$ & $\{0.263, 0.263, 0.260, 0.254\}$ & $\{0.280, 0.279,      0.276, 0.269\}$ \\[0.5cm]

P & $0.1$ $\symbol{92}$ parametric & $(0.567, 0.563, 0.551)$ & $(0.532, 0.541, 0.539)$ & $(0.453, 0.475, 0.477)$ \\
 
 & $\mathcal{L}=(6,12,18)$ \\ \midrule
 
SNU & $0.4$ $\symbol{92}$ parametric & $\{1.000, 1.000, 0.998, 0.951\}$ & $\{1.000, 1.000, 1.000, 0.995 \}$ & $\{1.000, 1.000, 1.000, 0.998 \}$ \\
 
& $0.4$ $\symbol{92}$ non-parametric & $\{1.000, 1.000, 1.000, 1.000\}$ & $\{1.000, 1.000, 1.000, 1.000\}$ & $\{1.000, 1.000, 1.000, 1.000\}$  \\[0.5cm]

de-biased MHM & $0.4$ $\symbol{92}$ parametric & $\{0.196, 0.196,       0.196, 0.195\}$ & $\{0.196, 0.196, 0.194, 0.189\}$ & $\{0.243, 0.242,      0.240, 0.235\}$ \\[0.5cm]

P & $0.4$ $\symbol{92}$ parametric & $(0.779, 0.748, 0.723)$ & $(0.955, 0.937, 0.920)$ & $(0.952, 0.929, 0.918)$ \\
 
  & $\mathcal{L}=(6,12,18)$ \\ \midrule \midrule
 
 & & \multicolumn{3}{c}{$r=1$} \\
SNU & $0.1$ $\symbol{92}$ parametric & $\{1.000, 1.000, 1.000, 1.000\}$ & $\{1.000, 1.000, 1.000, 0.999\}$ & $\{1.000, 1.000, 1.000, 1.000\}$ \\
 
& $0.1$ $\symbol{92}$ non-parametric & $\{1.000, 1.000, 1.000, 1.000\}$ & $\{1.000, 1.000, 1.000, 1.000\}$ & $\{1.000, 1.000, 1.000, 1.000\}$  \\[0.5cm]

de-biased MHM & $0.1$ $\symbol{92}$ parametric & $\{0.274, 0.274,       0.272, 0.266\}$ & $\{0.259, 0.259, 0.257, 0.253\}$ & $\{0.276, 0.276,      0.274, 0.269\}$ \\[0.5cm]

P & $0.1$ $\symbol{92}$ parametric & $(0.595, 0.583, 0.580)$ & $(0.575, 0.585, 0.581)$ & $(0.487, 0.509, 0.530)$ \\
 & $\mathcal{L}=(6,12,18)$ \\ \midrule
 
SNU & $0.4$ $\symbol{92}$ parametric & $\{1.000, 1.000, 0.997, 0.957\}$ & $\{1.000, 1.000, 1.000, 0.992\}$ & $\{1.000, 1.000, 1.000, 0.997\}$ \\
 
& $0.4$ $\symbol{92}$ non-parametric & $\{1.000, 1.000, 1.000, 1.000\}$ & $\{1.000, 1.000, 1.000, 1.000\}$ & $\{1.000, 1.000, 1.000, 1.000\}$  \\[0.5cm]

de-biased MHM & $0.4$ $\symbol{92}$ parametric & $\{0.187, 0.187,      0.187, 0.186\}$ & $\{0.204, 0.203, 0.201, 0.198\}$ & $\{0.242, 0.242,      0.240, 0.233\}$ \\[0.5cm]

P & $0.4$ $\symbol{92}$ parametric & $(0.795, 0.765, 0.746)$ & $(0.959, 0.941, 0.927)$ & $(0.954, 0.945, 0.930)$ \\
 & $\mathcal{L}=(6,12,18)$ \\ 
\bottomrule
\end{tabular}
\end{table}

\begin{table}[ht]
\footnotesize
\caption{Empirical size of the test statistics at $5\%$ level for the model given in expression (\ref{eq:integrablefs1}). The values in the curly brackets correspond to the 4 block sizes in the paper, and we assume $h=N^{-1/3}$.} \label{Tab.size_20}
\centering
\setlength{\tabcolsep}{1pt} 
\renewcommand{\arraystretch}{1.15} 
\begin{tabular}{@{} ccccc @{}}
\toprule
test & $d$ $\symbol{92}$ $\hat{f}(.)$ & LM & SLM $(\lambda=N^{-1/3})$ & SLM $(\lambda=N^{-1/6})$ \\\midrule
 && \multicolumn{3}{c}{$r=0.5$} \\
 SNU & $0.1$ $\symbol{92}$ parametric & $\{0.094, 0.071, 0.066, 0.072\}$ & $ \, \{0.075, 0.059, 0.056, 0.062\}$ & $\{0.091, 0.072, 0.066, 0.070\}$ \\
 
& $0.1$ $\symbol{92}$ non-parametric & $\{0.137, 0.075, 0.029, 0.009\}$ & $\{0.117, 0.062, 0.029, 0.010\}$ & $\{0.139, 0.071, 0.032, 0.012\}$  \\[0.5cm]

de-biased MHM & $0.1$ $\symbol{92}$ parametric & $\{0.210, 0.187,       0.106, 0.117\}$ & $\{0.190, 0.186, 0.097, 0.126\}$ & $\{0.176, 0.165,      0.092, 0.117\}$ \\[0.5cm]

P & $0.1$ $\symbol{92}$ parametric & $(0.050, 0.051, 0.056)$ & $(0.049, 0.051, 0.054)$ & $(0.050, 0.052, 0.056)$ \\
 
 & $\mathcal{L}=(6,12,18)$ \\ \midrule
 
SNU & $0.4$ $\symbol{92}$ parametric & $\{0.098, 0.069, 0.054, 0.064\}$ & $\{0.073, 0.059, 0.057, 0.061 \}$ & $\{0.083, 0.067, 0.065, 0.067\}$ \\
 
& $0.4$ $\symbol{92}$ non-parametric & $\{0.124, 0.060, 0.023, 0.006\}$ & $\{0.096, 0.050, 0.014, 0.003\}$ & $\{0.112, 0.053, 0.023, 0.004\}$  \\[0.5cm]

de-biased MHM & $0.4$ $\symbol{92}$ parametric & $\{0.292, 0.264,       0.169, 0.158\}$ & $\{0.219, 0.204, 0.114, 0.127\}$ & $\{0.200, 0.189,       0.111, 0.115\}$ \\[0.5cm]

P & $0.4$ $\symbol{92}$ parametric & $(0.047, 0.050, 0.056)$ & $(0.051, 0.050, 0.054)$ & $(0.049, 0.051, 0.054)$ \\
 
  & $\mathcal{L}=(6,12,18)$ \\ \midrule \midrule
 
 & & \multicolumn{3}{c}{$r=1$} \\
SNU & $0.1$ $\symbol{92}$ parametric & $\{0.102, 0.102, 0.094, 0.096\}$ & $\{0.074, 0.064, 0.072, 0.076\}$ & $\{0.062, 0.051, 0.051, 0.060\}$ \\
 
& $0.1$ $\symbol{92}$ non-parametric & $\{0.109, 0.035, 0.004, 0.001\}$ & $\{0.078, 0.030, 0.007, 0.001\}$ & $\{0.079, 0.034, 0.010, 0.002\}$  \\[0.5cm]

de-biased MHM & $0.1$ $\symbol{92}$ parametric & $\{0.311, 0.289,       0.188, 0.200 \}$ & $\{0.269, 0.253, 0.157, 0.180\}$ & $\{0.232, 0.225,       0.132, 0.160\}$ \\[0.5cm]

P & $0.1$ $\symbol{92}$ parametric & $(0.049, 0.051, 0.053)$ & $(0.049, 0.050, 0.054)$ & $(0.049, 0.050, 0.055)$ \\
 & $\mathcal{L}=(6,12,18)$ \\ \midrule
 
SNU & $0.4$ $\symbol{92}$ parametric & $\{0.245, 0.176, 0.146, 0.134\}$ & $\{0.432, 0.303, 0.213, 0.156\}$ & $\{0.280, 0.204, 0.164, 0.126\}$ \\
 
& $0.4$ $\symbol{92}$ non-parametric & $\{0.224, 0.108, 0.032, 0.003\}$ & $\{0.378, 0.189, 0.049, 0.006\}$ & $\{0.235, 0.099, 0.021, 0.001\}$  \\[0.5cm]

de-biased MHM & $0.4$ $\symbol{92}$ parametric & $\{0.346, 0.320, 0.226, 0.194\}$ & $\{0.320, 0.312, 0.226, 0.214\}$ & $\{0.305, 0.285, 0.197,       0.218\}$ \\[0.5cm]

P & $0.4$ $\symbol{92}$ parametric & $(0.051, 0.051, 0.054)$ & $(0.049, 0.051, 0.056)$ & $(0.048, 0.052, 0.056)$ \\
 & $\mathcal{L}=(6,12,18)$ \\ 
\bottomrule
\end{tabular}
\end{table}

\begin{table}[ht]
\footnotesize
\caption{Empirical power of the test statistics at $5\%$ level for the model  given in expression (\ref{eq:integrablefs2}). The values in the curly brackets correspond to the 4 block sizes in the paper, and we assume $h=N^{-1/3}$.} \label{Tab.power_21}
\centering
\setlength{\tabcolsep}{1pt} 
\renewcommand{\arraystretch}{1.15} 
\begin{tabular}{@{} ccccc @{}}
\toprule
test & $d$ $\symbol{92}$ $\hat{f}(.)$ & LM & SLM $(\lambda=N^{-1/3})$ & SLM $(\lambda=N^{-1/6})$ \\\midrule
 && \multicolumn{3}{c}{$r=0.5$} \\
 SNU & $0.1$ $\symbol{92}$ parametric & $\{0.943, 0.865, 0.732, 0.589\}$ & $ \, \{0.972, 0.911, 0.807, 0.688\}$ & 
 $\{0.980, 0.922, 0.823, 0.699\}$ \\
 
& $0.1$ $\symbol{92}$ non-parametric & $\{0.980, 0.968, 0.945, 0.918\}$ & $\{0.993, 0.984, 0.973, 0.958\}$ & $\{0.997, 0.992, 0.979, 0.964\}$  \\[0.5cm]

de-biased MHM & $0.1$ $\symbol{92}$ parametric & $\{0.359, 0.290, 0.135,      0.125\}$ & $\{0.314, 0.244, 0.124, 0.123\}$ & $\{0.296, 0.234, 0.125,       0.120 \}$ \\[0.5cm]

P & $0.1$ $\symbol{92}$ parametric & $(0.904, 0.943, 0.940)$ & $(0.919, 0.940, 0.945)$ & $(0.927, 0.950, 0.956)$ \\
 
 & $\mathcal{L}=(6,12,18)$ \\ \midrule
 
SNU & $0.4$ $\symbol{92}$ parametric & $\{0.774, 0.633, 0.476, 0.429\}$ & $\{0.936, 0.853, 0.739, 0.666\}$ & $\{0.964, 0.898, 0.801, 0.691\}$ \\
 
& $0.4$ $\symbol{92}$ non-parametric & $\{0.866, 0.796, 0.703, 0.616\}$ & $\{0.974, 0.947, 0.915, 0.864\}$ & $\{0.987, 0.974, 0.951, 0.931\}$   \\[0.5cm]

de-biased MHM & $0.4$ $\symbol{92}$ parametric & $\{0.684, 0.634, 0.430,       0.396\}$ & $\{0.419, 0.332, 0.171, 0.159\}$ & $\{0.343, 0.258, 0.136,       0.128 \}$ \\[0.5cm]

P & $0.4$ $\symbol{92}$ parametric & $(0.669, 0.736, 0.765)$ & $(0.725, 0.771, 0.785)$ & $(0.814, 0.864, 0.876)$ \\
 
  & $\mathcal{L}=(6,12,18)$ \\ \midrule \midrule
 
 & & \multicolumn{3}{c}{$r=1$} \\
SNU & $0.1$ $\symbol{92}$ parametric & $\{0.949, 0.874, 0.742, 0.607\}$ & $\{0.974, 0.915, 0.813, 0.688\}$ & $\{0.981, 0.922, 0.830, 0.692\}$ \\
 
& $0.1$ $\symbol{92}$ non-parametric & $\{0.981, 0.963, 0.943, 0.912\}$ & $\{0.989, 0.981, 0.970, 0.952\}$ & $\{0.996, 0.988, 0.980, 0.965\}$  \\[0.5cm]

de-biased MHM & $0.1$ $\symbol{92}$ parametric & $\{0.351, 0.292, 0.133,       0.122\}$ & $\{0.320, 0.250, 0.125, 0.111\}$ & $\{0.291, 0.235, 0.120,       0.106\}$ \\[0.5cm]

P & $0.1$ $\symbol{92}$ parametric & $(0.918, 0.941, 0.936)$ & $(0.926, 0.947, 0.952)$ & $(0.935, 0.957, 0.962)$ \\
 & $\mathcal{L}=(6,12,18)$ \\ \midrule
 
SNU & $0.4$ $\symbol{92}$ parametric & $\{0.812, 0.642, 0.480, 0.409\}$ & $\{0.950, 0.860, 0.729, 0.630\}$ & $\{0.970, 0.910, 0.796, 0.692\}$ \\
 
& $0.4$ $\symbol{92}$ non-parametric & $\{0.867, 0.782, 0.681, 0.564\}$ & $\{0.965, 0.938, 0.899, 0.849\}$ & $\{0.980, 0.966, 0.949, 0.919\}$  \\[0.5cm]

de-biased MHM & $0.4$ $\symbol{92}$ parametric & $\{0.684, 0.633, 0.416,       0.386\}$ & $\{0.408, 0.317, 0.172, 0.161\}$ & $\{0.359, 0.265, 0.129,        0.128 \}$ \\[0.5cm]

P & $0.4$ $\symbol{92}$ parametric & $(0.653, 0.738, 0.767)$ & $(0.716, 0.762, 0.788)$ & $(0.815, 0.860, 0.878)$ \\
 & $\mathcal{L}=(6,12,18)$ \\ 
\bottomrule
\end{tabular}
\end{table}

\begin{table}[ht]
\footnotesize
\caption{Empirical power of the test statistics at $5\%$ level for the model given in expression (\ref{eq:integrablefs3}). The values in the curly brackets correspond to the 4 block sizes in the paper, and we assume $h=N^{-1/3}$.} \label{Tab.power_22}
\centering
\setlength{\tabcolsep}{1pt} 
\renewcommand{\arraystretch}{1.15} 
\begin{tabular}{@{} ccccc @{}}
\toprule
test & $d$ $\symbol{92}$ $\hat{f}(.)$ & LM & SLM $(\lambda=N^{-1/3})$ & SLM $(\lambda=N^{-1/6})$ \\\midrule
 && \multicolumn{3}{c}{$r=0.5$} \\
 SNU & $0.1$ $\symbol{92}$ parametric & $\{0.833, 0.785, 0.742, 0.708\}$ & $ \, \{0.883, 0.849, 0.791, 0.752\}$ & 
 $\{0.886, 0.857, 0.804, 0.768\}$ \\
 
& $0.1$ $\symbol{92}$ non-parametric & $\{0.891, 0.847, 0.784, 0.710\}$ & $\{0.921, 0.896, 0.848, 0.788\}$ & $\{0.926, 0.903, 0.862, 0.810\}$  \\[0.5cm]

de-biased MHM & $0.1$ $\symbol{92}$ parametric & $\{0.697, 0.571, 0.433,       0.448\}$ & $\{0.640, 0.483, 0.353, 0.396\}$ & $\{0.627, 0.463, 0.342,       0.377 \}$ \\[0.5cm]

P & $0.1$ $\symbol{92}$ parametric & $(0.857, 0.889, 0.890)$ & $(0.852, 0.905, 0.902)$ & $(0.862, 0.917, 0.920)$ \\
 
 & $\mathcal{L}=(6,12,18)$ \\ \midrule
 
SNU & $0.4$ $\symbol{92}$ parametric & $\{0.575, 0.495, 0.423, 0.435\}$ & $\{0.780, 0.718, 0.655, 0.635 \}$ & $\{0.842, 0.804, 0.747, 0.708 \}$ \\
 
& $0.4$ $\symbol{92}$ non-parametric & $\{0.654, 0.538, 0.413, 0.288\}$ & $\{0.833, 0.761, 0.677, 0.581\}$ & $\{0.887, 0.841, 0.779, 0.697\}$  \\[0.5cm]

de-biased MHM & $0.4$ $\symbol{92}$ parametric & $\{0.862, 0.841, 0.694,       0.693\}$ & $\{0.718, 0.597, 0.444, 0.470\}$ & $\{0.668, 0.523, 0.380,       0.399\}$ \\[0.5cm]

P & $0.4$ $\symbol{92}$ parametric & $(0.742, 0.739, 0.719)$ & $(0.750, 0.785, 0.793)$ & $(0.779, 0.830, 0.834)$ \\
 
 & $\mathcal{L}=(6,12,18)$ \\ \midrule \midrule
 
 & & \multicolumn{3}{c}{$r=1$} \\
SNU & $0.1$ $\symbol{92}$ parametric & $\{0.848, 0.814, 0.757, 0.715\}$ & $\{0.888, 0.864, 0.816, 0.771\}$ & $\{0.901, 0.875, 0.826, 0.780\}$ \\
 
& $0.1$ $\symbol{92}$ non-parametric & $\{0.871, 0.822, 0.765, 0.693\}$ & $\{0.912, 0.872, 0.826, 0.774\}$ & $\{0.923, 0.892, 0.843, 0.790\}$  \\[0.5cm]

de-biased MHM & $0.1$ $\symbol{92}$ parametric & $\{0.710, 0.561, 0.432,       0.454 \}$ & $\{0.650, 0.487, 0.367, 0.401\}$ & $\{0.619, 0.458, 0.341,       0.392\}$ \\[0.5cm]

P & $0.1$ $\symbol{92}$ parametric & $(0.849, 0.895, 0.896)$ & $(0.851, 0.893, 0.893)$ & $(0.852, 0.907, 0.906)$ \\
 & $\mathcal{L}=(6,12,18)$ \\ \midrule
 
SNU & $0.4$ $\symbol{92}$ parametric & $\{0.606, 0.502, 0.411, 0.403\}$ & $\{0.814, 0.742, 0.652, 0.597\}$ & $\{0.879, 0.828, 0.762, 0.712\}$ \\
 
& $0.4$ $\symbol{92}$ non-parametric & $\{0.620, 0.494, 0.378, 0.272\}$ & $\{0.808, 0.735, 0.648, 0.565\}$ & $\{0.872, 0.821, 0.758, 0.685\}$  \\[0.5cm]

de-biased MHM & $0.4$ $\symbol{92}$ parametric & $\{0.855, 0.826, 0.685,       0.673\}$ & $\{0.727, 0.605, 0.446, 0.466\}$ & $\{0.674, 0.522, 0.391,       0.431\}$ \\[0.5cm]

P & $0.4$ $\symbol{92}$ parametric & $(0.739, 0.752, 0.732)$ & $(0.751, 0.795, 0.793)$ & $(0.792, 0.846, 0.850)$ \\
 & $\mathcal{L}=(6,12,18)$ \\ 
\bottomrule
\end{tabular}
\end{table}

\begin{table}[ht]
\footnotesize
\caption{Empirical power of the test statistics at $5\%$ level for the model given in expression (\ref{eq:integrablefs4}). The values in the curly brackets correspond to the 4 block sizes in the paper, and we assume $h=N^{-1/3}$.} \label{Tab.power_23}
\centering
\setlength{\tabcolsep}{1pt} 
\renewcommand{\arraystretch}{1.15} 
\begin{tabular}{@{} ccccc @{}}
\toprule
test & $d$ $\symbol{92}$ $\hat{f}(.)$ & LM & SLM $(\lambda=N^{-1/3})$ & SLM $(\lambda=N^{-1/6})$ \\\midrule
 && \multicolumn{3}{c}{$r=0.5$} \\
 SNU & $0.1$ $\symbol{92}$ parametric & $\{1.000, 1.000, 1.000, 0.997\}$ & $ \, \{1.000, 0.999, 0.998, 0.994\}$ & $\{1.000, 1.000, 1.000, 0.993\}$ \\
 
& $0.1$ $\symbol{92}$ non-parametric & $\{1.000, 1.000, 1.000, 1.000\}$ & $\{1.000, 0.999, 0.999, 0.999\}$ & $\{1.000, 1.000, 1.000, 1.000\}$  \\[0.5cm]

de-biased MHM & $0.1$ $\symbol{92}$ parametric & $\{0.254, 0.237, 0.192,       0.164\}$ & $\{0.286, 0.263, 0.215, 0.173\}$ & $\{0.302, 0.285, 0.235,       0.193\}$ \\[0.5cm]

P & $0.1$ $\symbol{92}$ parametric & $(0.854, 0.924, 0.943)$ & $(0.999, 1.000, 0.999)$ & $(1.000, 1.000, 0.999)$ \\
 
 & $\mathcal{L}=(6,12,18)$ \\ \midrule
 
SNU & $0.4$ $\symbol{92}$ parametric & $\{1.000, 1.000, 0.992, 0.929\}$ & $\{1.000, 1.000, 1.000, 1.000 \}$ & $\{1.000, 1.000, 1.000, 1.000 \}$ \\
 
& $0.4$ $\symbol{92}$ non-parametric & $\{1.000, 1.000, 1.000, 1.000\}$ & $\{1.000, 1.000, 1.000, 1.000\}$ & $\{1.000, 1.000, 1.000, 1.000\}$  \\[0.5cm]

de-biased MHM & $0.4$ $\symbol{92}$ parametric & $\{0.102, 0.098, 0.083,      0.071\}$ & $\{0.198, 0.178, 0.144, 0.120\}$ & $\{0.268, 0.244, 0.192,       0.166\}$ \\[0.5cm]

P & $0.4$ $\symbol{92}$ parametric & $(0.999, 0.999, 0.999)$ & $(1.000, 1.000, 0.998)$ & $(0.999, 1.000, 1.000)$ \\
 
  & $\mathcal{L}=(6,12,18)$ \\ \midrule \midrule
 
 & & \multicolumn{3}{c}{$r=1$} \\
SNU & $0.1$ $\symbol{92}$ parametric &  $\{1.000, 1.000, 1.000, 0.998 \}$ & $\{1.000, 1.000, 1.000, 1.000\}$ & $\{1.000, 1.000, 0.999, 0.994\}$ \\
 
& $0.1$ $\symbol{92}$ non-parametric & $\{1.000, 1.000, 1.000, 1.000\}$ & $\{1.000, 1.000, 1.000, 1.000\}$ & $\{1.000, 1.000, 1.000, 1.000\}$  \\[0.5cm]

de-biased MHM & $0.1$ $\symbol{92}$ parametric & $\{0.227, 0.211, 0.178,       0.140\}$ & $\{0.270, 0.246, 0.204, 0.167\}$ & $\{0.293, 0.269, 0.221,       0.187\}$ \\[0.5cm]

P & $0.1$ $\symbol{92}$ parametric & $(0.884, 0.938, 0.951)$ & $(0.999, 1.000, 1.000)$ & $(0.998, 0.999, 0.999)$ \\
 & $\mathcal{L}=(6,12,18)$ \\ \midrule
 
SNU & $0.4$ $\symbol{92}$ parametric & $\{1.000, 1.000, 0.995, 0.935\}$ & $\{1.000, 1.000, 1.000, 1.000\}$ & $\{1.000, 1.000, 1.000, 0.999\}$ \\
 
& $0.4$ $\symbol{92}$ non-parametric & $\{1.000, 1.000, 1.000, 1.000\}$ & $\{1.000, 1.000, 1.000, 1.000\}$ & $\{1.000, 1.000, 0.999, 0.999\}$  \\[0.5cm]

de-biased MHM & $0.4$ $\symbol{92}$ parametric & $\{0.099, 0.095, 0.085,      0.073\}$ & $\{0.173, 0.159, 0.121, 0.103\}$ & $\{0.235, 0.202, 0.170,       0.141\}$ \\[0.5cm]

P & $0.4$ $\symbol{92}$ parametric & $(0.999, 1.000, 1.000)$ & $(0.998, 0.998, 0.997)$ & $(0.998, 0.998, 0.998)$ \\
 & $\mathcal{L}=(6,12,18)$ \\ 
\bottomrule
\end{tabular}
\end{table}

\begin{table}[ht]
\footnotesize
\caption{Empirical power of the test statistics at $5\%$ level for the model given in expression (\ref{eq:integrablefs5}). The values in the curly brackets correspond to the 4 block sizes in the paper, and we assume $h=N^{-1/3}$.} \label{Tab.power_24}
\centering
\setlength{\tabcolsep}{1pt} 
\renewcommand{\arraystretch}{1.15} 
\begin{tabular}{@{} ccccc @{}}
\toprule
test & $d$ $\symbol{92}$ $\hat{f}(.)$ & LM & SLM $(\lambda=N^{-1/3})$ & SLM $(\lambda=N^{-1/6})$ \\\midrule
 && \multicolumn{3}{c}{$r=0.5$} \\
 SNU & $0.1$ $\symbol{92}$ parametric & $\{1.000, 1.000, 0.991, 0.926\}$ & $ \, \{1.000, 1.000, 0.995, 0.945\}$ & $\{1.000, 1.000, 0.995, 0.951\}$ \\
 
& $0.1$ $\symbol{92}$ non-parametric &  $\{1.000, 1.000, 1.000, 1.000\}$ & $\{1.000, 1.000, 1.000, 1.000\}$ & $\{1.000, 1.000, 1.000, 1.000\}$  \\[0.5cm]

de-biased MHM & $0.1$ $\symbol{92}$ parametric & $\{0.201, 0.185, 0.142,       0.105\}$ & $\{0.209, 0.179, 0.130, 0.104\}$ & $\{0.207, 0.176, 0.128,       0.103\}$ \\[0.5cm]

P & $0.1$ $\symbol{92}$ parametric & $(0.921, 0.959, 0.971)$ & $(0.899, 0.952, 0.956)$ & $(0.891, 0.940, 0.946)$ \\
 
 & $\mathcal{L}=(6,12,18)$ \\ \midrule
 
SNU & $0.4$ $\symbol{92}$ parametric & $\{0.994, 0.964, 0.880, 0.694\}$ & $\{1.000, 0.999, 0.990, 0.920\}$ & $\{1.000, 1.000, 0.995, 0.937 \}$ \\
 
& $0.4$ $\symbol{92}$ non-parametric & $\{1.000, 1.000, 0.995, 0.995\}$ & $\{1.000, 1.000, 1.000, 1.000\}$ & $\{1.000, 1.000, 1.000, 1.000\}$  \\[0.5cm]

de-biased MHM & $0.4$ $\symbol{92}$ parametric & $\{0.157, 0.152, 0.135,   0.106 \}$ & $\{0.216, 0.196, 0.142, 0.108 \}$ & $\{0.219, 0.194, 0.139,      0.108\}$ \\[0.5cm]

P & $0.4$ $\symbol{92}$ parametric & $(1.000, 1.000, 1.000)$ & $(0.997, 0.999, 0.999)$ & $(0.978, 0.984, 0.985)$ \\
 
  & $\mathcal{L}=(6,12,18)$ \\ \midrule \midrule
 
 & & \multicolumn{3}{c}{$r=1$} \\
SNU & $0.1$ $\symbol{92}$ parametric & $\{1.000, 0.999, 0.992, 0.917\}$ & $\{1.000, 1.000, 0.997, 0.948\}$ & $\{1.000, 1.000, 0.997, 0.952\}$ \\
 
& $0.1$ $\symbol{92}$ non-parametric & $\{1.000, 1.000, 1.000, 1.000\}$ & $\{1.000, 1.000, 1.000, 1.000\}$ & $\{1.000, 1.000, 1.000, 1.000\}$  \\[0.5cm]

de-biased MHM & $0.1$ $\symbol{92}$ parametric & $\{0.201, 0.185, 0.144,     0.115\}$ & $\{0.209, 0.185, 0.143, 0.119\}$ & $\{0.209, 0.178, 0.139,       0.116\}$ \\[0.5cm]

P & $0.1$ $\symbol{92}$ parametric & $(0.940, 0.971, 0.971)$ & $(0.919, 0.957, 0.961)$ & $(0.898, 0.945, 0.948)$ \\
 & $\mathcal{L}=(6,12,18)$ \\ \midrule
 
SNU & $0.4$ $\symbol{92}$ parametric & $\{0.997, 0.972, 0.874, 0.692\}$ & $\{1.000, 1.000, 0.994, 0.924\}$ & $\{1.000, 1.000, 0.994, 0.933\}$ \\
 
& $0.4$ $\symbol{92}$ non-parametric &  $\{1.000, 1.000, 0.998, 0.996\}$ & $\{1.000, 1.000, 1.000, 1.000\}$ & $\{1.000, 1.000, 1.000, 1.000\}$  \\[0.5cm]

de-biased MHM & $0.4$ $\symbol{92}$ parametric & $\{0.159, 0.153, 0.131,    0.101\}$ & $\{0.219, 0.197, 0.153, 0.122\}$ & $\{0.222, 0.199, 0.151,       0.120\}$ \\[0.5cm]

P & $0.4$ $\symbol{92}$ parametric & $(1.000, 1.000, 1.000)$ & $(0.997, 0.999, 0.999)$ & $(0.983, 0.987, 0.989)$ \\
 & $\mathcal{L}=(6,12,18)$ \\ 
\bottomrule
\end{tabular}
\end{table}

\begin{table}[ht]
\footnotesize
\caption{Empirical size of the test statistics at $5\%$ level for the model  given in expression (\ref{eq:integrablefs1}), where we assume $u_k$'s follow an MA(1) process. The values in the curly brackets correspond to the 4 block sizes in the paper, and we assume $h=N^{-1/3}$.} \label{Tab.size_20_MA}
\centering
\setlength{\tabcolsep}{1pt} 
\renewcommand{\arraystretch}{1.15} 
\begin{tabular}{@{} ccccc @{}}
\toprule
test & $d$ $\symbol{92}$ $\hat{f}(.)$ & LM & SLM $(\lambda=N^{-1/3})$ & SLM $(\lambda=N^{-1/6})$ \\\midrule
 && \multicolumn{3}{c}{$r=0.5$} \\
 SNU & $0.1$ $\symbol{92}$ parametric & $\{0.103, 0.065, 0.059, 0.062\}$ & $ \, \{0.085, 0.059, 0.049, 0.056\}$ & $\{0.092, 0.064, 0.053, 0.055\}$ \\
 
& $0.1$ $\symbol{92}$ non-parametric & $\{0.202, 0.134, 0.072, 0.023\}$ & $\{0.193, 0.117, 0.058, 0.024\}$ & $\{0.207, 0.130, 0.065, 0.023\}$  \\[0.5cm]

P & $0.1$ $\symbol{92}$ parametric & $(1.000, 1.000, 1.000)$ & $(1.000, 1.000, 1.000)$ & $(1.000, 1.000, 1.000)$ \\
 
 & $\mathcal{L}=(6,12,18)$ \\ \midrule
 
SNU & $0.4$ $\symbol{92}$ parametric & $\{0.154, 0.098, 0.071, 0.064\}$ & $\{0.081, 0.056, 0.046, 0.056\}$ & $\{0.087, 0.067, 0.056, 0.059\}$ \\
 
& $0.4$ $\symbol{92}$ non-parametric & $\{0.256, 0.163, 0.077, 0.028\}$ & $\{0.164, 0.090, 0.037, 0.012\}$ & $\{0.158, 0.096, 0.050, 0.018\}$  \\[0.5cm]

P & $0.4$ $\symbol{92}$ parametric & $(1.000, 1.000, 1.000)$ & $(1.000, 1.000, 1.000)$ & $(1.000, 1.000, 1.000)$ \\
 
  & $\mathcal{L}=(6,12,18)$ \\ \midrule \midrule
 
 & & \multicolumn{3}{c}{$r=1$} \\
SNU & $0.1$ $\symbol{92}$ parametric & $\{0.051, 0.047, 0.061, 0.079\}$ & $\{0.059, 0.050, 0.050, 0.057\}$ & $\{0.058, 0.041, 0.038, 0.049\}$ \\
 
& $0.1$ $\symbol{92}$ non-parametric & $\{0.108, 0.042, 0.014, 0.003\}$ & $\{0.117, 0.054, 0.020, 0.003\}$ & $\{0.120, 0.061, 0.023, 0.008\}$  \\[0.5cm]

P & $0.1$ $\symbol{92}$ parametric & $(1.000, 1.000, 1.000)$ & $(1.000, 1.000, 1.000)$ & $(1.000, 1.000, 1.000)$ \\
 & $\mathcal{L}=(6,12,18)$ \\ \midrule
 
SNU & $0.4$ $\symbol{92}$ parametric & $\{0.133, 0.100, 0.082, 0.083\}$ & $\{0.304, 0.240, 0.192, 0.148\}$ & $\{0.168, 0.145, 0.132, 0.123\}$ \\
 
& $0.4$ $\symbol{92}$ non-parametric & $\{0.149, 0.065, 0.019, 0.002\}$ & $\{0.285, 0.137, 0.024, 0.002\}$ & $\{0.166, 0.070, 0.011, 0.001\}$  \\[0.5cm]

P & $0.4$ $\symbol{92}$ parametric & $(1.000, 1.000, 1.000)$ & $(1.000, 1.000, 1.000)$ & $(1.000, 1.000, 1.000)$ \\
 & $\mathcal{L}=(6,12,18)$ \\ 
\bottomrule
\end{tabular}
\end{table}

\begin{table}[ht]
\footnotesize
\caption{Empirical size of the SNU and MHM test statistics at $5\%$ level for the models given in expressions (\ref{eq:nonintegrablefs1}) and  (\ref{eq:integrablefs1}). For all the cases, the number of blocks is  thinned to $M=N^{0.9}$, and the values in the curly brackets correspond to the 4 block sizes in the paper. Additionally, we assume $r=0.5$ and $h=N^{-1/3}$.} \label{Tab.size_thin}
\centering
\setlength{\tabcolsep}{1pt} 
\renewcommand{\arraystretch}{1.15} 
\begin{tabular}{@{} ccccc @{}}
\toprule
test & $d$ $\symbol{92}$ $\hat{f}(.)$ & LM & SLM $(\lambda=N^{-1/3})$ & SLM $(\lambda=N^{-1/6})$ \\\midrule
 & & \multicolumn{3}{c}{expression (\ref{eq:nonintegrablefs1})} \\
 SNU & $0.1$ $\symbol{92}$ parametric & $\{0.149, 0.123, 0.110, 0.121 \}$ & $ \, \{0.132, 0.109, 0.089, 0.096 \}$ & $\{0.149, 0.116, 0.106, 0.102 \}$ \\
 
& $0.1$ $\symbol{92}$ non-parametric & $\{0.147, 0.122, 0.060, 0.019 \}$ & $\{0.128, 0.106, 0.059, 0.018 \}$ & $\{0.142, 0.129, 0.068, 0.021 \}$  \\[0.5cm]

de-biased MHM & $0.1$ $\symbol{92}$ parametric & $\{0.120, 0.080, 0.060,      0.069\}$ & $\{0.112, 0.082, 0.067, 0.071 \}$ & $\{0.117, 0.082, 0.066,      0.078 \}$ \\ \midrule

SNU & $0.4$ $\symbol{92}$ parametric & $\{0.146, 0.108, 0.089, 0.090 \}$ & $\{0.112, 0.086, 0.079, 0.084 \}$ & $\{0.123, 0.101, 0.087, 0.091 \}$ \\
 
& $0.4$ $\symbol{92}$ non-parametric & $\{0.151, 0.106, 0.046, 0.013 \}$ & $\{0.114, 0.081, 0.030, 0.005 \}$ & $\{0.124, 0.091, 0.041, 0.011 \}$  \\[0.5cm]

de-biased MHM & $0.4$ $\symbol{92}$ parametric & $\{0.131, 0.106, 0.087,      0.083 \}$ & $\{0.126, 0.089, 0.079, 0.079 \}$ & $\{0.112, 0.079, 0.069,      0.071 \}$ \\ \midrule \midrule

 & & \multicolumn{3}{c}{expression (\ref{eq:integrablefs1})} \\
SNU & $0.1$ $\symbol{92}$ parametric & $\{0.100, 0.074, 0.085, 0.097 \}$ & $\{0.082, 0.071, 0.068, 0.086 \}$ & $\{0.099, 0.084, 0.081, 0.095 \}$ \\
 
& $0.1$ $\symbol{92}$ non-parametric & $\{0.140, 0.077, 0.029, 0.010 \}$ & $\{0.115, 0.066, 0.032, 0.011 \}$ & $\{0.137, 0.075, 0.033, 0.014 \}$  \\[0.5cm]

de-biased MHM & $0.1$ $\symbol{92}$ parametric & $\{0.233, 0.207, 0.171,       0.182 \}$ & $\{0.202, 0.189, 0.144, 0.165 \}$ & $\{0.195, 0.177, 0.139,       0.165 \}$ \\ \midrule

SNU & $0.4$ $\symbol{92}$ parametric & $\{0.103, 0.078, 0.070, 0.082 \}$ & $\{0.082, 0.069, 0.075, 0.086 \}$ & $\{0.093, 0.078, 0.081, 0.090 \}$ \\
 
& $0.4$ $\symbol{92}$ non-parametric & $\{0.130, 0.062, 0.023, 0.008 \}$ & $\{0.095, 0.050, 0.016, 0.003 \}$ & $\{0.113, 0.051, 0.025, 0.007 \}$  \\[0.5cm]

de-biased MHM & $0.4$ $\symbol{92}$ parametric & $\{0.310, 0.288, 0.228,       0.221 \}$ & $\{0.248, 0.228, 0.180, 0.189 \}$ & $\{0.211, 0.204, 0.155,       0.171 \}$ \\
 
\bottomrule
\end{tabular}
\end{table}

\newpage
\clearpage

\subsection{B.5 \quad Density and Size of Tests for Different Values of $r$ and $N$}

\begin{figure}[!h]
\centering
\begin{tabular}{ c }
 \includegraphics[width=1\textwidth]{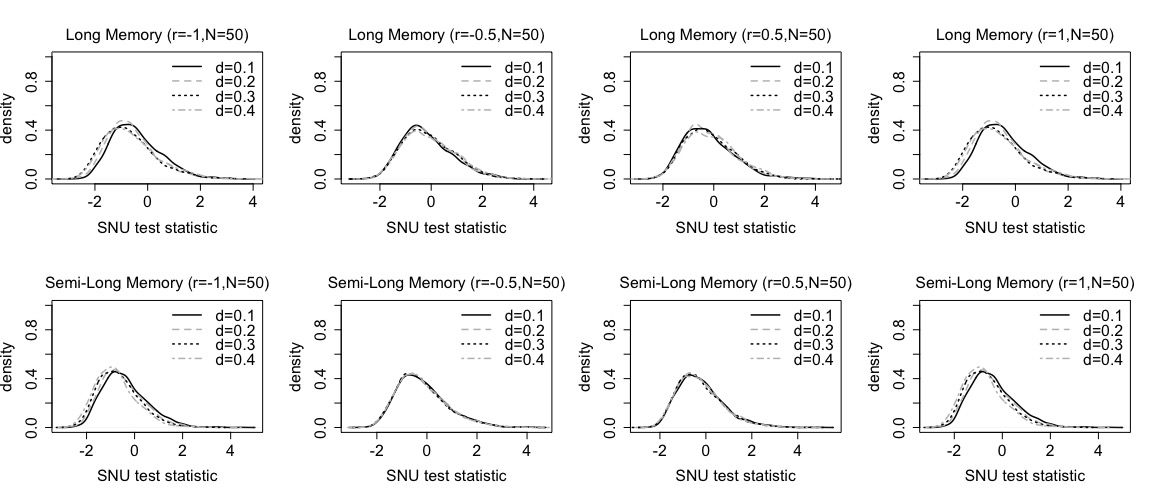} 
 \end{tabular}
\caption{Monte Carlo densities of SNU test statistic with $N=50$.}
\label{Fig.dist2012_N50}
\end{figure}

\begin{figure}[!h]
\centering
\begin{tabular}{ c }
 \includegraphics[width=1\textwidth]{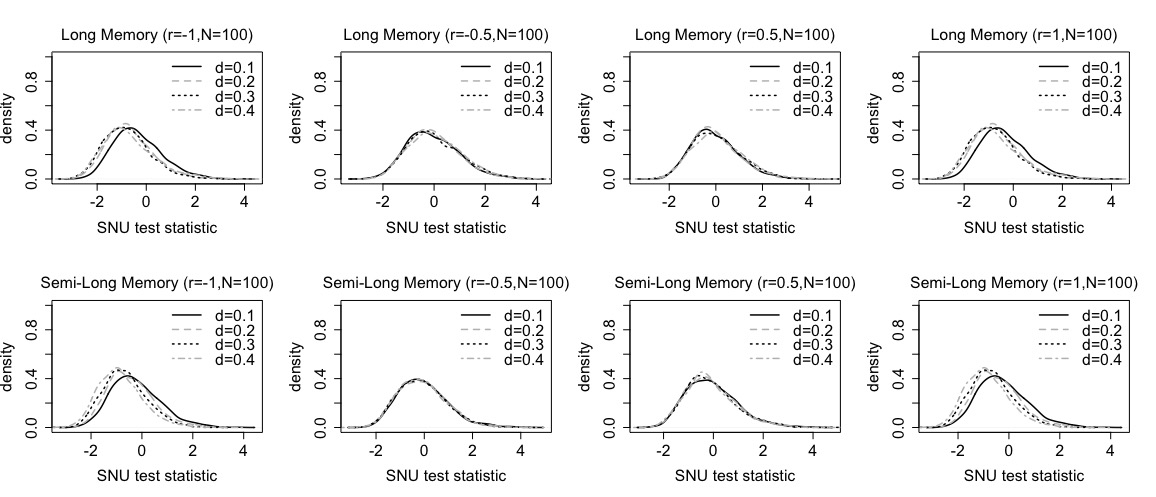} 
 \end{tabular}
\caption{Monte Carlo densities of SNU test statistic with $N=100$.}
\label{Fig.dist2012_N100}
\end{figure}

\begin{figure}[!h]
\centering
\begin{tabular}{ c }
 \includegraphics[width=1\textwidth]{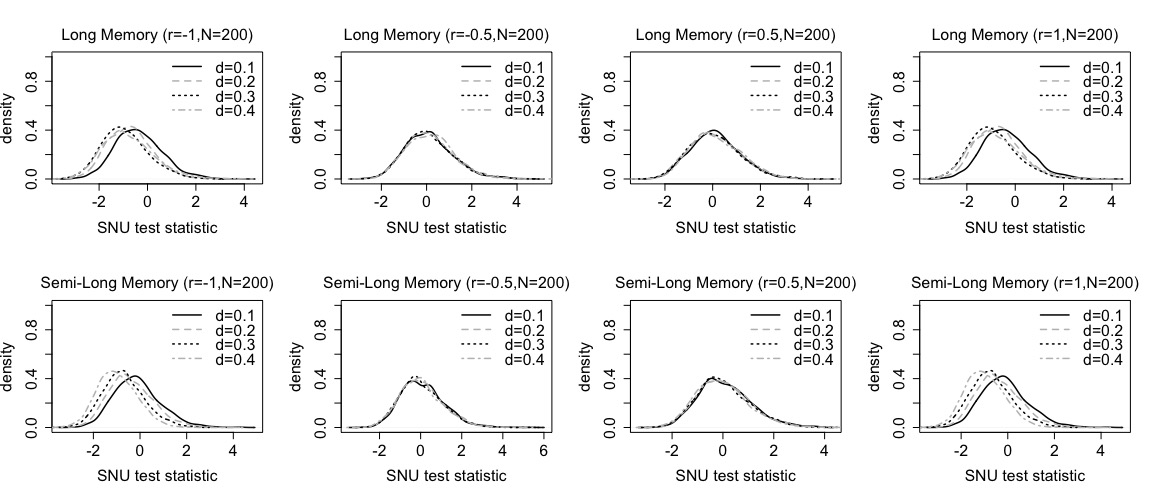} 
 \end{tabular}
\caption{Monte Carlo densities of SNU test statistic with $N=200$.}
\label{Fig.dist2012_N200}
\end{figure}

\begin{figure}[!h]
\centering
\begin{tabular}{ c }
 \includegraphics[width=1\textwidth]{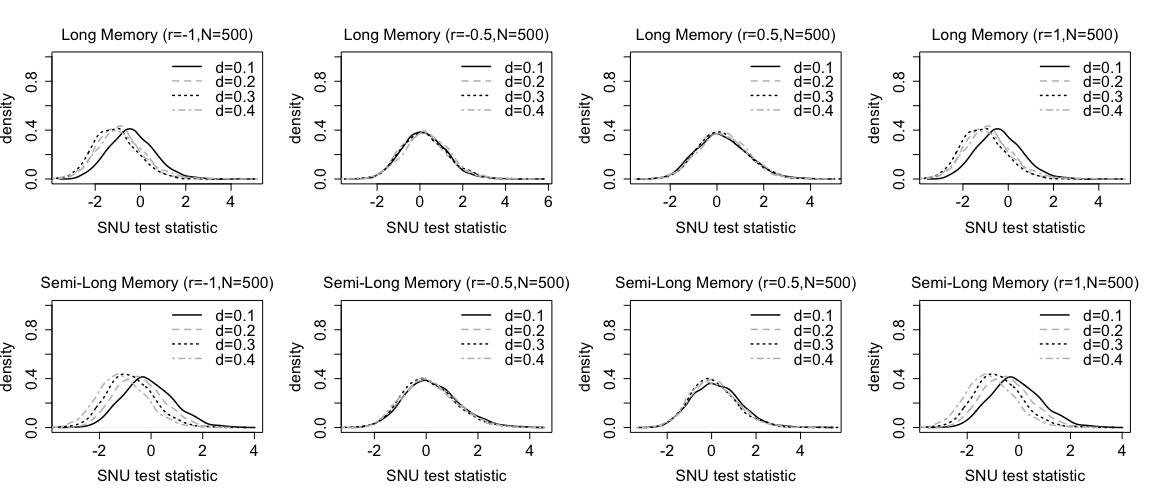} 
 \end{tabular}
\caption{Monte Carlo densities of SNU test statistic with $N=500$.}
\label{Fig.dist2012_N500}
\end{figure}

\begin{figure}[!h]
\centering
\begin{tabular}{ c }
 \includegraphics[width=1\textwidth]{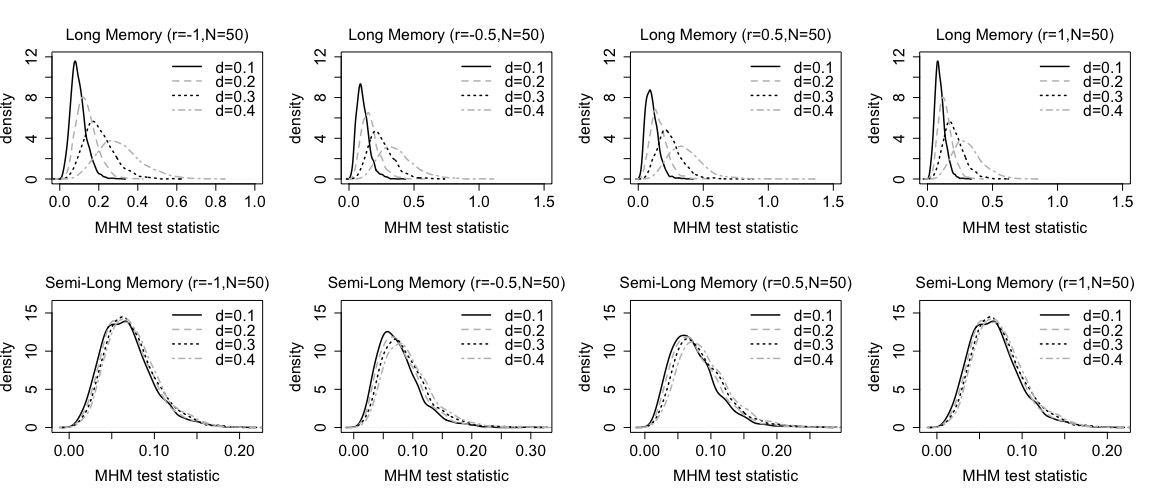} 
 \end{tabular}
\caption{Monte Carlo densities of MHM test statistic with $N=50$.}
\label{Fig.dist2016_N50}
\end{figure}

\begin{figure}[!h]
\centering
\begin{tabular}{ c }
 \includegraphics[width=1\textwidth]{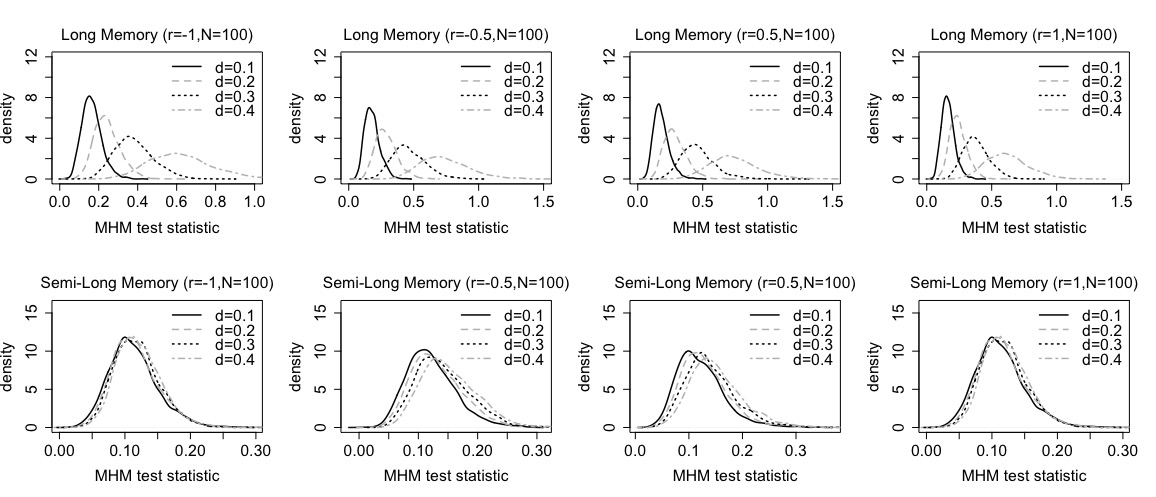} 
 \end{tabular}
\caption{Monte Carlo densities of MHM test statistic with $N=100$.}
\label{Fig.dist2016_N100}
\end{figure}

\begin{figure}[!h]
\centering
\begin{tabular}{ c }
 \includegraphics[width=1\textwidth]{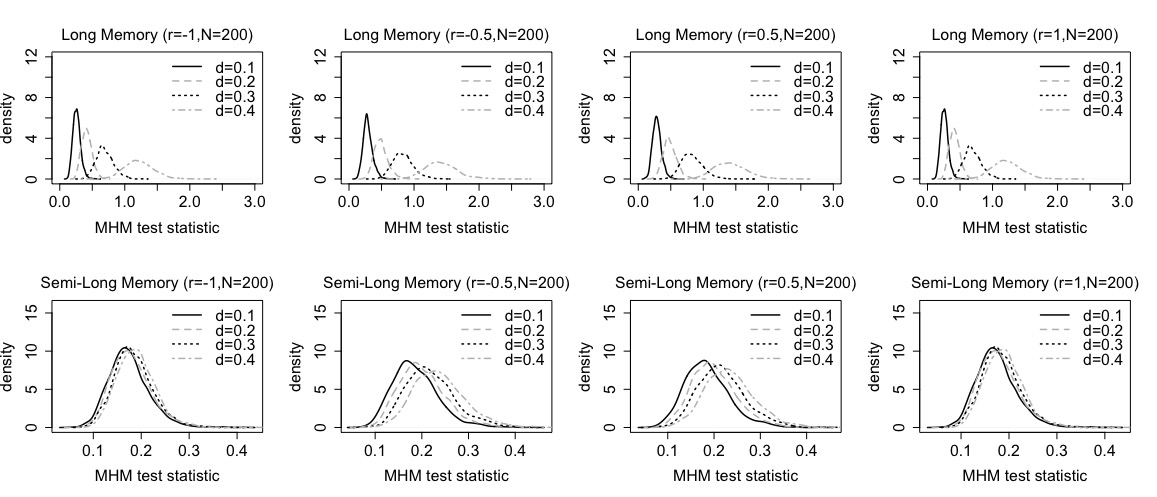} 
 \end{tabular}
\caption{Monte Carlo densities of MHM test statistic with $N=200$.}
\label{Fig.dist2016_N200}
\end{figure}

\begin{figure}[!h]
\centering
\begin{tabular}{ c }
 \includegraphics[width=1\textwidth]{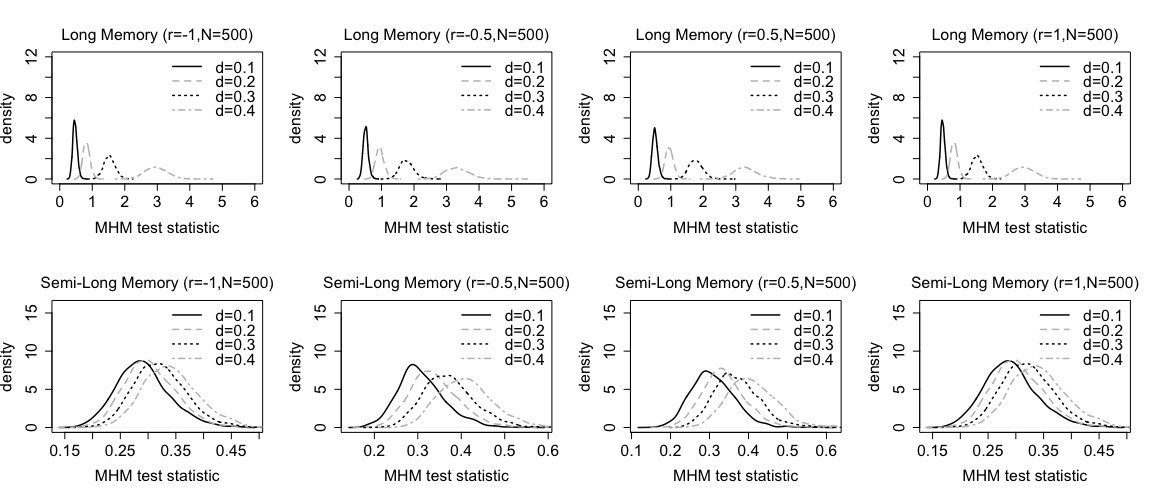} 
 \end{tabular}
\caption{Monte Carlo densities of MHM test statistic with $N=500$.}
\label{Fig.dist2016_N500}
\end{figure}

Firstly, we provide Monte Carlo densities of SNU and MHM test statistics for different values of endogeneity $r=\{-1, -0.5, 0.5, 1\}$ and different sample sizes $N=\{50, 100, 200, 500 \}$. We assume $h=N^{-1/3}$, and the tempering parameter $\lambda$ is $N^{-1/6}$ for the SLM process. The results are listed in Figures \ref{Fig.dist2012_N50}--\ref{Fig.dist2016_N500}. 

\begin{figure}[!h]
\centering
\begin{tabular}{ c }
 \includegraphics[width=0.7\textwidth]{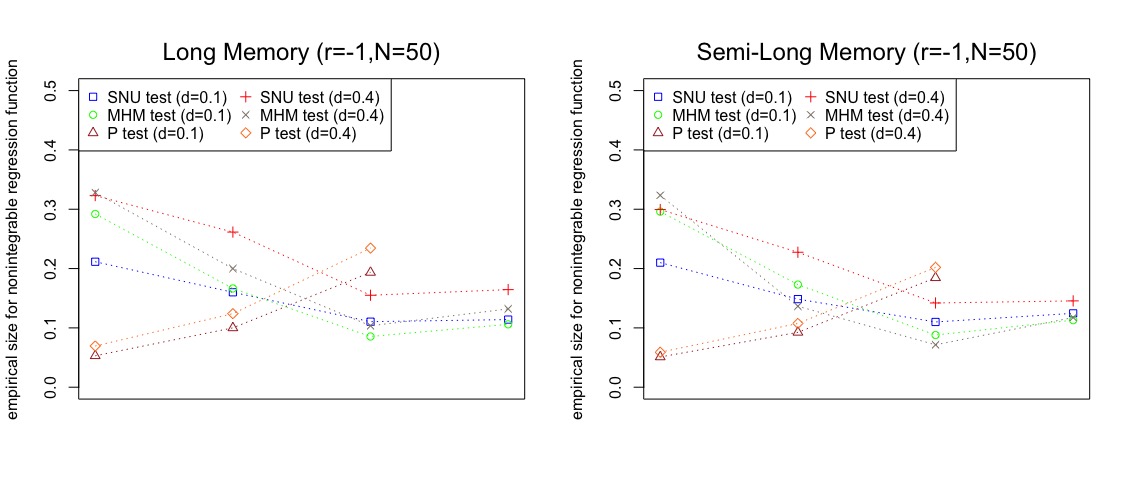} \\
 \includegraphics[width=0.7\textwidth]{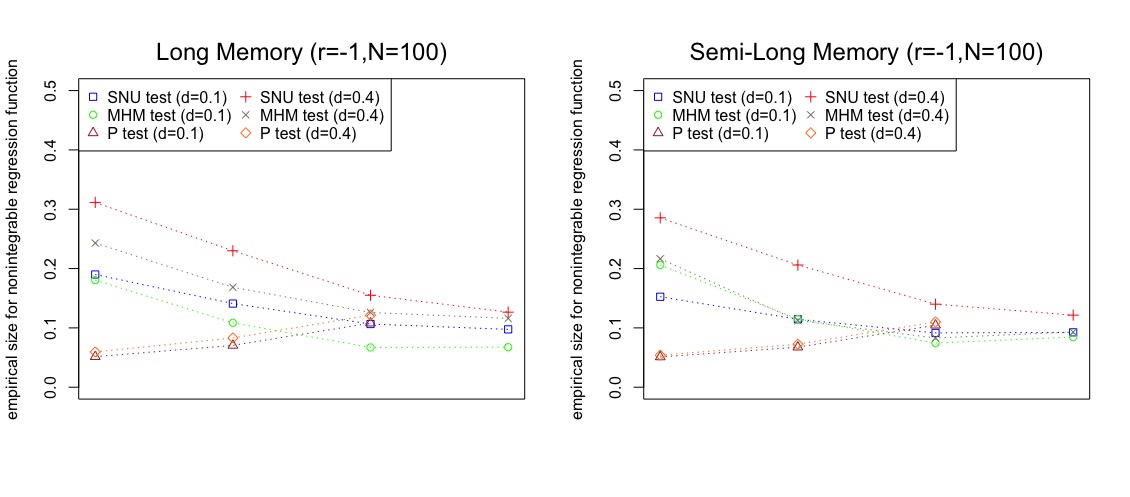}
 \\
 \includegraphics[width=0.7\textwidth]{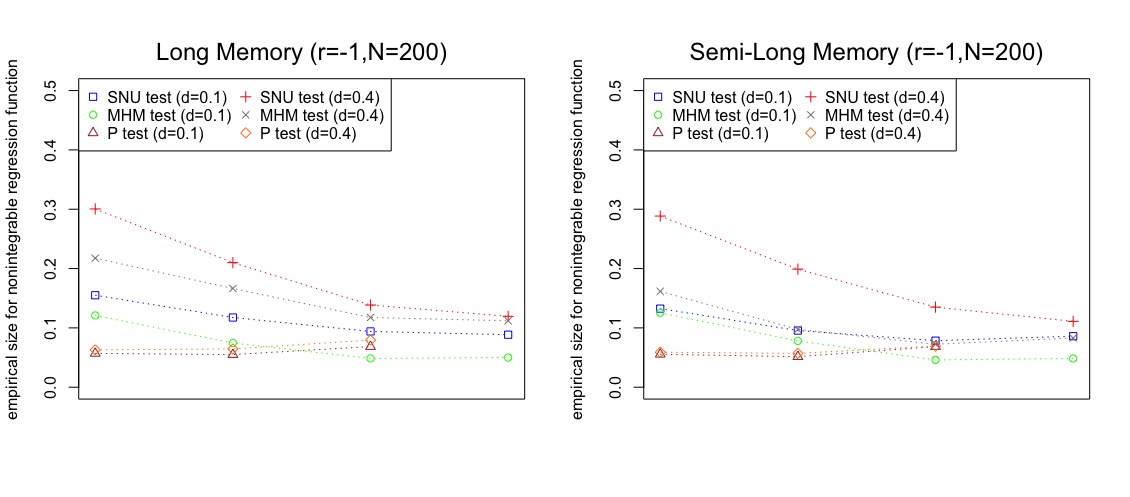}
 \\
 \includegraphics[width=0.7\textwidth]{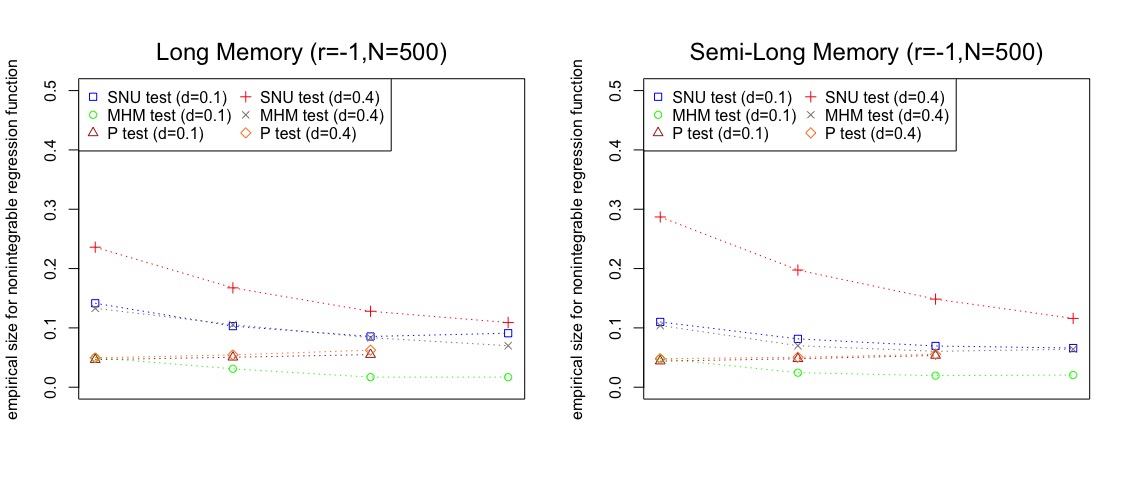}
 \end{tabular}
\caption{Empirical size of test statistics for nonintegrable regression function with $r=-1$ and different values of $N$. For  SNU/MHM statistics, sizes are plotted for four   block lengths  $b=[c N^{1/2}]$, $c=0.5,1,2,4$ (left to right); for the P statistic, three sizes are plotted left to right representing values of $\mathcal{L}=(6,12,18)$. }
\label{Fig.size_rm1_nonintegrable}
\end{figure}

\begin{figure}[!h]
\centering
\begin{tabular}{ c }
 \includegraphics[width=0.7\textwidth]{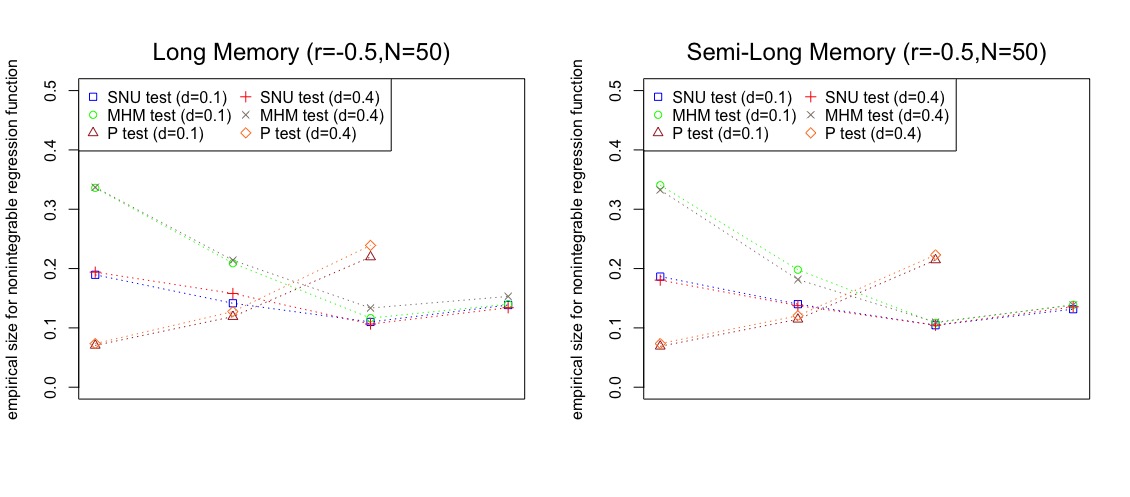} \\
 \includegraphics[width=0.7\textwidth]{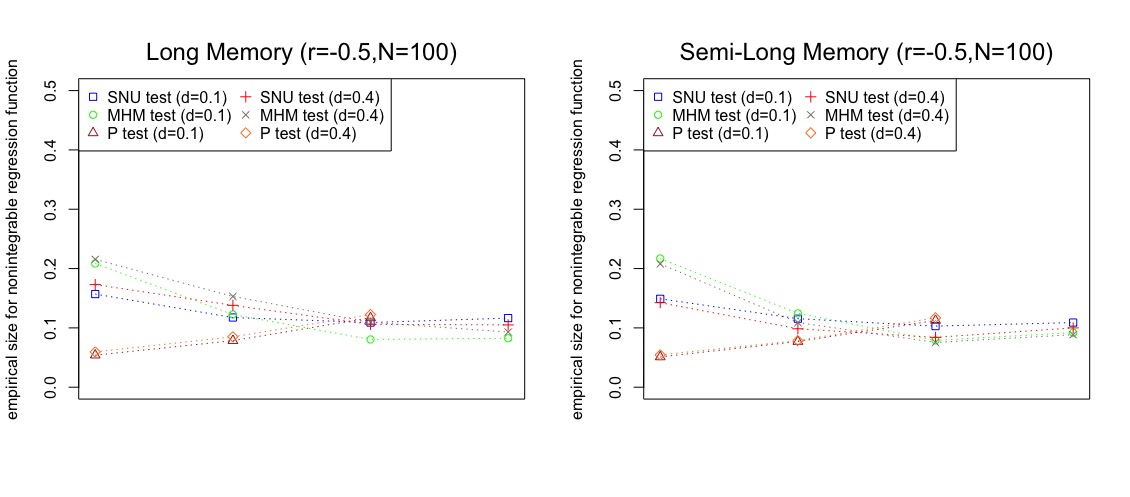}
 \\
 \includegraphics[width=0.7\textwidth]{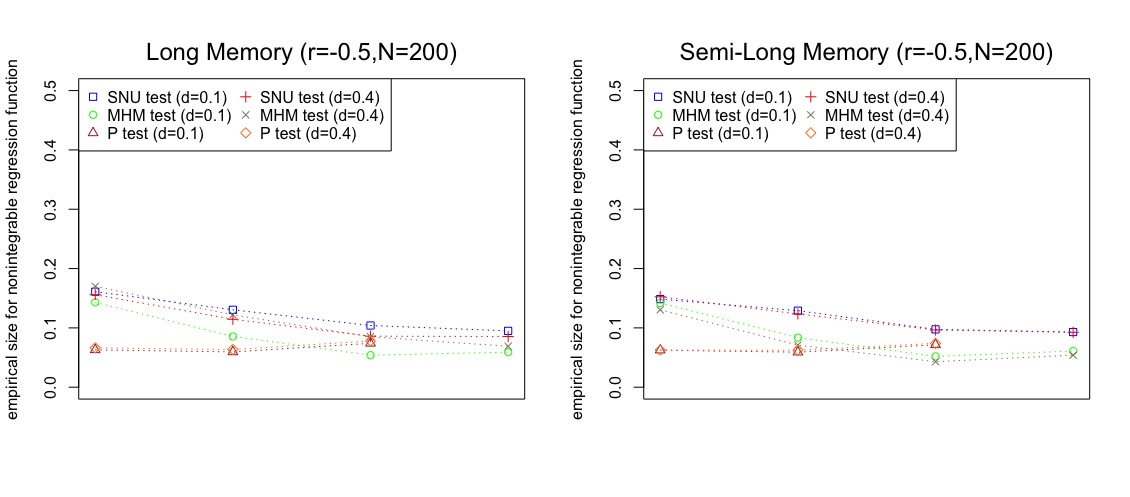}
 \\
 \includegraphics[width=0.7\textwidth]{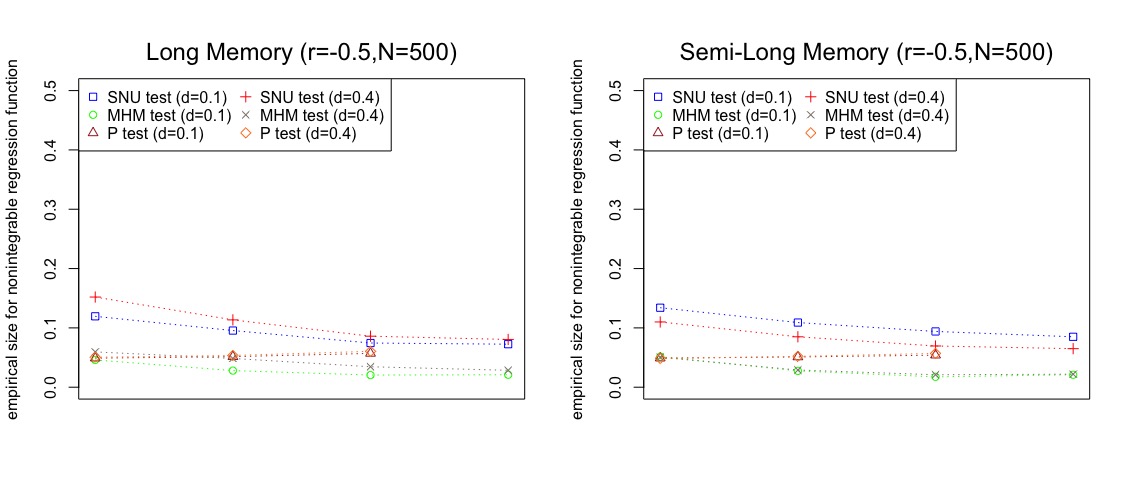}
 \end{tabular}
\caption{Empirical size of test statistics for nonintegrable regression function with $r=-0.5$ and different values of $N$. For SNU/MHM statistics, sizes are plotted for four   block lengths  $b=[c N^{1/2}]$, $c=0.5,1,2,4$ (left to right); for the P statistic, three sizes are plotted left to right representing values of $\mathcal{L}=(6,12,18)$. }
\label{Fig.size_rm05_nonintegrable}
\end{figure}

\begin{figure}[!h]
\centering
\begin{tabular}{ c }
 \includegraphics[width=0.7\textwidth]{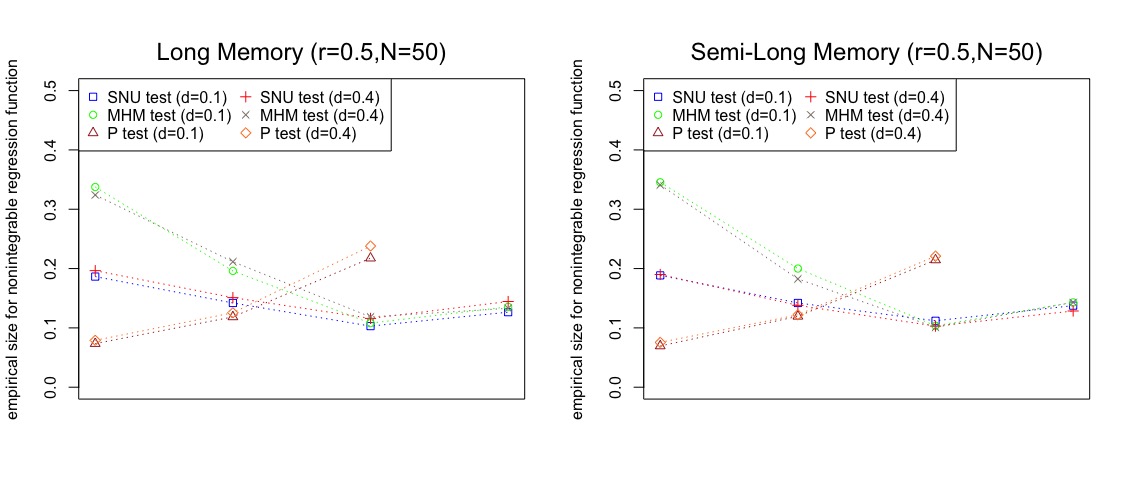} \\
 \includegraphics[width=0.7\textwidth]{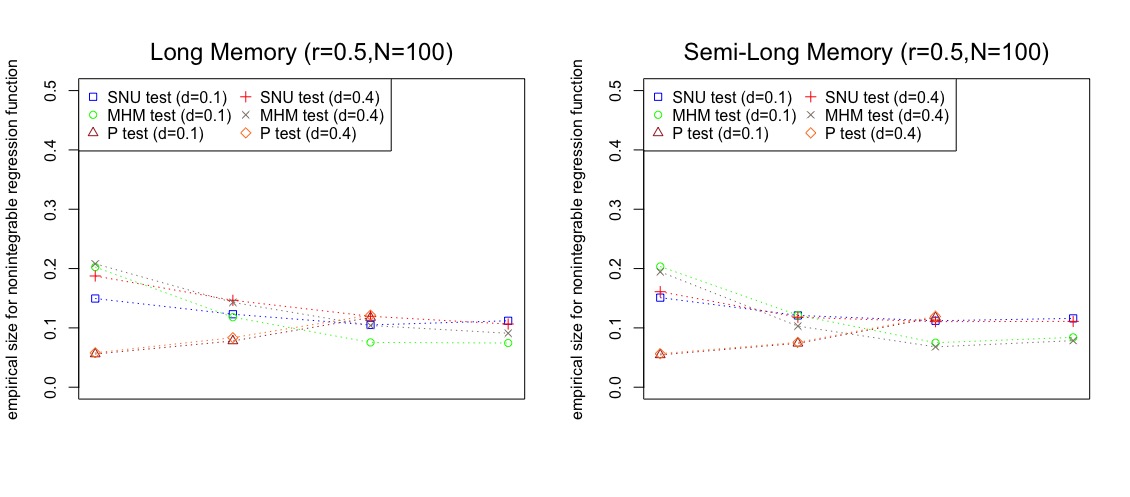}
 \\
 \includegraphics[width=0.7\textwidth]{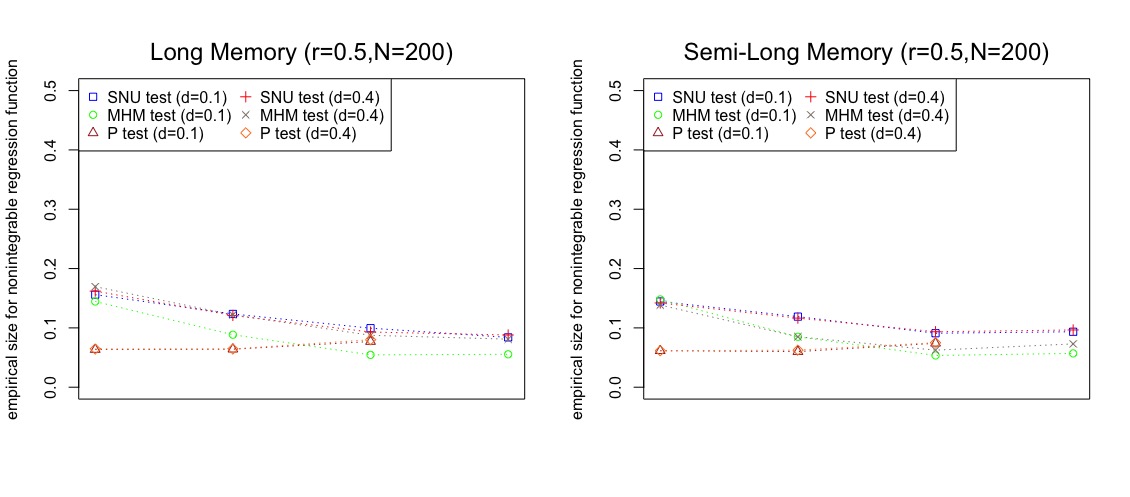}
 \\
 \includegraphics[width=0.7\textwidth]{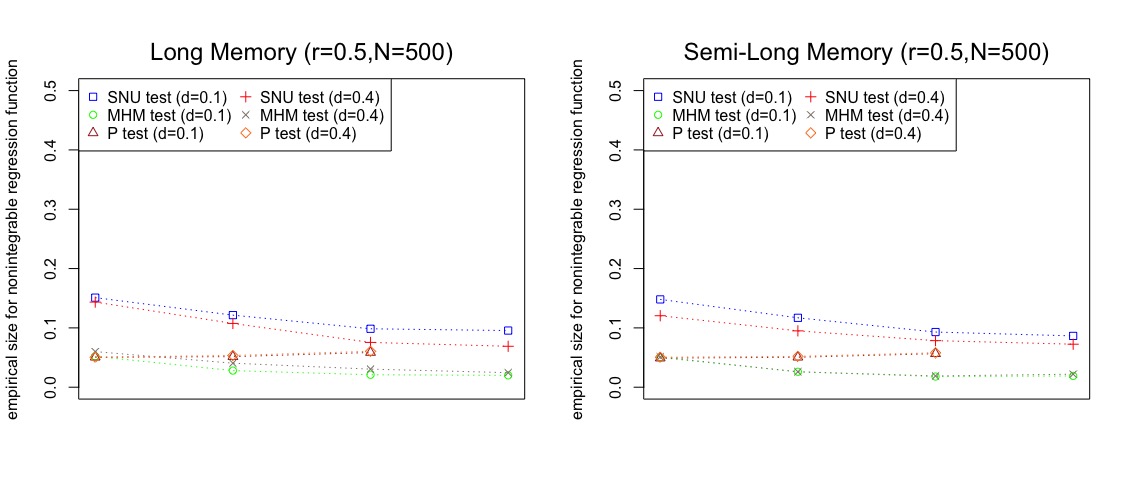}
 \end{tabular}
\caption{Empirical size of test statistics for nonintegrable regression function with $r=0.5$ and different values of $N$. For  SNU/MHM statistics, sizes are plotted for four   block lengths  $b=[c N^{1/2}]$, $c=0.5,1,2,4$ (left to right); for the P statistic, three sizes are plotted left to right representing values of $\mathcal{L}=(6,12,18)$.}
\label{Fig.size_rp05_nonintegrable}
\end{figure}

\begin{figure}[!h]
\centering
\begin{tabular}{ c }
 \includegraphics[width=0.7\textwidth]{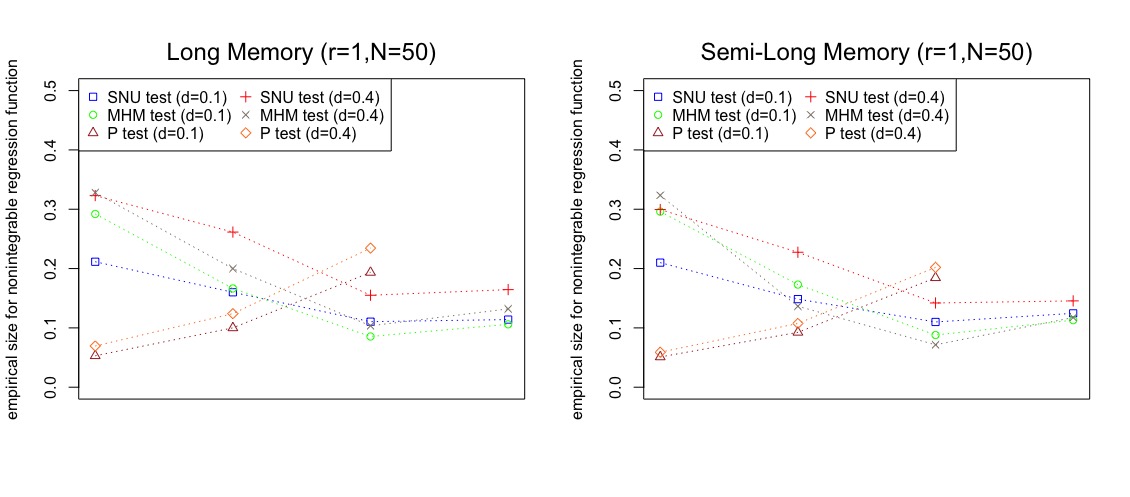} \\
 \includegraphics[width=0.7\textwidth]{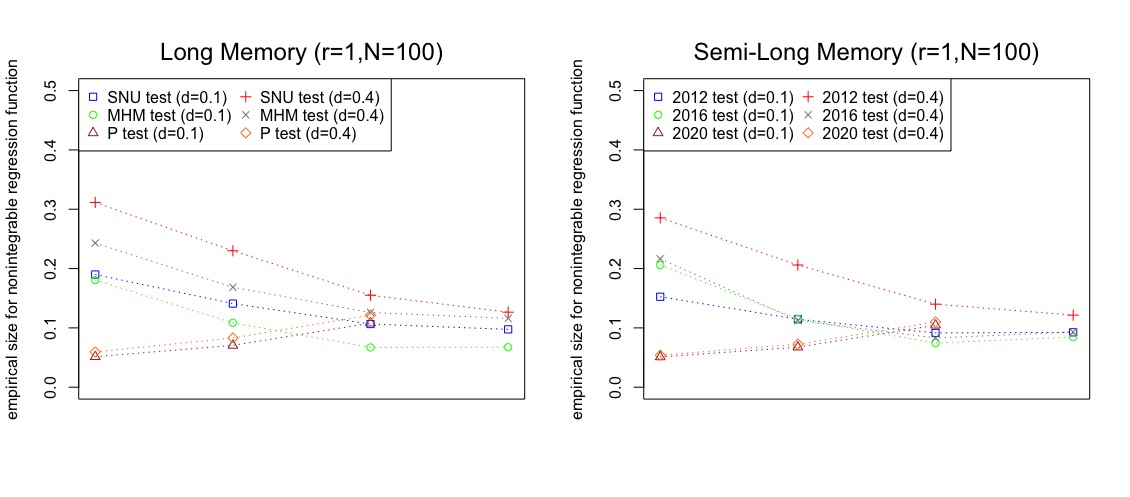}
 \\
 \includegraphics[width=0.7\textwidth]{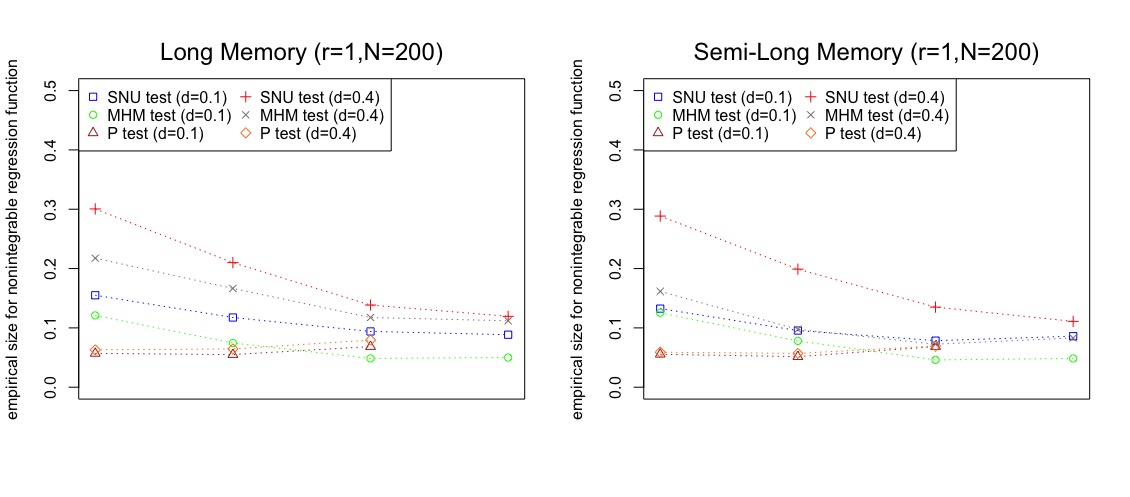}
 \\
 \includegraphics[width=0.7\textwidth]{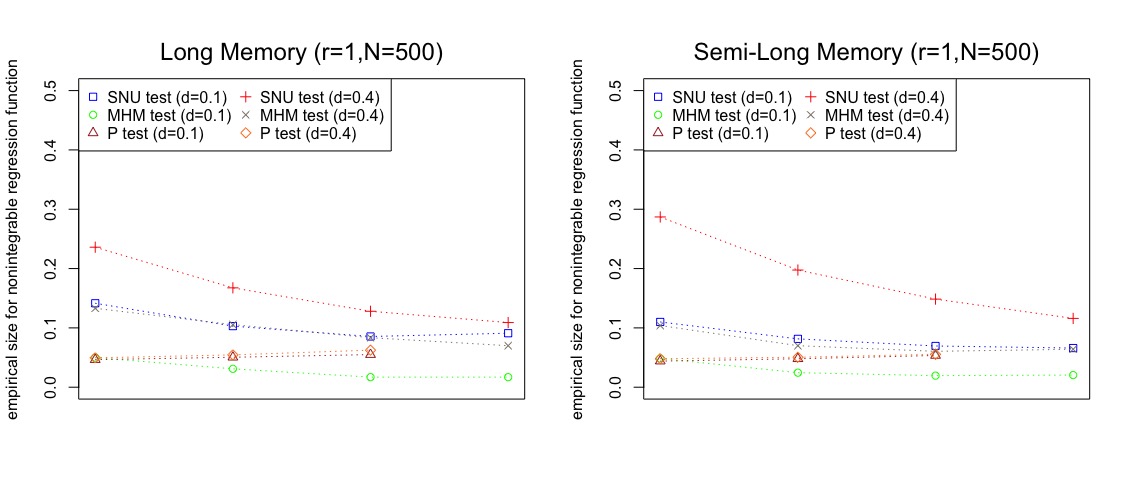}
 \end{tabular}
\caption{Empirical size of test statistics for nonintegrable regression function with $r=1$ and different values of $N$. For  SNU/MHM statistics, sizes are plotted for four   block lengths  $b=[c N^{1/2}]$, $c=0.5,1,2,4$ (left to right); for the P  statistic, three sizes are plotted left to right representing values of $\mathcal{L}=(6,12,18)$.}
\label{Fig.size_rp1_nonintegrable}
\end{figure}

\begin{figure}[!h]
\centering
\begin{tabular}{ c }
 \includegraphics[width=0.7\textwidth]{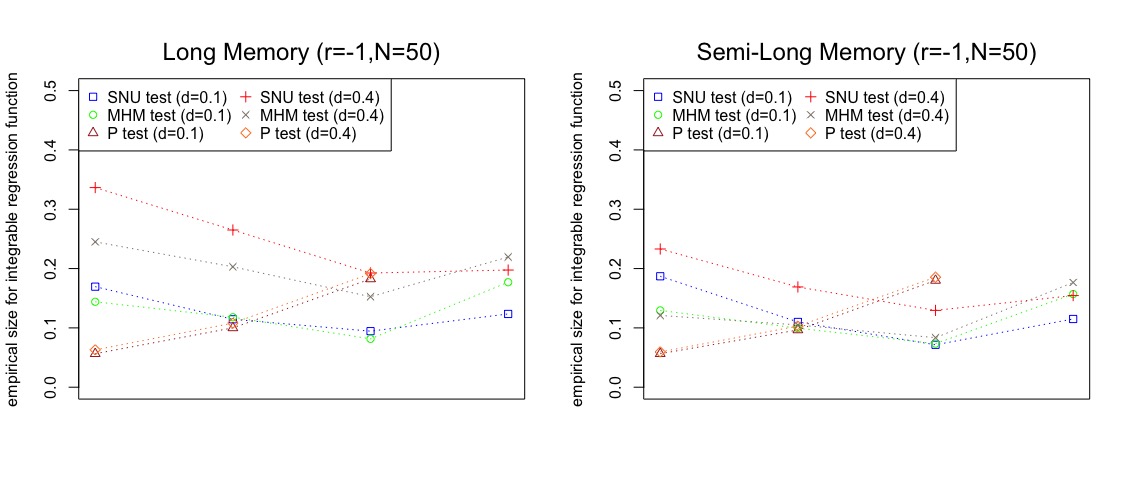} \\
 \includegraphics[width=0.7\textwidth]{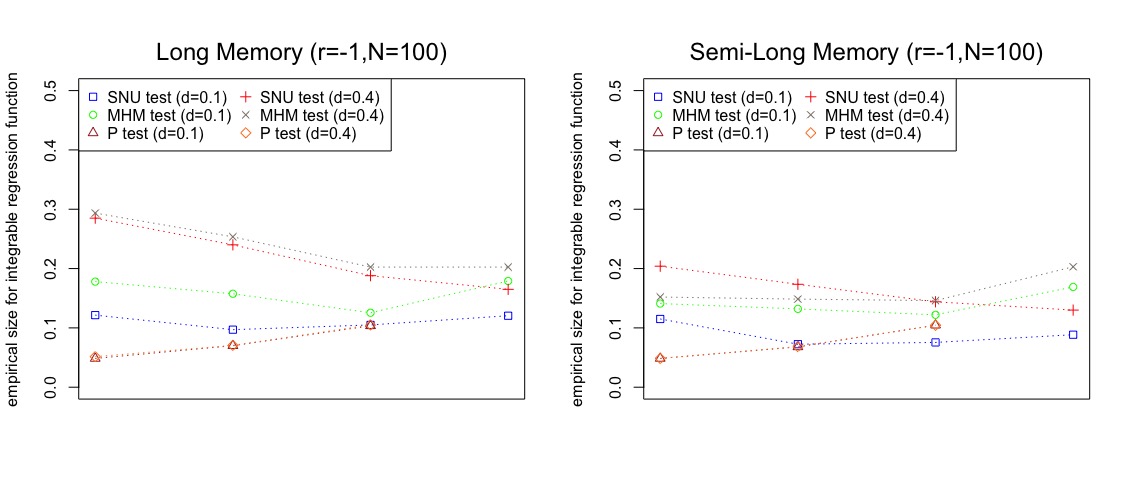}
 \\
 \includegraphics[width=0.7\textwidth]{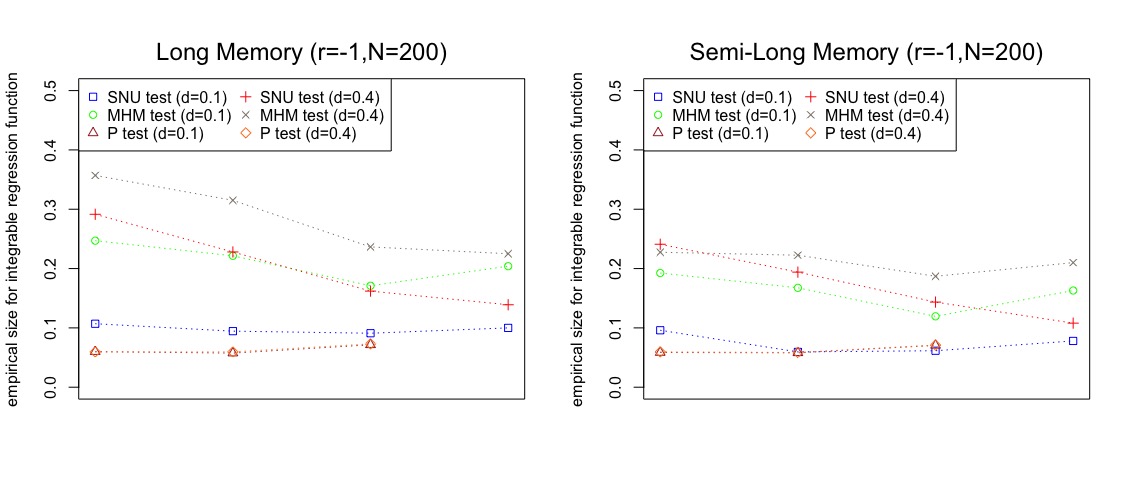}
 \\
 \includegraphics[width=0.7\textwidth]{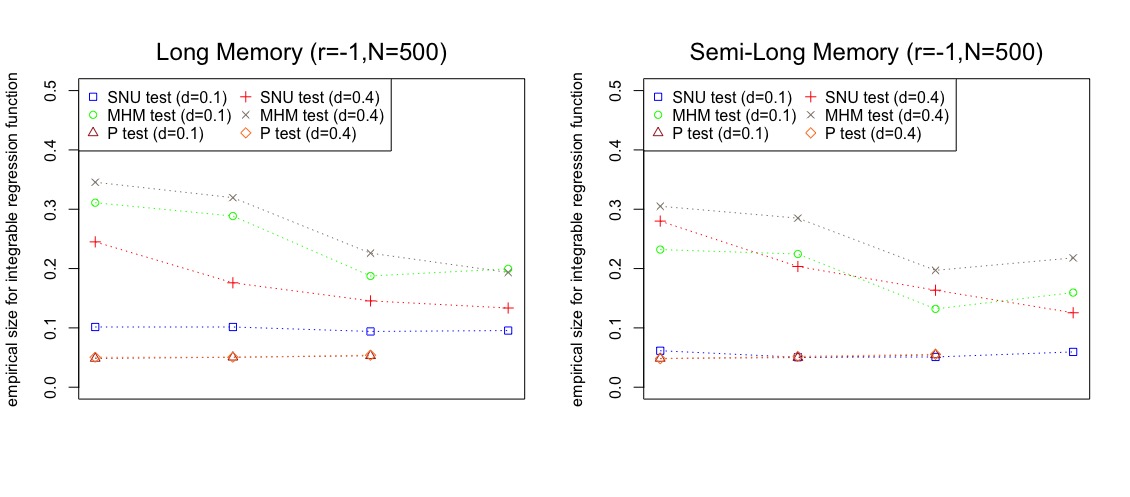}
 \end{tabular}
\caption{Empirical size of test statistics for integrable regression function with $r=-1$ and different values of $N$. For  SNU/MHM statistics, sizes are plotted for four   block lengths  $b=[c N^{1/2}]$, $c=0.5,1,2,4$ (left to right); for the P statistic, three sizes are plotted left to right representing values of $\mathcal{L}=(6,12,18)$.}
\label{Fig.size_rm1_integrable}
\end{figure}

\begin{figure}[!h]
\centering
\begin{tabular}{ c }
 \includegraphics[width=0.7\textwidth]{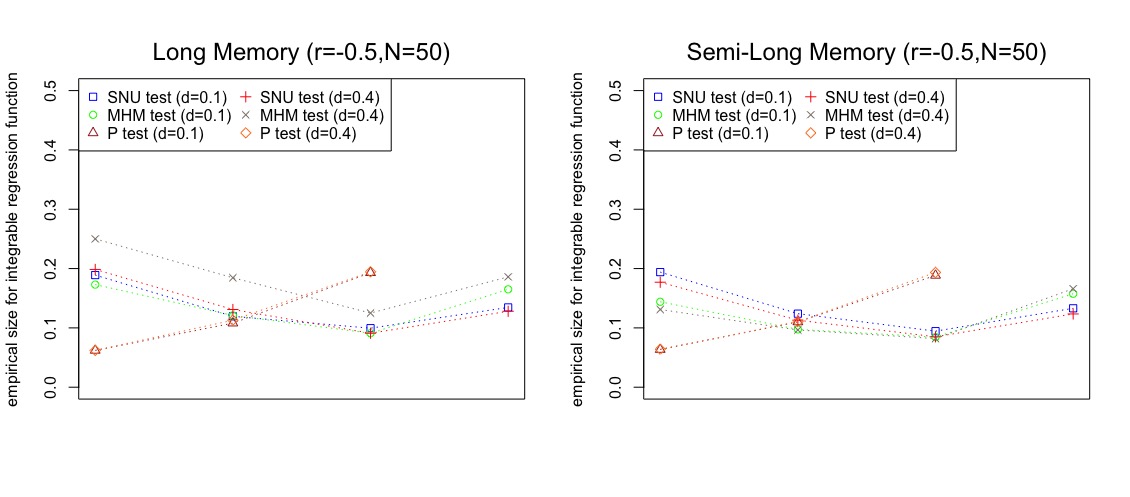} \\
 \includegraphics[width=0.7\textwidth]{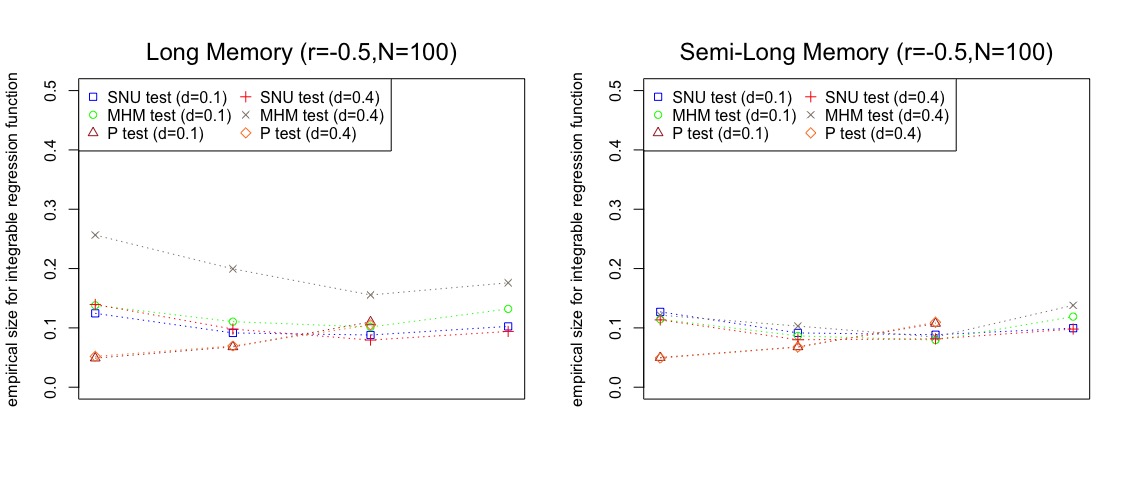}
 \\
 \includegraphics[width=0.7\textwidth]{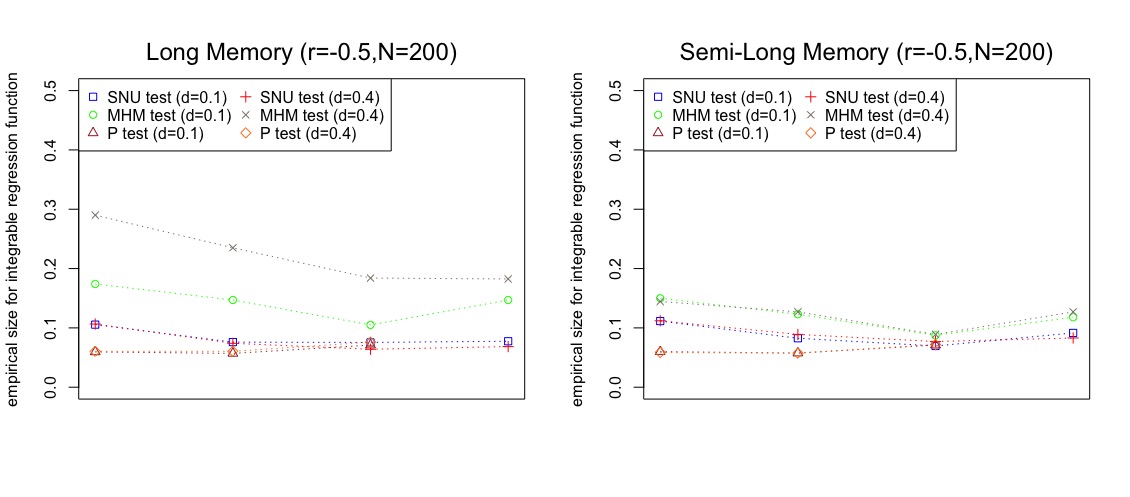}
 \\
 \includegraphics[width=0.7\textwidth]{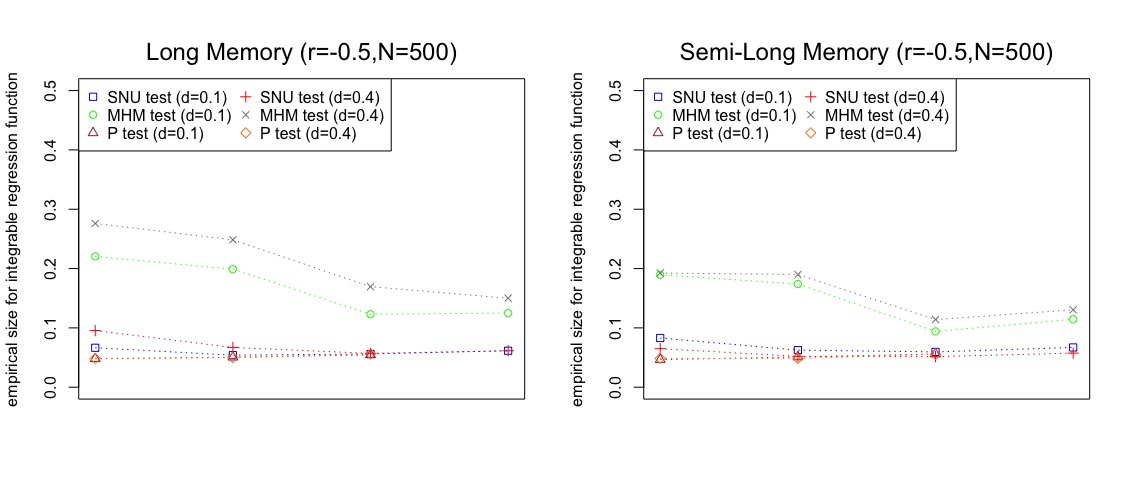}
 \end{tabular}
\caption{Empirical size of test statistics for integrable regression function with $r=-0.5$ and different values of $N$. For  SNU/MHM statistics, sizes are plotted for four   block lengths  $b=[c N^{1/2}]$, $c=0.5,1,2,4$ (left to right); for the P  statistic, three sizes are plotted left to right representing values of $\mathcal{L}=(6,12,18)$.}
\label{Fig.size_rm05_integrable}
\end{figure}

\begin{figure}[!h]
\centering
\begin{tabular}{ c }
 \includegraphics[width=0.7\textwidth]{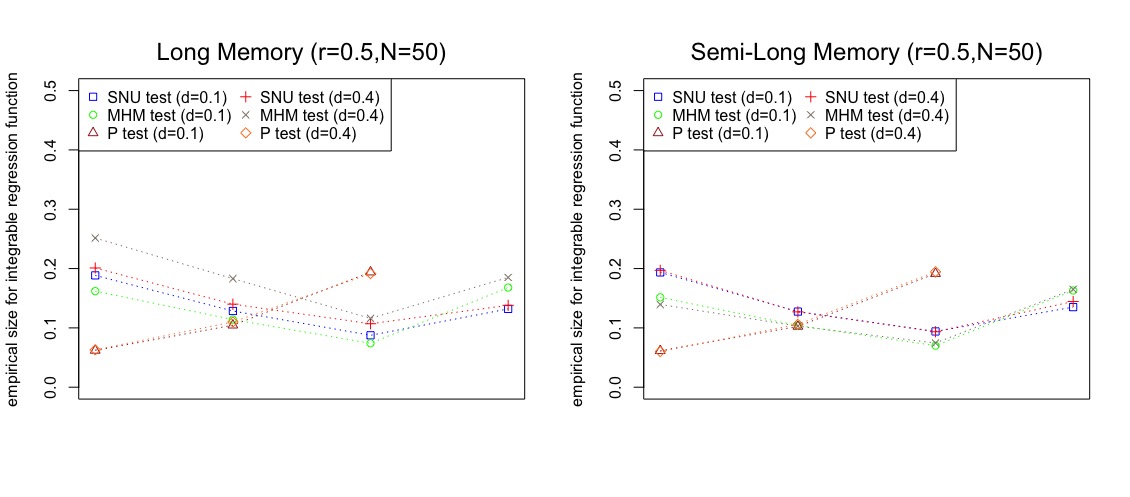} \\
 \includegraphics[width=0.7\textwidth]{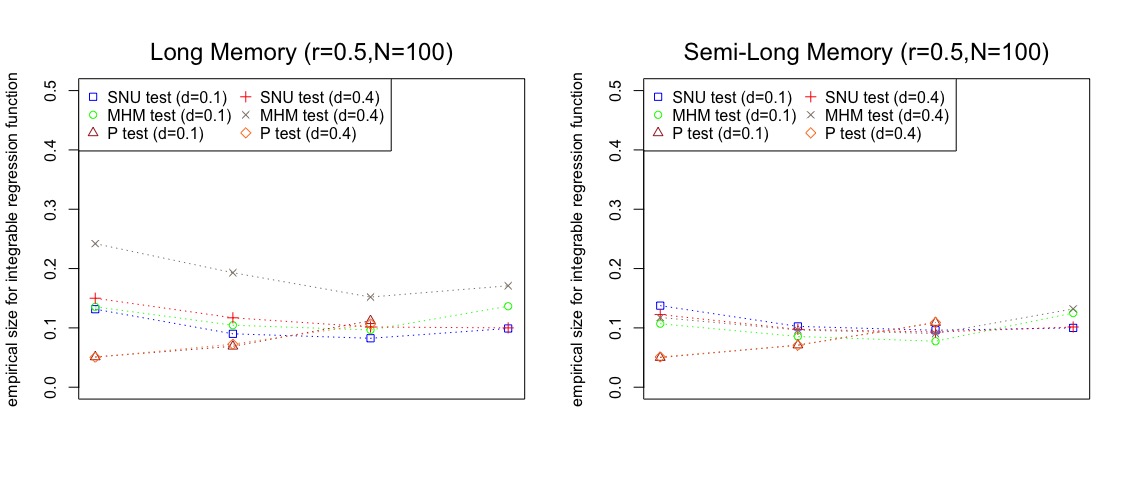}
 \\
 \includegraphics[width=0.7\textwidth]{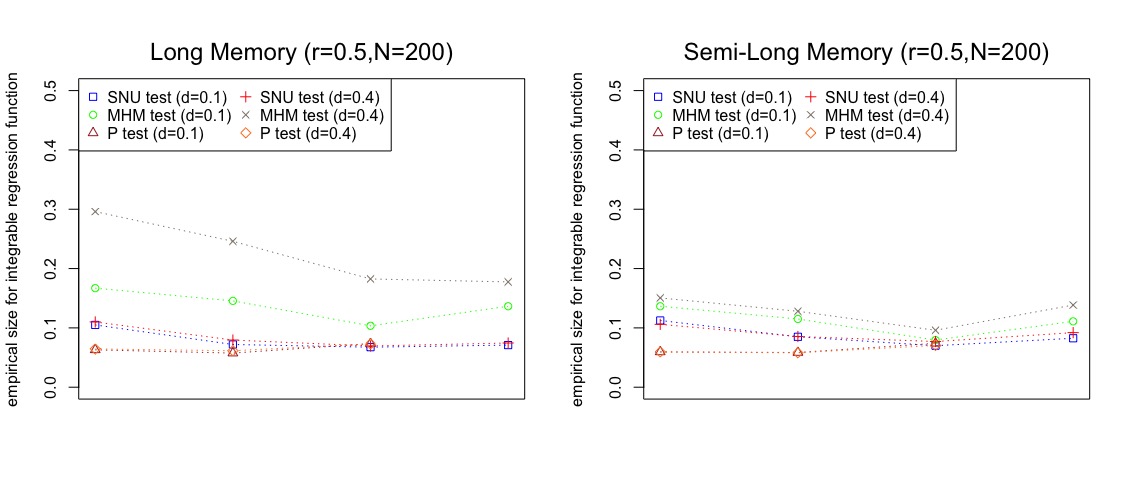}
 \\
 \includegraphics[width=0.7\textwidth]{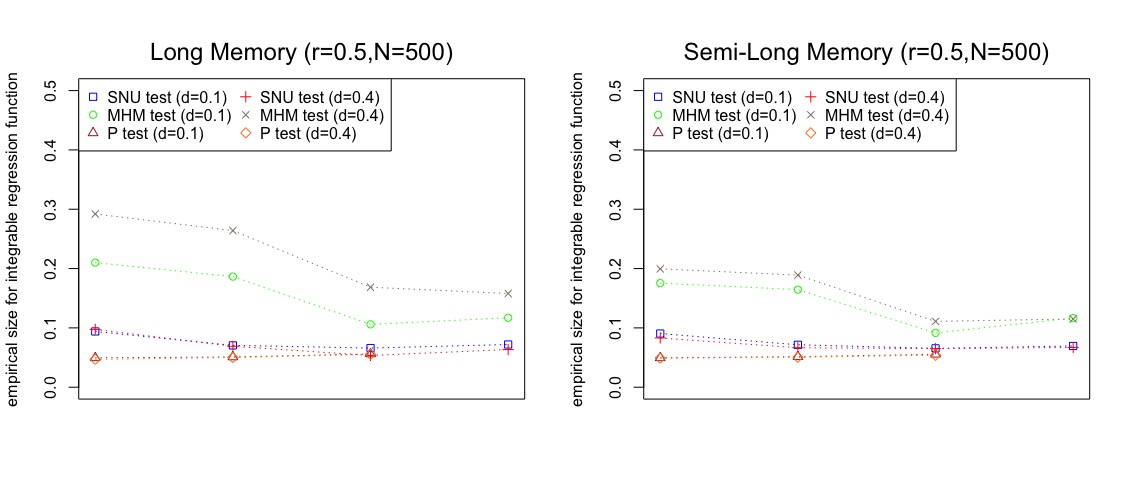}
 \end{tabular}
\caption{Empirical size of test statistics for integrable regression function with $r=0.5$ and different values of $N$. For  SNU/MHM statistics, sizes are plotted for four   block lengths  $b=[c N^{1/2}]$, $c=0.5,1,2,4$ (left to right); for the P  statistic, three sizes are plotted left to right representing values of $\mathcal{L}=(6,12,18)$.}
\label{Fig.size_rp05_integrable}
\end{figure}

\begin{figure}[!h]
\centering
\begin{tabular}{ c }
 \includegraphics[width=0.7\textwidth]{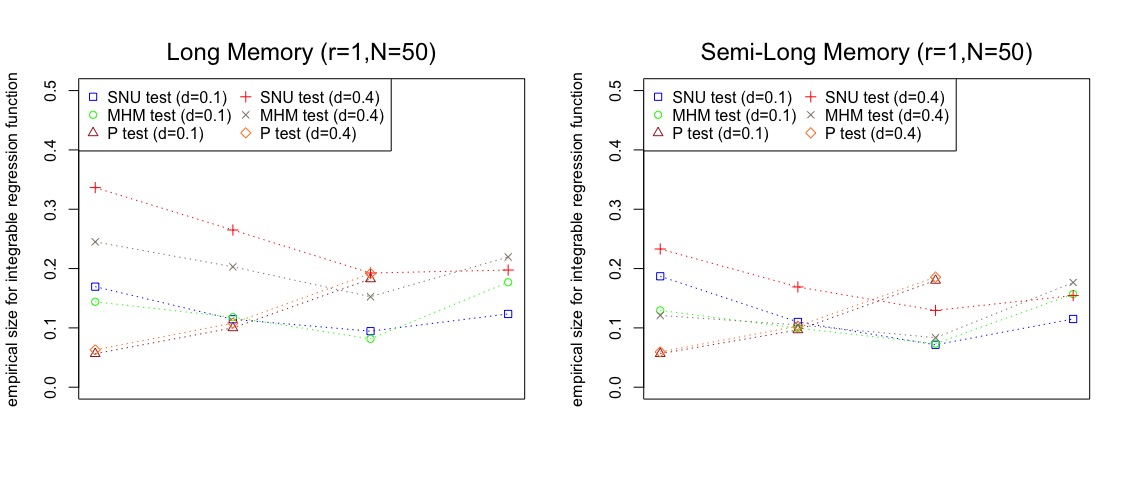} \\
 \includegraphics[width=0.7\textwidth]{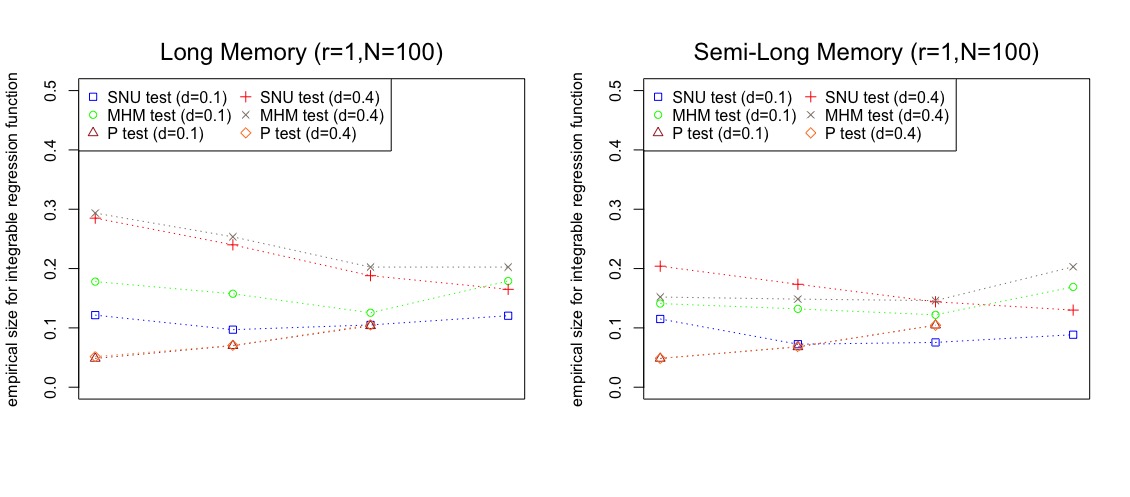}
 \\
 \includegraphics[width=0.7\textwidth]{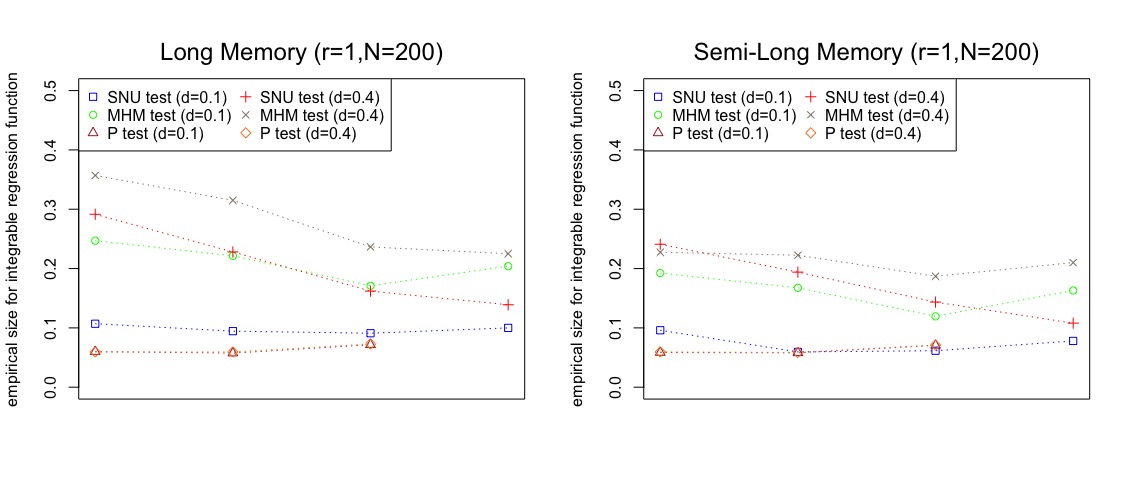}
 \\
 \includegraphics[width=0.7\textwidth]{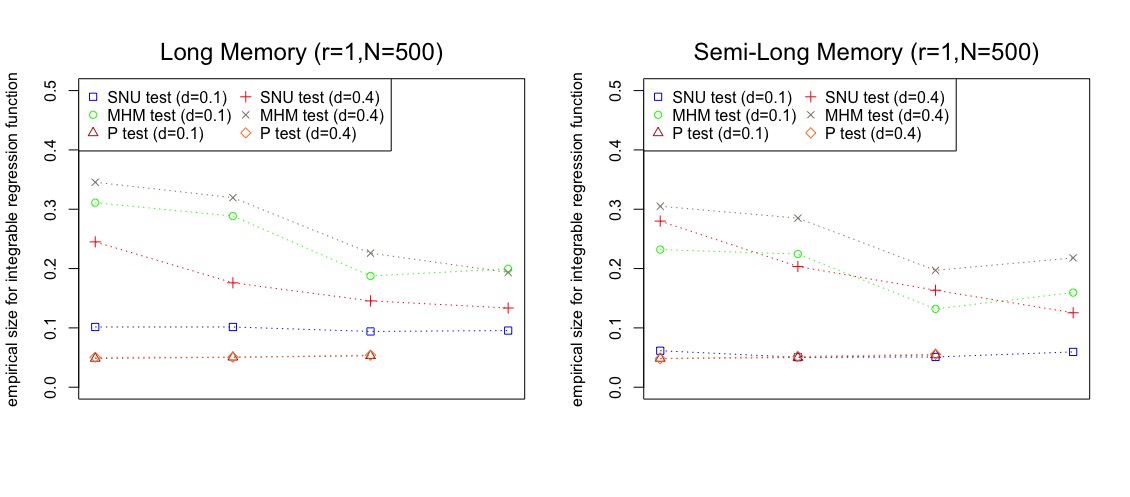}
 \end{tabular}
\caption{Empirical size of test statistics for integrable regression function with $r=1$ and different values of $N$. For SNU/MHM statistics, sizes are plotted for four   block lengths  $b=[c N^{1/2}]$, $c=0.5,1,2,4$ (left to right); for the P  statistic, three sizes are plotted left to right representing values of $\mathcal{L}=(6,12,18)$.}
\label{Fig.size_rp1_integrable}
\end{figure}

Secondly, we calculate the size of all test statistics discussed in this paper for nonintegrable regression function (\ref{eq:nonintegrablefs1}) and integrable regression function (\ref{eq:integrablefs1}) for different values of $r$ and sample size $N$. We again assume $h=N^{-1/3}$, and the tempering parameter $\lambda$ is $N^{-1/6}$ for the SLM process. Additionally, we allow the
residuals for the subsamples to follow the parametric forms. The results are graphically displayed in Figures \ref{Fig.size_rm1_nonintegrable}--\ref{Fig.size_rp1_integrable}. 

Note that the four values for the SNU and de-biased MHM test statistics from left to right correspond to the 4 block sizes in the paper, and the three values for the P test statistic from left to right  correspond to the 3 values of $\mathcal{L}=(6, 12, 18)$. As the sample size and block size increase, the values of size for all the test statistics decrease. The nominal size level of $\alpha$ is assumed to be $0.05$ for all the cases. We  investigated the power of test statistics, and they are relatively large (except for the de-biased MHM test statistic). Overall, the patterns of results are aligned with our other results given in both the main paper and the supplementary material.
We also examined the size and power of all test statistics for higher order values of $d$ such as $d \in (0.5,2)$ and in particular for the SLM structure. We found that the results consistently remain within a reasonable range but have not given here for the brevity.

\newpage
\clearpage

\newpage
\clearpage

\newpage
\clearpage

\newpage
\clearpage

\newpage
\clearpage

\newpage
\clearpage

\newpage
\clearpage

\newpage
\clearpage

\newpage
\clearpage

\newpage
\clearpage

\newpage
\clearpage

\newpage
\clearpage

\newpage
\clearpage

\subsection{B.6 \quad Simulated Examples for Minimal Volatility Rule}

\begin{figure}[ht]
\centering
\begin{tabular}{ cc }
\includegraphics[width=0.5\textwidth]{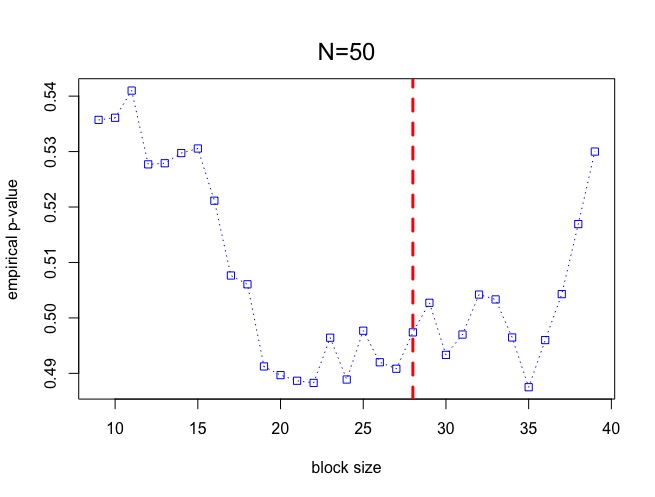} & \includegraphics[width=0.5\textwidth]{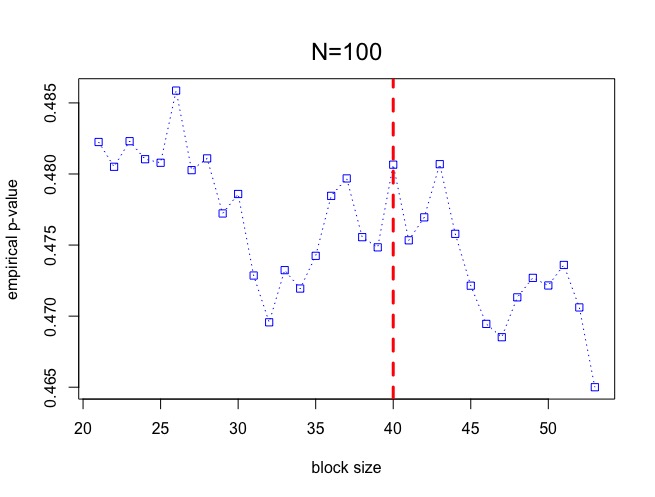} \\
\includegraphics[width=0.5\textwidth]{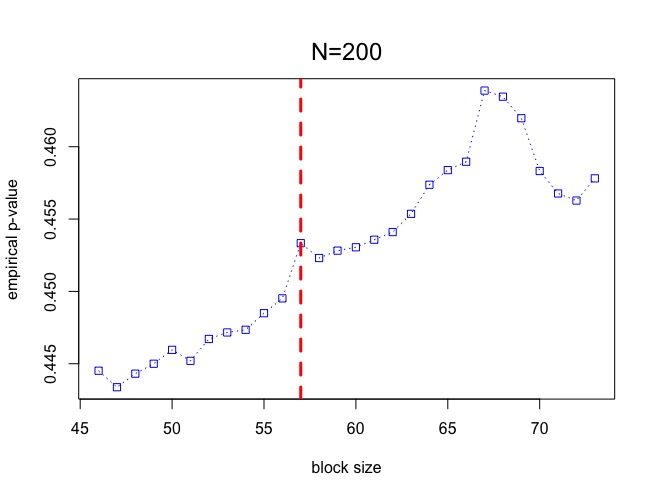} & \includegraphics[width=0.5\textwidth]{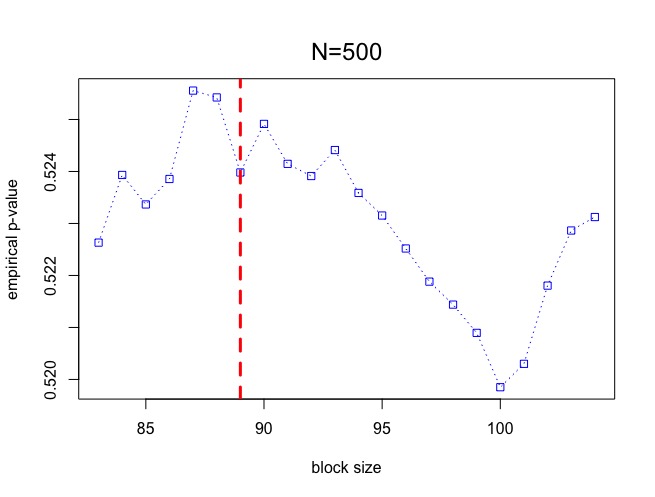} 
 \end{tabular}
 \caption{Empirical p-value versus block size when data is generated under the null hypothesis of $f(x)=x$. The vertical lines correspond to the block size of $[4N^{1/2}]$.}
\label{Fig.minvol_null}
\end{figure}

\begin{figure}[ht]
\centering
\begin{tabular}{ cc }
\includegraphics[width=0.5\textwidth]{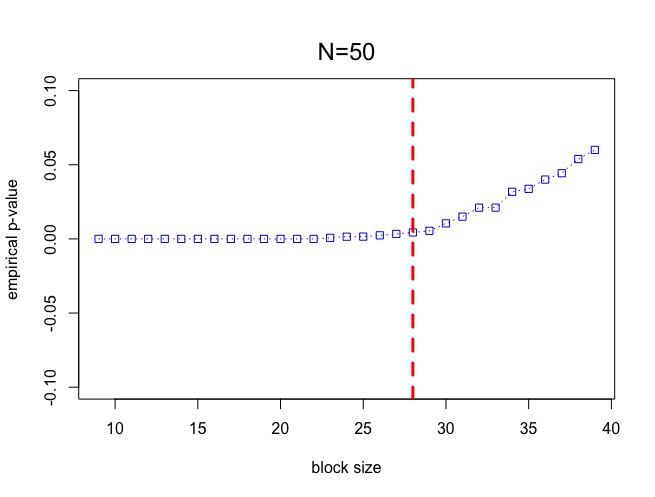} & \includegraphics[width=0.5\textwidth]{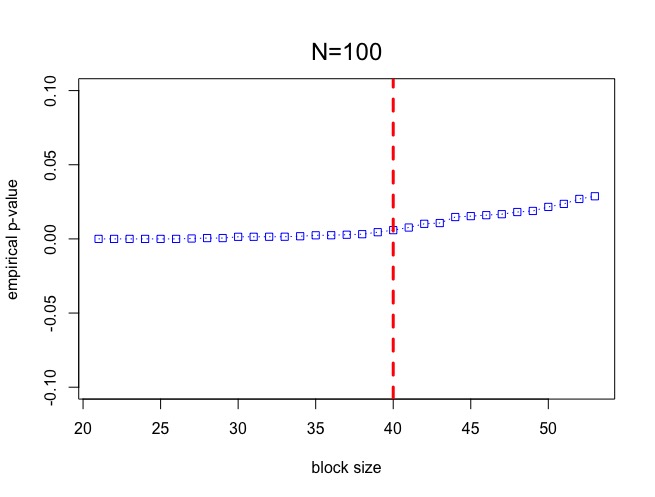} \\
\includegraphics[width=0.5\textwidth]{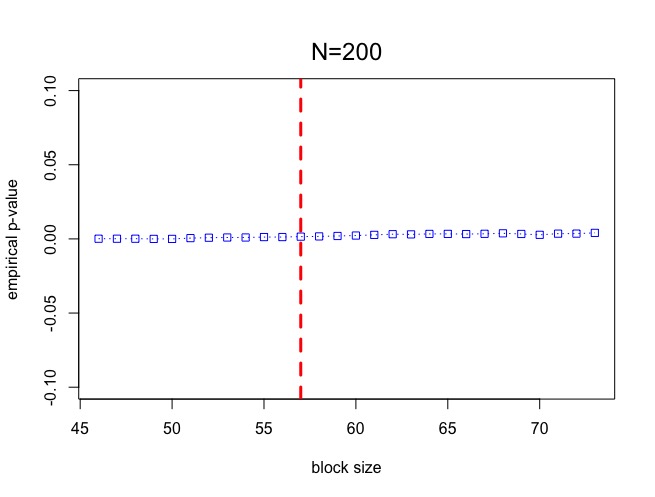} & \includegraphics[width=0.5\textwidth]{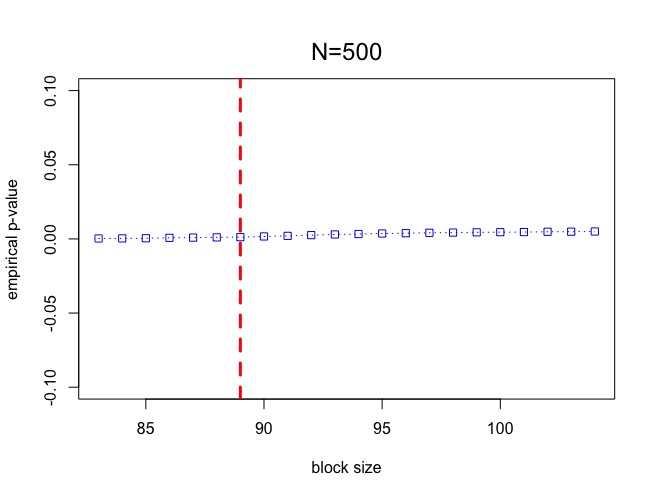} 
 \end{tabular}
 \caption{Empirical p-value versus block size when data is generated under the alternative hypothesis of $f(x)=x+\rho_N |x|^{\nu}$. The vertical lines correspond to the block size of $[4N^{1/2}]$.}
\label{Fig.minvol_alter}
\end{figure}

Here, we give some examples to illustrate the minimal volatility rule with simulated data sets. In the demonstration, we generated 50 time series with sample sizes $N=\{50, 100, 200, 500 \}$, $d=0.1$, and $r=0.5$ under the null ($f(x)=x$) and alternative ($f(x)=x+\rho_N |x|^{\nu}$) hypotheses for trend functions.  For each generated time series, we applied the subsampling method to produce an empirical p-value for the SNU test statistic at each block length in a range of consecutive block sizes.  The averages of these empirical p-values over 50 time series, per block length $b$, are depicted in Figure~\ref{Fig.minvol_null}  (null data generation) and Figure~\ref{Fig.minvol_alter} (alternative data generation).

We observe that when the null hypothesis holds, the empirical p-value curves from subsampling vary away from zero in Figure~\ref{Fig.minvol_null}, and also tend to exhibit local regions of stable fluctuation, where the latter are reasonably in line with a block length of $b=[4N^{1/2}]$ as a block  form  suggested from  simulated Tables \ref{Tab.size_nonintegrable} and \ref{Tab.power_nonintegrable}  in the manuscript when $\hat{f}(.)$ follows a parametric form. On the other hand, when the alternative is true, the empirical p-value curves from subsampling tend to be centered around zero in Figure~\ref{Fig.minvol_alter}; further, block lengths $b=[4N^{1/2}]$ again appear reasonable. These patterns  in subsampling p-values over block size  align with behavior   observed in our data application concerning the CKC hypothesis and support the use of block choice by minimal volatility.

\newpage
\clearpage

\section{C \quad Steps for the De-biased MHM Test Statistic} \label{sec:App3}
\renewcommand{\theequation}{C.\arabic{equation}}
\setcounter{equation}{0}
\renewcommand{\thelemma}{C.\arabic{lemma}}
\renewcommand{\thedefinition}{C.\arabic{definition}}
\setcounter{table}{0}
\renewcommand{\thetable}{C.\arabic{table}}
\setcounter{figure}{0}
\renewcommand{\thefigure}{C.\arabic{figure}}

Empirically, the MHM test statistic $\tau_N^{-1} T_N$ appears to exhibit some finite-sample bias or drift which can potentially be addressed with subsampling.  The idea is to formulate a de-biased version $\tau_N^{-1} T_N -\hat{B}_N$ of the test statistic, for some bias correction $\hat{B}_N$, and approximate its sampling distribution with a counterpart estimated by subsampling.  
\begin{itemize}
\item For a given block size $b$ and associated  
subsample test statistics $\{\tau_b^{-1} T_{i,b}\}_{i=1}^M$, a bias adjusted version of subsampling is readily given by the subsampling distribution from  $\{\tau_b^{-1} T_{i,b} - \hat{B}_b\}_{i=1}^M$, where $\hat{B}_b \equiv \sum_{i=1}^M \tau_b^{-1} T_{i,b}/M $ is the sample average of subsampling statistics.  This can be used to approximate the distribution of $\tau_N^{-1} T_N -\hat{B}_N$, where we may also use subsampling to approximate
the correction $\hat{B}_N$.

\item While the form of bias $B_N$ is unknown, this is linear on the log scale in log sample size, when the bias behaves as $B_N\sim C N^a$ on original scale for constants $C>0,a \in \mathbb{R}$.  

\item We  take two block lengths, say $b_1 = \lfloor 3 N^{1/2} \rfloor$ and  $b_2 = \lfloor 4 N^{1/2} \rfloor$, then compute subsample test statistic averages as 
$\hat{B}_{b_1}$ and $\hat{B}_{b_2}$ as bias estimates for these sample sizes, and determine a straight line through the points 
$(\log b_i, \log \hat{B}_{b_i})$, $i=1,2$. 

\item A prediction, say $\hat{P}_N$, from this line at $\log N$ estimates   
 $\log \hat{B}_N$ and we may define $\hat{B}_N = \exp[\hat{P}_N]$.  
 
 \item P-values can be approximated by the proportion of de-biased subsampling statistics $\{\tau_b^{-1} T_{i,b} - \hat{B}_b\}_{i=1}^M$, for a given block size $b$, which exceed   $\tau_N^{-1} T_{N} - \hat{B}_N$.   
\end{itemize}

\section{D \quad Practical Guidance on the Selection of Bandwidth} \label{sec:App4}
\renewcommand{\theequation}{D.\arabic{equation}}
\setcounter{equation}{0}
\renewcommand{\thelemma}{D.\arabic{lemma}}
\renewcommand{\thedefinition}{D.\arabic{definition}}
\setcounter{table}{0}
\renewcommand{\thetable}{D.\arabic{table}}
\setcounter{figure}{0}
\renewcommand{\thefigure}{D.\arabic{figure}}

We select a suitable bandwidth based on the \textit{cross-validation} method  described by \cite{hardlebook}, \cite{park1990comparison}, and \cite{rice1984bandwidth}. Each observation is temporarily removed from the data set, the regression is fitted using the remaining observations, and the deleted observation is predicted from that regression. The average of squared deviations between deleted observations and predictions is then used as a selection criterion for bandwidth.  Running the procedure for a gradient of bandwidth values allows us to select the bandwidth value that results in the minimum mean squared prediction error. 
In a standard kernel regression framework without outliers in the data, this cross-validation procedure has been shown to produce bandwidths that are asymptotically consistent; see for instance  \cite{hardle1992regression} and \cite{wong1983consistency}.

In particular, we use least squares leave-one-out cross-validation as our \textit{objective} function defined as follows
\begin{equation} \label{eq:LCV}
\text{LCV}(h):=\frac{1}{N} \sum_{k=1}^{N} (y_k- \hat{f}_{-k}(x_k))^2,
\end{equation}
where $\hat{f}_{-k}(x)$ is the estimate of $\hat{f}(x_k)$ computed from the data omitting the $k$-th observation $(x_k,y_k)$; i.e.

\begin{equation*}
\hat{f}_{-k}(x)= \Big\{ \sum_{j \neq k} K \Big[\frac{x_j-x}{h} \Big] \Big\}^{-1} \sum_{j \neq k} K \Big[ \frac{x_j-x}{h} \Big] y(x_j).
\end{equation*}
An optimal bandwidth according to this criterion can be chosen as $\hat{h}_{\text{LCV}} := \text{argmin}_{h>0} \text{LCV}(h)$.

For our CKC data set and for a range of bandwidths, the values of LCV (objective function) from (\ref{eq:LCV}) are calculated and plotted on the left hand side of  Figure \ref{Fig.countries} for both countries. 
The optimal bandwidths and their related LCV are given in Table \ref{Tab.LCV}. 

\begin{table}[ht]
\footnotesize
\caption{Optimal bandwidths and their LCV values for two countries.} \label{Tab.LCV}
\centering
\setlength{\tabcolsep}{10pt} 
\renewcommand{\arraystretch}{1.25} 
\begin{tabular}{ccc}
\toprule
Country & Optimal Bandwidth & Related LCV \\ 
\midrule
Spain & 0.151 & 0.005  \\
France &  0.073 & 0.013 \\
\bottomrule
\end{tabular}
\end{table}

\section{E \quad Simulated Example without an AR Structure for $u_k$'s} \label{sec:App4}
\renewcommand{\theequation}{E.\arabic{equation}}
\setcounter{equation}{0}
\renewcommand{\thelemma}{E.\arabic{lemma}}
\renewcommand{\thedefinition}{E.\arabic{definition}}
\setcounter{table}{0}
\renewcommand{\thetable}{E.\arabic{table}}
\setcounter{figure}{0}
\renewcommand{\thefigure}{E.\arabic{figure}}

\begin{figure}[ht]
\centering
\begin{tabular}{ c }
\includegraphics[width=0.6\textwidth]{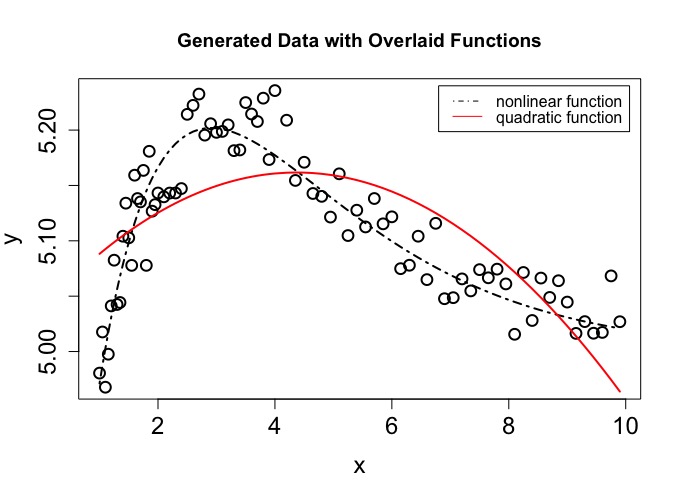} 
\end{tabular}
\caption{Generated data without an AR structure for the residuals along with the nonlinear and quadratic regression functions.}
\label{Fig.Data_woAR}
\end{figure}

To illustrate the potential difficulty in using the P test statistic that was identified in the carbon curve application for France, 
we simulated a data set where the error process $u_k$ is white noise but the regression function is highly nonlinear,
\begin{equation} \label{nonlinear_reg}
g(x_k; \alpha,\beta,\gamma)=\gamma + (\alpha x_k - \beta) \exp(- \alpha x_k) \quad \text{for} \	\quad k=1,...,80,
\end{equation}
where $(\alpha,\beta,\gamma)'=(0.55,0.60,5)$. The regressor $x_k$ is a range of constants (see Figure \ref{Fig.Data_woAR}). The error term $u_k$ simply follows a white noise process of Normal$(\mu=0,\sigma=0.025)$. See the available code ``\texttt{Data$\_$wo$\_$AR.gss}" for the details of generated data set. The data points are displayed in Figure \ref{Fig.Data_woAR}, where we overlay the nonlinear regression function from (\ref{nonlinear_reg}) and a quadratic regression function from (\ref{quad_noAR}).

Now, suppose that we did not know the data follows the regression function from equation (\ref{nonlinear_reg}), and we fit a quadratic regression function with the form of 
\begin{equation} \label{quad_noAR}
f(x_k)= \theta_0+\theta_1 x_k+\theta_2 x_k^2.
\end{equation}

\noindent By investigating the residuals from fitting model (\ref{quad_noAR}), these seemingly follow an AR(2) process with $p=2$. We apply the P test statistic, and the results are given in Table \ref{Tab.2020test_woAR}.

\begin{table}[ht]
\footnotesize
\caption{Results of P test statistic for the generated data.} \label{Tab.2020test_woAR}
\centering
\setlength{\tabcolsep}{10pt} 
\renewcommand{\arraystretch}{1.25} 
\begin{tabular}{ccc}
\toprule
$\mathcal{L}$ & $\tilde{U}_N(\mathcal{L})$ & p-value \\ 
\midrule
6 & 2.151 & 0.708 \\
12 & 2.986  & 0.982 \\
18 & 3.817 & 0.999 \\
\bottomrule
\end{tabular}
\end{table}

Despite the fact that the quadratic form is a quite poor description of the actual expectation function in the data, it is not rejected using the P test statistic for any value of $\mathcal{L}$. We conclude that the P test statistic is highly dependent on the assumed process structure for the $u$-terms, and one cannot necessarily depend on residuals from the hypothesized model to be tested in determining whether the error process has AR structure (and its order). In this example the p-values for the same null hypothesis of a quadratic regression curve are zero based on the SNU test statistic with subsampling and for all block sizes.

\section*{Bibliography}

H{\"a}rdle, W., Hall, P. and Marron, J. (1992) Regression smoothing parameters that are not 

far from their optimum. {\it Journal of the American Statistical Association} 87, 227–233.

\vspace{0.5cm}

\noindent H{\"a}rdle, W. and Marron, J. (1983) {\it Optimal bandwidth selection in nonparametric function}

{\it estimation}. Institute of Statistics Mimeo Series No. 1530. Univ. North Carolina, 

Chapel Hill.

\vspace{0.5cm}

\noindent Mosaferi, S. and Kaiser, M. S. (2022) Nonparametric cointegrating regression functions

with endogeneity and semi-long memory. {\it arXiv:2111.00972}.

\vspace{0.5cm}

\noindent Park, B. U. and Marron, J. S. (1990) Comparison of data-driven bandwidth selectors.

{\it Journal of the American Statistical Association} 85, 66–72.

\vspace{0.5cm}

\noindent Rice, J. (1984) Bandwidth choice for nonparametric regression. {\it The Annals of Statistics}

12, 1215–1230.

\vspace{0.5cm}

\noindent Wang, Q. and Phillips, P. C. B. (2016) Nonparametric cointegrating regression with endo-

geneity and long memory. {\it Econometric Theory} 32, 359–401.

\vspace{0.5cm}

\noindent Wong, W. H. (1983) On the consistency of cross-validation in kernel nonparametric regres-

sion. {\it The Annals of Statistics} 11, 1136–1141.

\end{document}